\documentclass[11pt]{article} 
\pdfoutput=1
\usepackage[left=1.1in,top=1.1in,bottom=1.1in,right=1.1in,nohead]{geometry}

\usepackage{lineno}
\usepackage{cite}
\usepackage{authblk}
\usepackage{units}
\usepackage{amsmath}
\usepackage{url}
\usepackage{graphicx}
\usepackage{epstopdf}   
\usepackage{textcomp}
\usepackage[font={small,it}]{caption}
\usepackage{lscape}
\usepackage{multirow} 
\usepackage{footnote}
\usepackage[usenames,dvipsnames,svgnames,table]{xcolor}
\usepackage{enumitem}

\newcommand{\latitude}{47.58N}
\newcommand{\longitude}{92.13W}
\newcommand{\attenlength}{\unit[40]{m}}

\newcommand{\tonethree}{\ensuremath{\theta_{13}}}

\newcommand{\dcp}{$\delta_{CP}$}

\newcommand{\neutrino}{\ensuremath{\nu}}
\newcommand{\numu}{\ensuremath{\nu_{\mu}}}
\newcommand{\nue}{\ensuremath{\nu_{e}}}
\newcommand{\nutau}{\ensuremath{\nu_{\tau}}}

\newcommand{\nubar}{\ensuremath{\overline{\neutrino}}}

\newcommand{\numue}     {\ensuremath{\nu_{\mu} \rightarrow \nu_{e}}}

\newcommand{\degree}{\ensuremath{^{\circ}}}
\newcommand{\dg}{\degree}

\newcommand{\CER}{Cherenkov}

\newcommand{\nova}{{\rm NO}v{\rm A}}
\newcommand{\chips}{CHIPS}
\newcommand{\globes}{GLoBES}
\newcommand{\inch}{\ensuremath{^{\prime\prime}}}

\begin{document}
\date{\today}
\title{\bf CHerenkov detectors In mine PitS (\chips{})\\ Letter of Intent to FNAL}
\renewcommand\Affilfont{\itshape\footnotesize}

\newcommand{\caltech}{1}
\newcommand{\cinci}{2}
\newcommand{\fnal}{3}
\newcommand{\iowa}{4}
\newcommand{\ucl}{5}
\newcommand{\manchester}{6}
\newcommand{\umn}{7}
\newcommand{\duluth}{8}
\newcommand{\pitt}{9}
\newcommand{\stan}{10}
\newcommand{\sussex}{11}
\newcommand{\ut}{12}
\newcommand{\tufts}{13}
\newcommand{\wm}{14}
\newcommand{\wisc}{15}

\author[\fnal]{P.~Adamson}
\author[\ut]{S.~V.~Cao}
\author[\tufts]{J.~A.~B.~Coelho}
\author[\iowa]{G.~S.~Davies}
\author[\manchester]{J.~J.~Evans}
\author[\manchester]{P.~Guzowski}
\author[\duluth]{A.~Habig}
\author[\sussex]{J.~Hartnell}
\author[\ucl]{A.~Holin}
\author[\ut]{J. Huang}
\author[\fnal]{A.~Kreymer}
\author[\wm]{M.~Kordosky}
\author[\ut]{K.~Lang}
\author[\umn]{M.~L.~Marshak}
\author[\ut]{R.~Mehdiyev}
\author[\umn]{J.~Meier}
\author[\umn]{W.~Miller}
\author[\pitt]{D.~Naples}
\author[\wm]{J.~K.~Nelson}
\author[\ucl]{R.~J.~Nichol}
\author[\pitt]{V.~Paolone}
\author[\caltech]{R.~B.~Patterson}
\author[\umn]{G.~Pawloski}
\author[\ucl]{A.~Perch}
\author[\ucl]{M.~Pf\"utzner}
\author[\ut]{M.~Proga}
\author[\ucl]{A.~Radovic}
\author[\iowa]{M.~C.~Sanchez}
\author[\umn]{S.~Schreiner}
\author[\manchester]{S.~S\"oldner-Rembold}
\author[\cinci]{A.~Sousa}
\author[\ucl]{J.~Thomas}
\author[\wm]{P.~Vahle}
\author[\wisc]{C.~Wendt}
\author[\ucl]{L.~H.~Whitehead}
\author[\stan]{S.~Wojcicki}

\affil[\caltech]{Lauritsen Laboratory, California Institute of Technology, Pasadena, CA 91125, USA}
\affil[\cinci]{Department of Physics, University of Cincinnati, Cincinnati, OH 45221, USA}
\affil[\fnal]{Fermi National Accelerator Laboratory, Batavia, IL 60510, USA}
\affil[\iowa]{Department of Physics and Astronomy, Iowa State University, Ames, IA 50011, USA}
\affil[\ucl]{Department of Physics and Astronomy, UCL, Gower Street, London WC1E 6BT, UK}
\affil[\manchester]{School of Physics and Astronomy, University of Manchester, Oxford Road, Manchester M13 9PL, UK}
\affil[\umn]{University of Minnesota, Minneapolis, MN 55455, USA}
\affil[\duluth]{Department of Physics, University of Minnesota -- Duluth, Duluth, MN 55812, USA}
\affil[\pitt]{Department of Physics and Astronomy, University of Pittsburgh, Pittsburgh, PA 15260, USA}
\affil[\stan]{Department of Physics, Stanford University, Stanford, CA 94305, USA}
\affil[\sussex]{Department of Physics and Astronomy, University of Sussex, Falmer, Brighton BN1 9QH, UK}
\affil[\ut]{Department of Physics, University of Texas at Austin, 1 University Station C1600, Austin, TX 78712, USA}
\affil[\tufts]{Physics Department, Tufts University, Medford, MA 02155, USA}
\affil[\wm]{Department of Physics, College of William \& Mary, Williamsburg, VA 23187, USA}
\affil[\wisc]{Physics Department, University of Wisconsin, Madison, WI 53706, USA}

\maketitle
\clearpage
\tableofcontents
\clearpage
 
\section{Introduction}
\subsection{Motivation}
Recent observations of a large \tonethree{} mixing angle have refocussed the next generation of long baseline experiments towards resolving the mass hierarchy, determining the octant of $\theta_{23}$, and measuring \dcp{}. Degeneracies among the remaining oscillation parameters mean that, unless nature has chosen extremely favorable values, \nova{} may not be able to satisfactorily measure all the remaining unknowns.  Other planned experiments are unlikely to significantly improve our knowledge of these unknowns until 2023 when the first LBNE experiment, the \unit[10]{kton} LAr detector, is planned to be operational and taking initial beam data.  An additional 10 years of data is required to fully realize the projected sensitivity.  This leaves a long drought of physics output from the Fermilab long-baseline neutrino program. 

For the U.S. long-baseline neutrino program to continue to be an attractive and vibrant endeavor, it is essential to have a phased program that can achieve new physics results on both short and long time scales.  To achieve that aim, we advocate for enhanced exploitation of the NuMI beam, as part of a new plan to develop an experimental long-baseline neutrino program that can lead the world in delivering new neutrino insights.  Fermilab's NuMI beam line has been the workhorse of the U.S. neutrino program over the past seven years.  After upgrades, NuMI will run at double its original intensity and will be the most powerful neutrino beam in the world.  With its flexible running configurations and its suite of near detectors, the beam will be the best understood neutrino beam ever constructed, and it is a resource that creates unprecedented opportunities.  As an initial stage of the new long-baseline program, detectors could be developed and run in the NuMI beam, delivering world class constraints on $\delta_{CP}$, even while the new LBNE beam line is being built. 
 
\subsection{\chips{} Concept} 
This Letter of Intent outlines a proposal to build a large, yet cost-effective, \unit[100]{kton} fiducial mass water Cherenkov detector that will initially run in the NuMI beam line.  The \chips{} detector ({\bf CH}erenkov detector {\bf I}n Mine {\bf P}it{\bf S}) will be deployed in a flooded mine pit, removing the necessity and expense of a substantial external structure capable of supporting a large detector mass.  There are a number of mine pits in northern Minnesota along the NuMI beam that could be used to deploy such a detector.  In particular, the Wentworth Pit 2W is \unit[7]{mrad} away from the central axis of the beam, a position which optimizes rate and background rejection.  The pit is also one of the deepest in the area, allowing for a water overburden of several tens of meters.   The detector is designed so that it can be moved to a mine pit in the LBNE beam line once that becomes operational.  

While one can not achieve the ideal baseline to measure the mass hierarchy in the NuMI beam, studies performed by the eNuMI working group~\cite{enumiweb} show that detectors in the NuMI beamline can constrain the value of $\delta_{CP}$.  The \chips{} experiment will probe $\delta_{CP}$ by measuring electron neutrino appearance in the NuMI muon neutrino beam.  Assuming the nominal beam power that NuMI will achieve in the \nova{} era, the nominal \nova{} beam configuration, and a \unit[100]{kton} fiducial mass \chips{} detector deployed in the Wentworth Pit, on the order of 340 (190) \nue{}-CC events would be observed in the normal (inverted) hierarchy above a background of approximately 640 events in a three year run with the beam in neutrino mode.  In antineutrino mode, about 200 (150) \nue{}-CC events should be observed on a background of about 350 events.  With these event rates, the combination of \chips{}, NOvA and T2K provide an error on \dcp{} better than \unit[25]{\degree{}} for all values of \dcp{}, assuming the mass hierarchy and other degeneracies are resolved.  Being close to the beam axis, \chips{} sees a relatively wide energy distribution and high flux, and thus provides complementary information to the off-axis experiments.  Combining \chips{} data with the off-axis results can further constrain $\delta_{CP}$, improve the significance of a discovery of CP violation in the neutrino sector, and help resolve ambiguities in the mixing parameters.  

Even a modest target mass (\unit[10]{kton}) can improve the resolution in $\delta_{CP}$ over \nova{}, indicating a prototype detector in the NuMI beam can deliver meaningful contributions to neutrino physics on a short time scale.  This document also describes an R\&D plan to prove the CHIPS concept and to study ways to reduce the cost per kiloton of building such a detector, at the same time delivering additive results on \dcp{}.  This R\&D effort will encourage a new, vibrant detector development community centered on the FNAL neutrino program.  A nationwide consortium of laboratory and university groups are already collaborating to focus on the development of new and innovative photodetector technologies~\cite{pdwork}.  U.S. companies are beginning to develop other photodetector technologies, providing competition that will drive down the cost of instrumentation.  Leveraging these efforts will enable U.S. leadership in the construction of megaton size neutrino detectors.

\section{Physics Reach}
\label{physsim}
\begin{figure}[h]
\begin{center}
\begin{minipage}{0.49\textwidth}
\includegraphics[width=\textwidth]{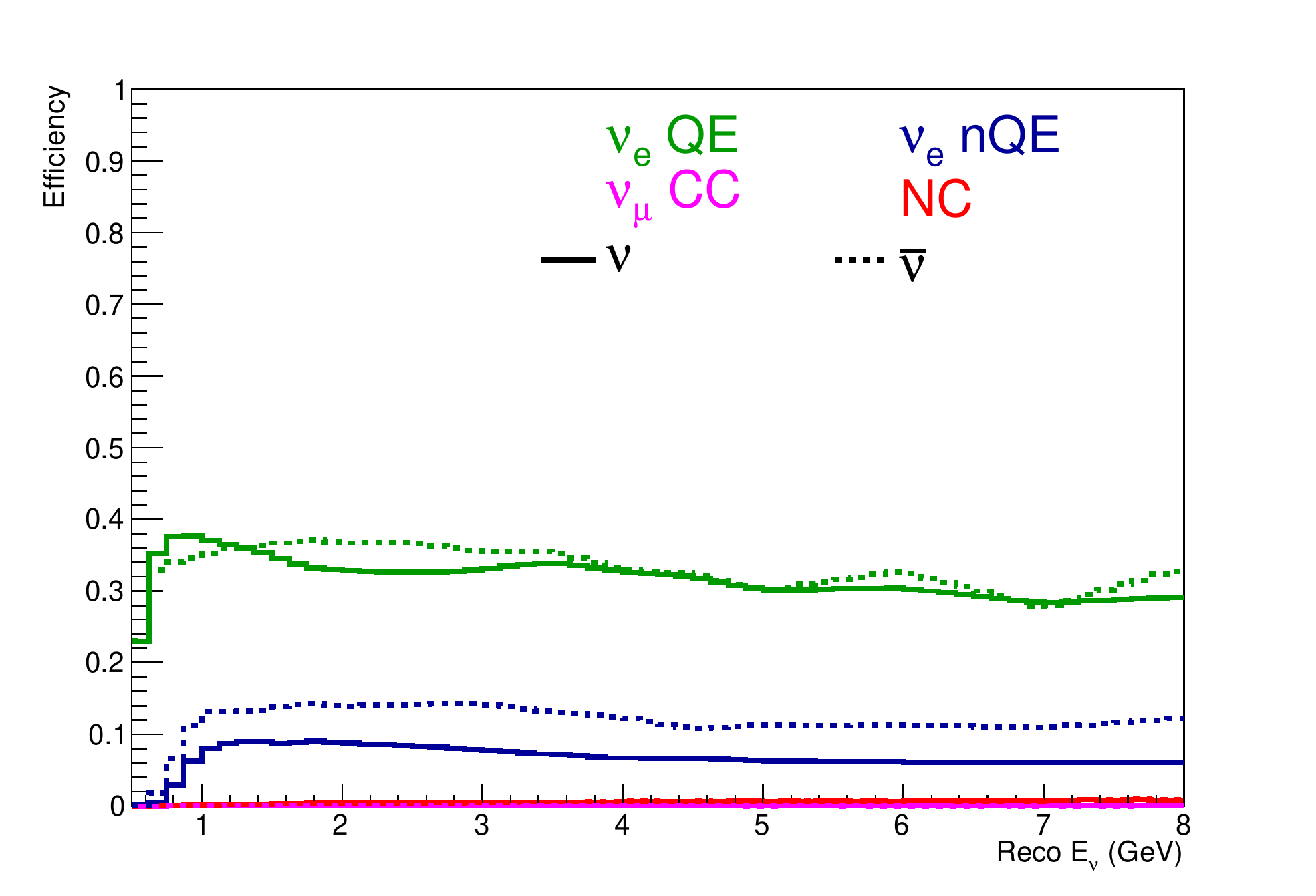}
\end{minipage}
\begin{minipage}{0.49\textwidth}
\includegraphics[width=\textwidth]{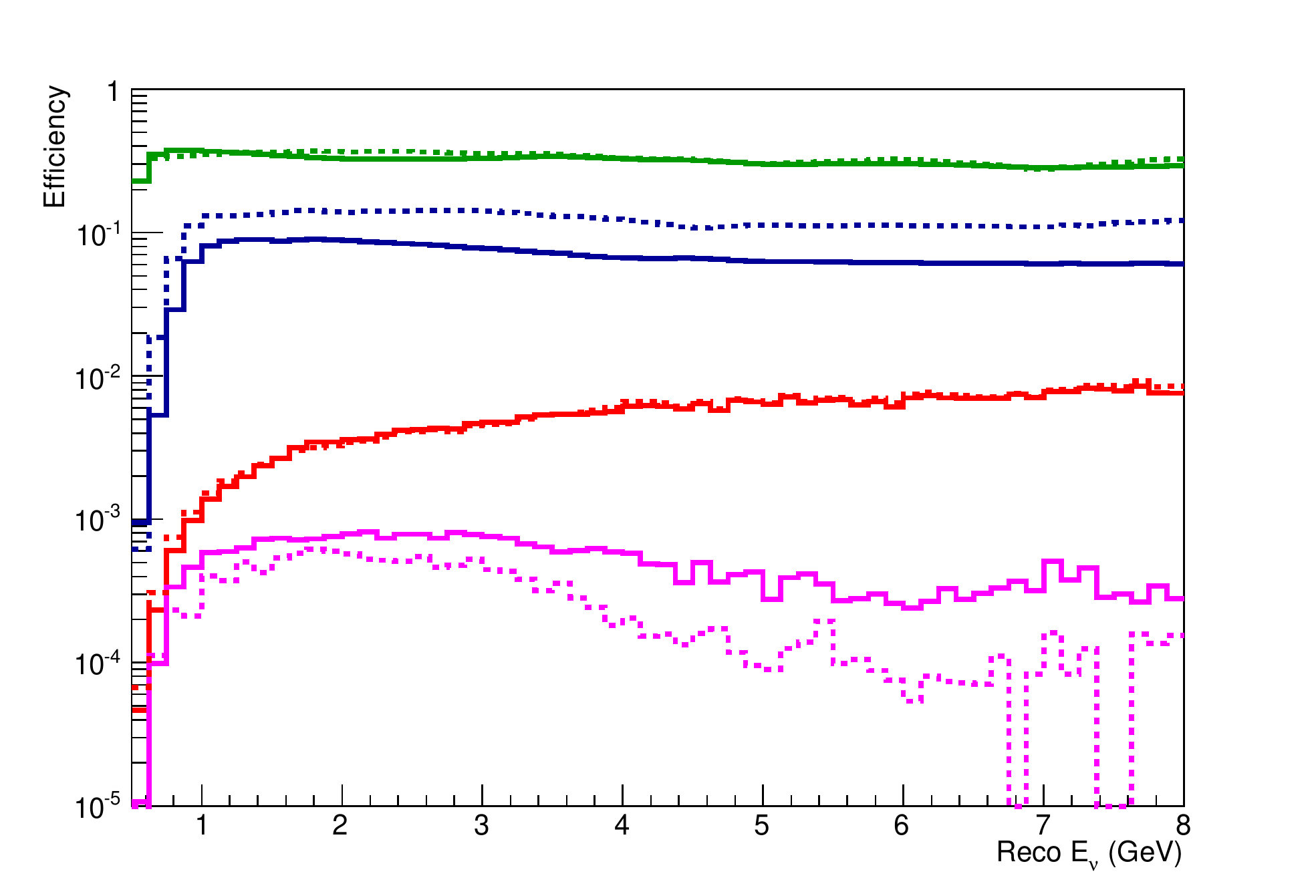}
\end{minipage}
\end{center}
\caption{Final assumed efficiency for each event type as a function of reconstructed energy in linear (Left) and log (Right) scales.}
\label{efficiencies} 
\end{figure}

\begin{figure}[htbp]
\begin{center}
\begin{minipage}{0.49\textwidth}
\includegraphics[width=\textwidth]{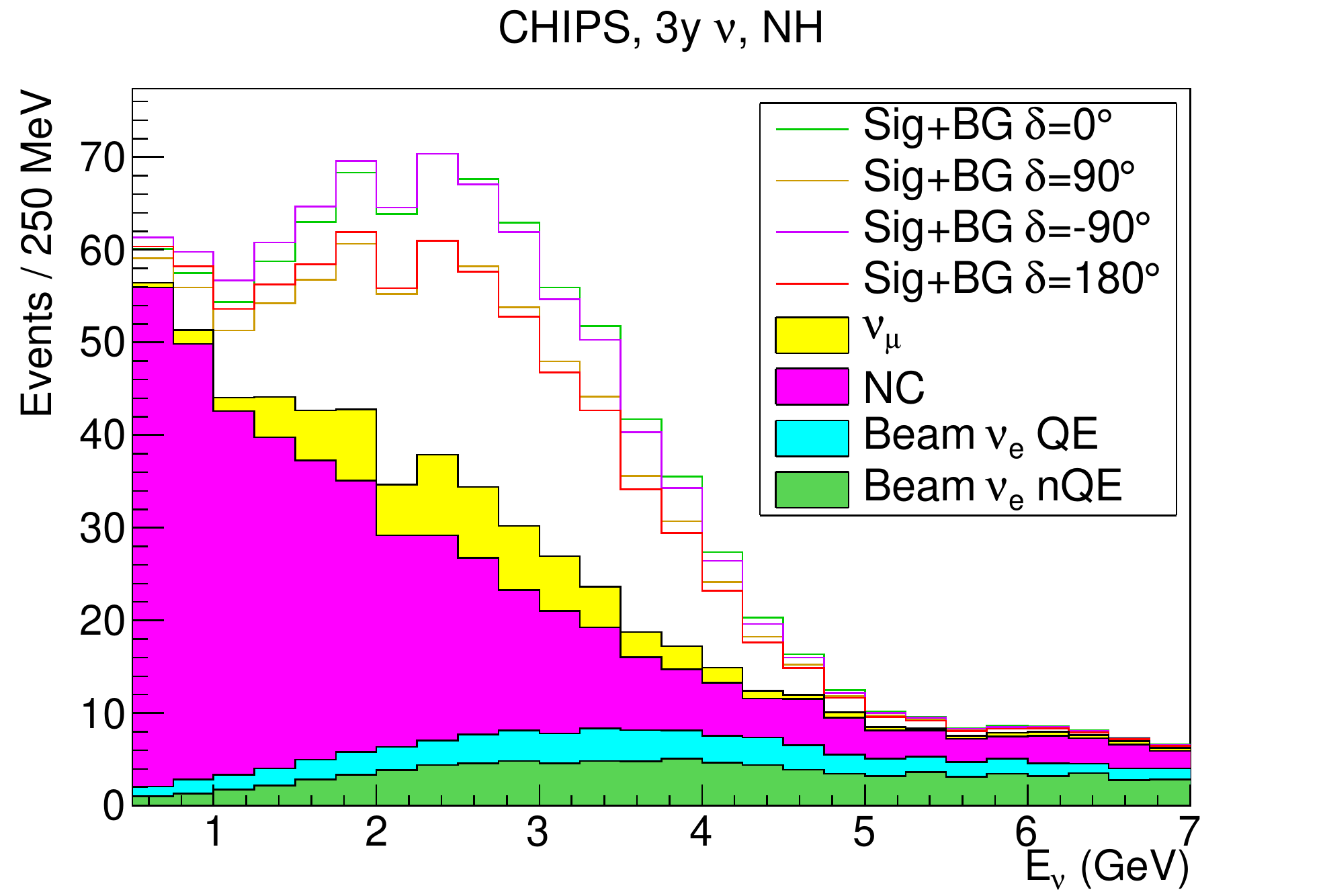}
\end{minipage}
\begin{minipage}{0.49\textwidth}
\includegraphics[width=\textwidth]{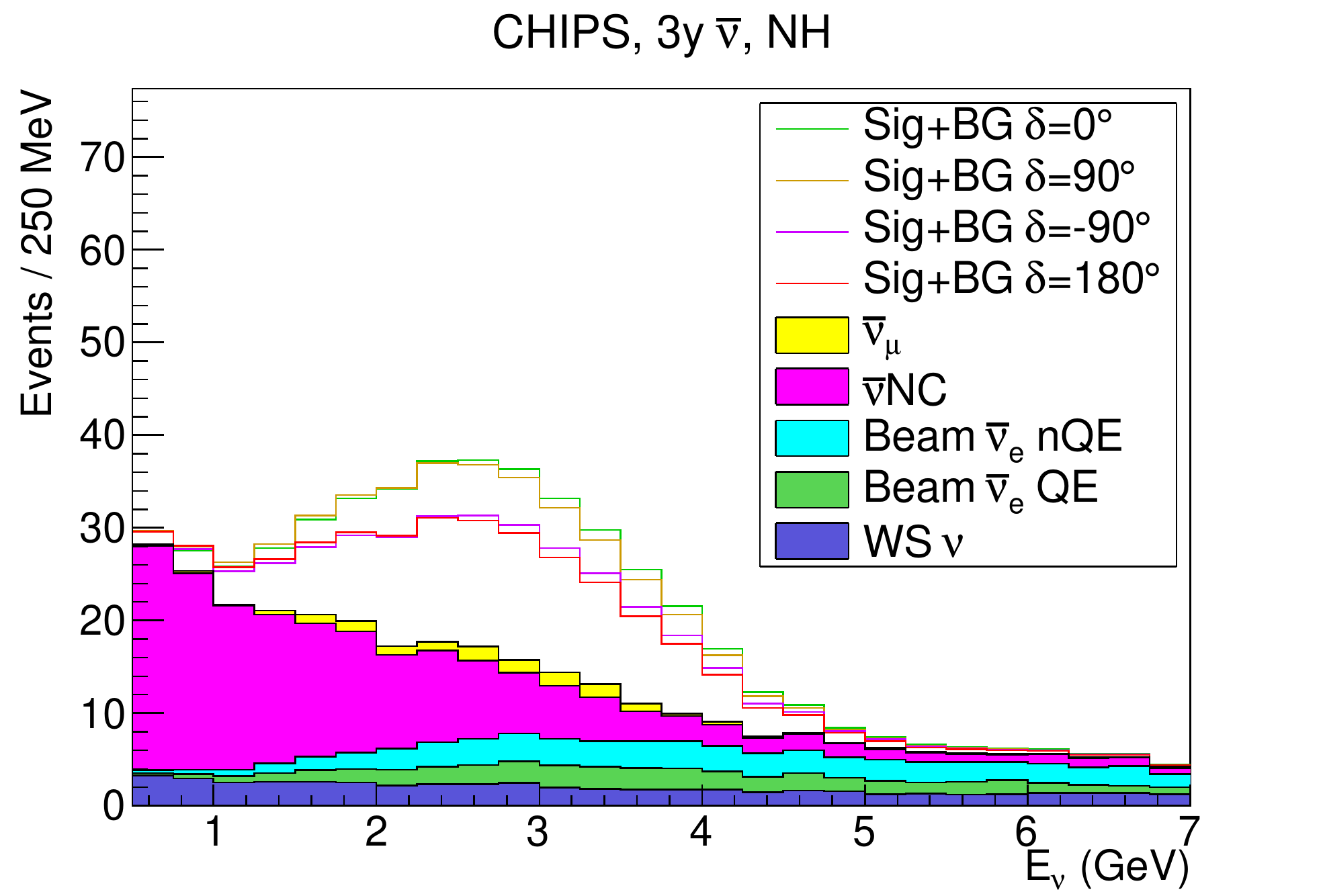}
\end{minipage}\\
\begin{minipage}{0.49\textwidth}
\includegraphics[width=\textwidth]{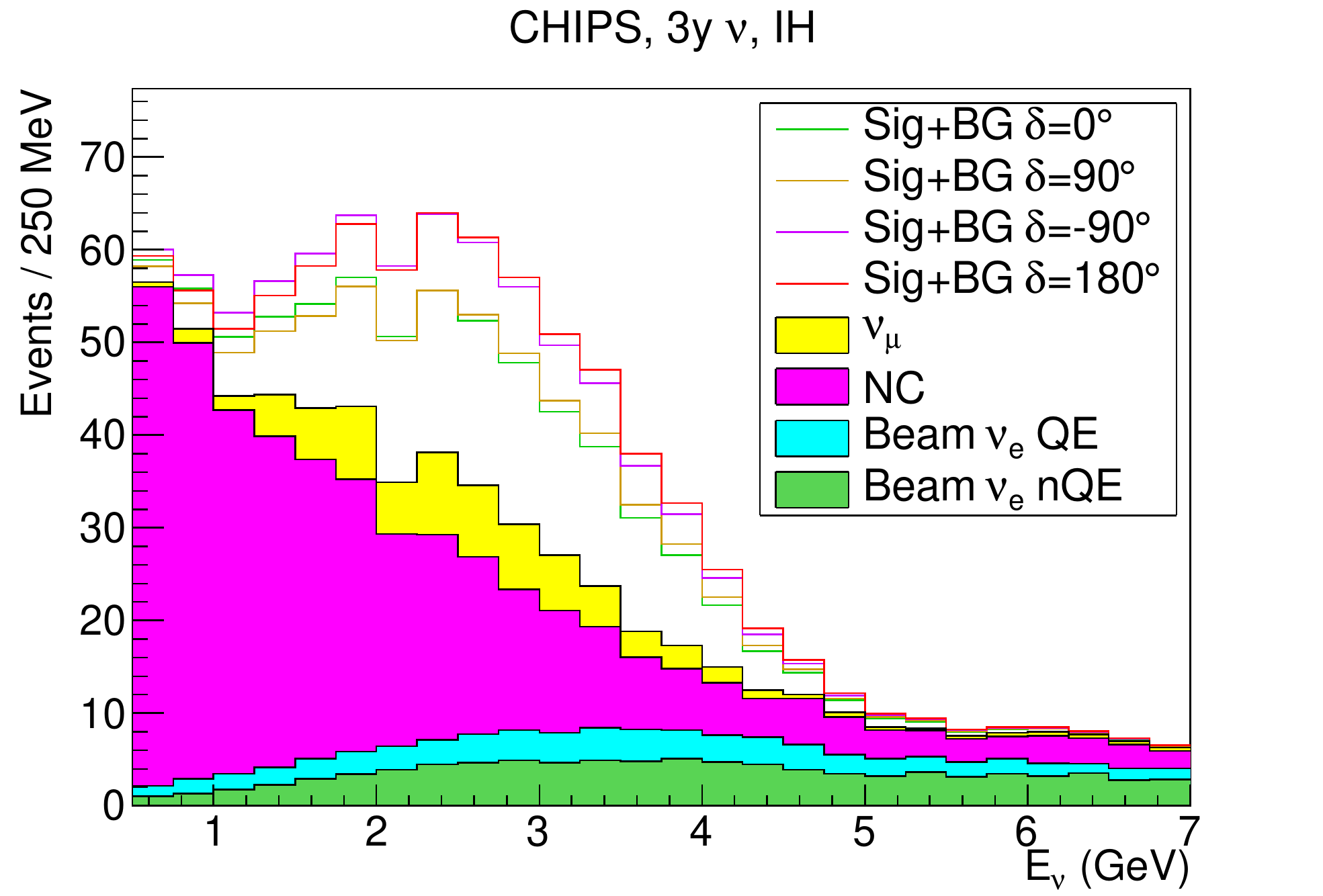}
\end{minipage}
\begin{minipage}{0.49\textwidth}
\includegraphics[width=\textwidth]{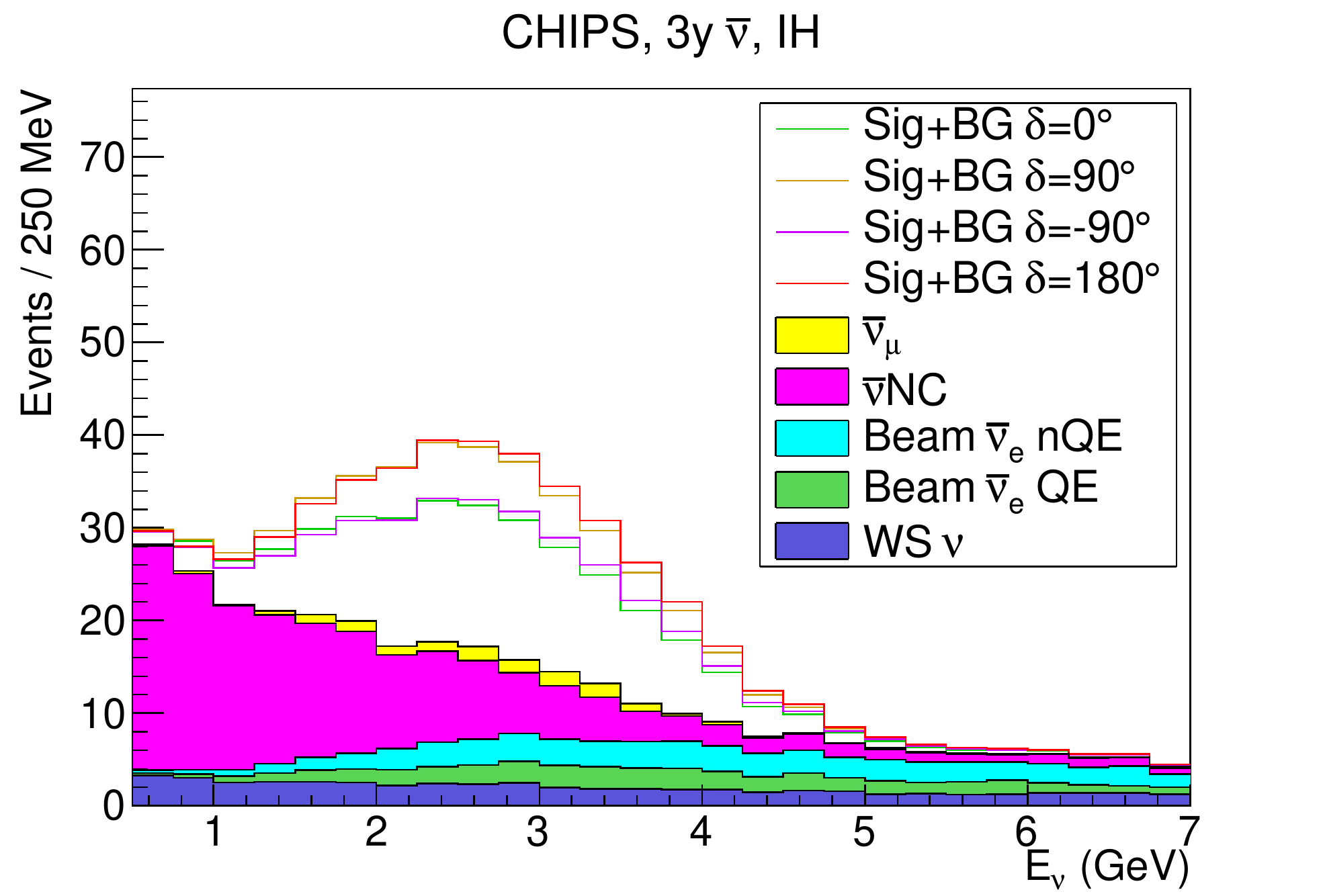}
\end{minipage}
\caption{Event rates when running \chips{} in 3 years of neutrino beam (left) and three years of antineutrino beam (right) for Normal Hierarchy (top) and Inverted Hierarchy (bottom).  Beam \nue{} events are divided into quasielastic (QE) and non-quasielastic (nQE) samples.  The wrong sign (WS) neutrino sample is separated in the antineutrino beam plots.}
\label{fig:evrates}
\end{center}
\end{figure}

The physics capabilities of \chips{} have been studied using \globes{}~\cite{GLOBES}.  The nominal experimental setup assumes a \unit[100]{kton} fiducial mass detector with an exposure of \unit[$6\times 10^{20}$]{POT/year}, which is the NOvA expected yearly exposure.  The medium energy (ME) flux described in Section~\ref{fluxsec} is used as input to these simulations.  Cross sections used are standard to \globes{}.  Three flavor neutrino oscillations are incorporated into event rate predictions; the known oscillation parameters are fixed at the values given in reference~\cite{fogli} and are summarized in Table~\ref{table:osc-params}.  Selection efficiencies for each event type as a function of energy included in \globes{} are based on Super-K experience, using a 20\% photodetector coverage.  Two levels of selection are applied.  First, a pre-smearing efficiency (vs. true energy) is applied, which represents a cut based on what fraction of each event type looks like a single electron.  True energy is then converted to a reconstructed energy using migration matrices, again from Super-K.  Then a post-smearing efficiency (vs. reconstructed energy) is applied, based on the Super-K log-likelihood cut.  The resulting efficiency for each event type as a function of reconstructed energy are shown in Figure~\ref{efficiencies}.  The energy distribution of each event type, for each mass hierarchy is shown in Figure~\ref{fig:evrates}.  Integrated event counts are given in Table~\ref{tab:evtrates}. 

\begin{table}[htbp]
\begin{center}
\begin{tabular}{|rl|}
\hline 
Parameter&Value\\
\hline
$\sin^2 \theta_{12}$ & $0.312$ \\
$\sin^2 2\theta_{13}$ & $0.096$ \\
$\sin^2 \theta_{23}$ & $0.39$ \\
$\theta_{23}$ octant & $\theta_{23} < {\pi/4}$ \\ 
$\Delta m^2_{21}$ & $7.6\times10^{-5}$ eV$^2$ \\
$\Delta m^2_{31}$ (NH) & $2.45\times10^{-3}$ eV$^2$ \\
$\Delta m^2_{31}$ (IH) & $-2.31\times10^{-3}$ eV$^2$ \\
\hline
\end{tabular}
\caption{Neutrino oscillation parameters used in this study. Taken from~\cite{fogli}.}
\label{table:osc-params}
\end{center}
\end{table}

\begin{table}[htbp]
\begin{center}
\begin{tabular}{|lrrrr|}
\hline
Event Type&\multicolumn{2}{c}{$\nu$ Mode}&\multicolumn{2}{c|}{\nubar{} Mode}\\
&NH&IH&NH&IH\\
\hline
Appeared \nue{}&341&186&199&154\\
\numu{}-CC&72&74&13&13\\
NC&401&401&175&175\\
Beam \nue{}&162&163&100&99\\
Wrong Sign $\nu$&&&54&54\\
\hline
\end{tabular}
\caption{Number of selected events in \unit[100]{kton} fiducial mass \chips{} detector after 3 years in each mode, for both the normal hierarchy (NH) and the inverted hierarchy (IH).}
\label{tab:evtrates}
\end{center}
\end{table}

Events from $\nu_{\tau}$ appearance are not included in the \globes{} simulations.  Independent calculations indicate there will be 3.7 $\nu_{\tau}$-CC interactions per kton per year, integrated over all energies.  An estimate of how many of these events would pass the $\nue{}$ selection was made using the selection efficiencies from \globes{} for each tau decay mode: the \nue{} selection efficiency is applied to the electron decay mode, the \numu{} selection efficiency is applied to the muon decay mode, and the NC selection efficiency is applied to the hadronic decay mode.  
The event counts from each decay mode are weighted by the branching fractions and summed, to produce a prediction of 0.35 additional background events per kton per year from \nutau{} appearance.  This estimate will be further refined once a full simulation and event reconstruction suite is available.

\begin{figure}[p]
\begin{center}
\begin{minipage}{0.49\textwidth}
\includegraphics[width=\textwidth]{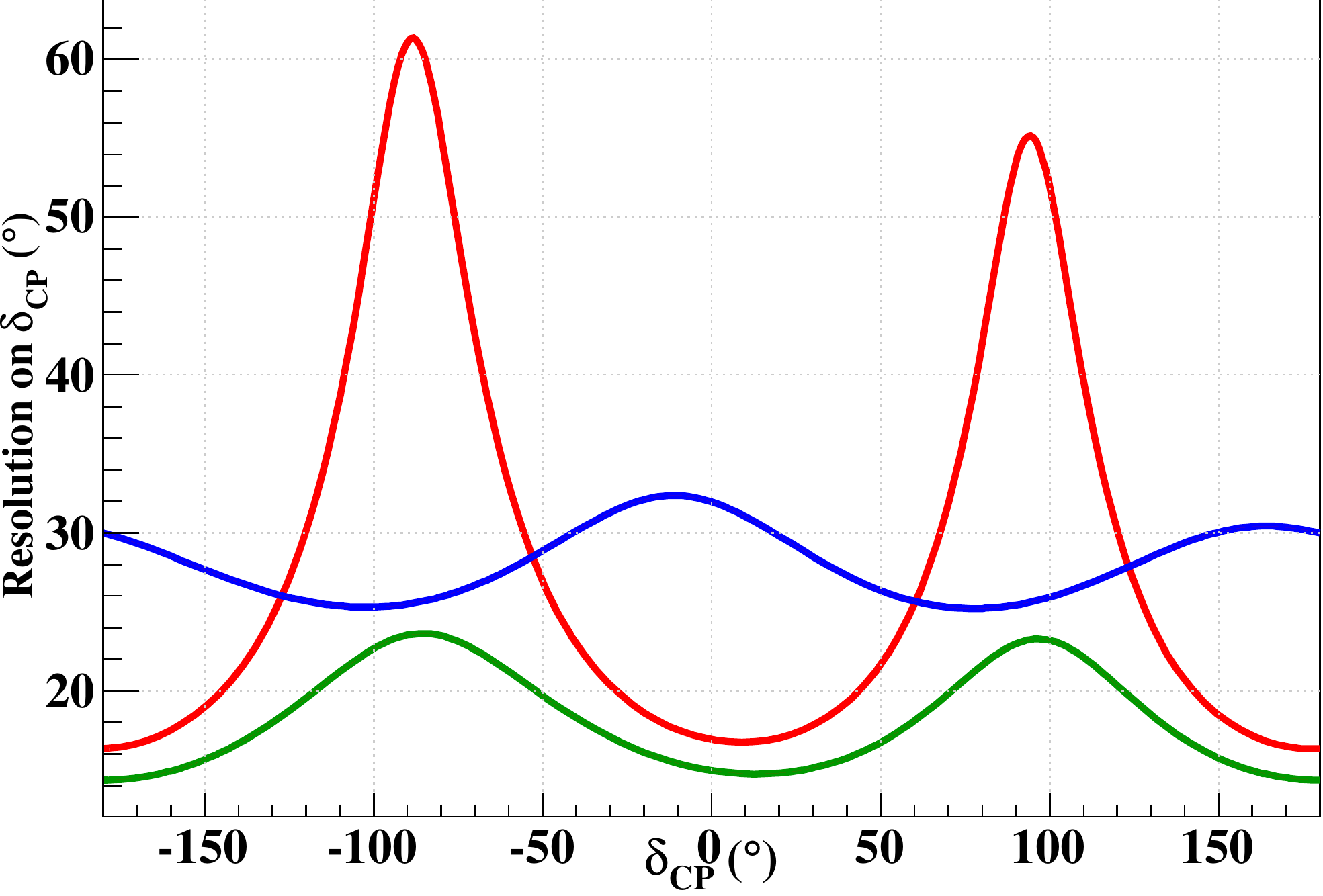}
\end{minipage}
\begin{minipage}{0.49\textwidth} 
\includegraphics[width=\textwidth]{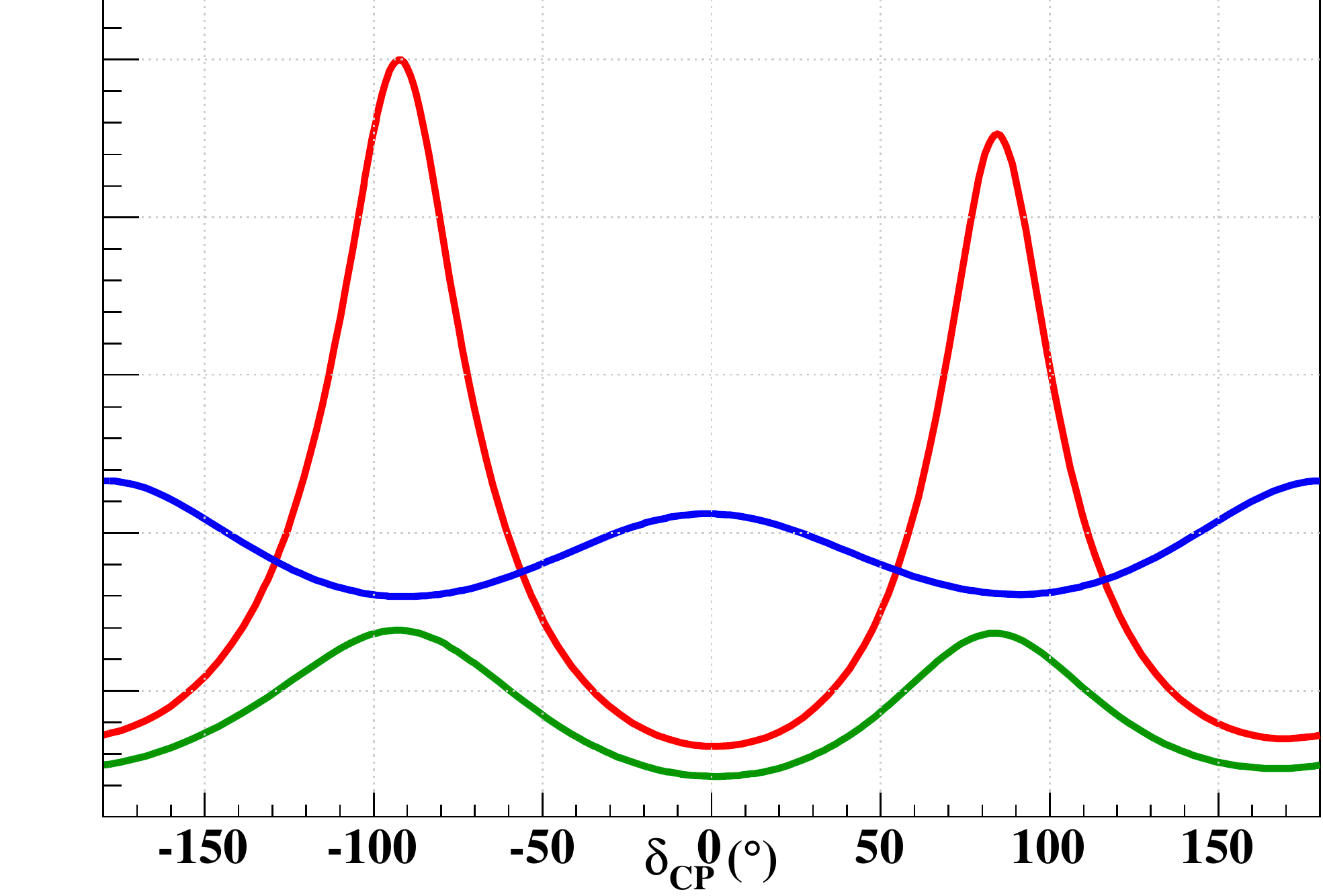}
\end{minipage}\\
\begin{minipage}{0.49\textwidth} 
\includegraphics[width=\textwidth]{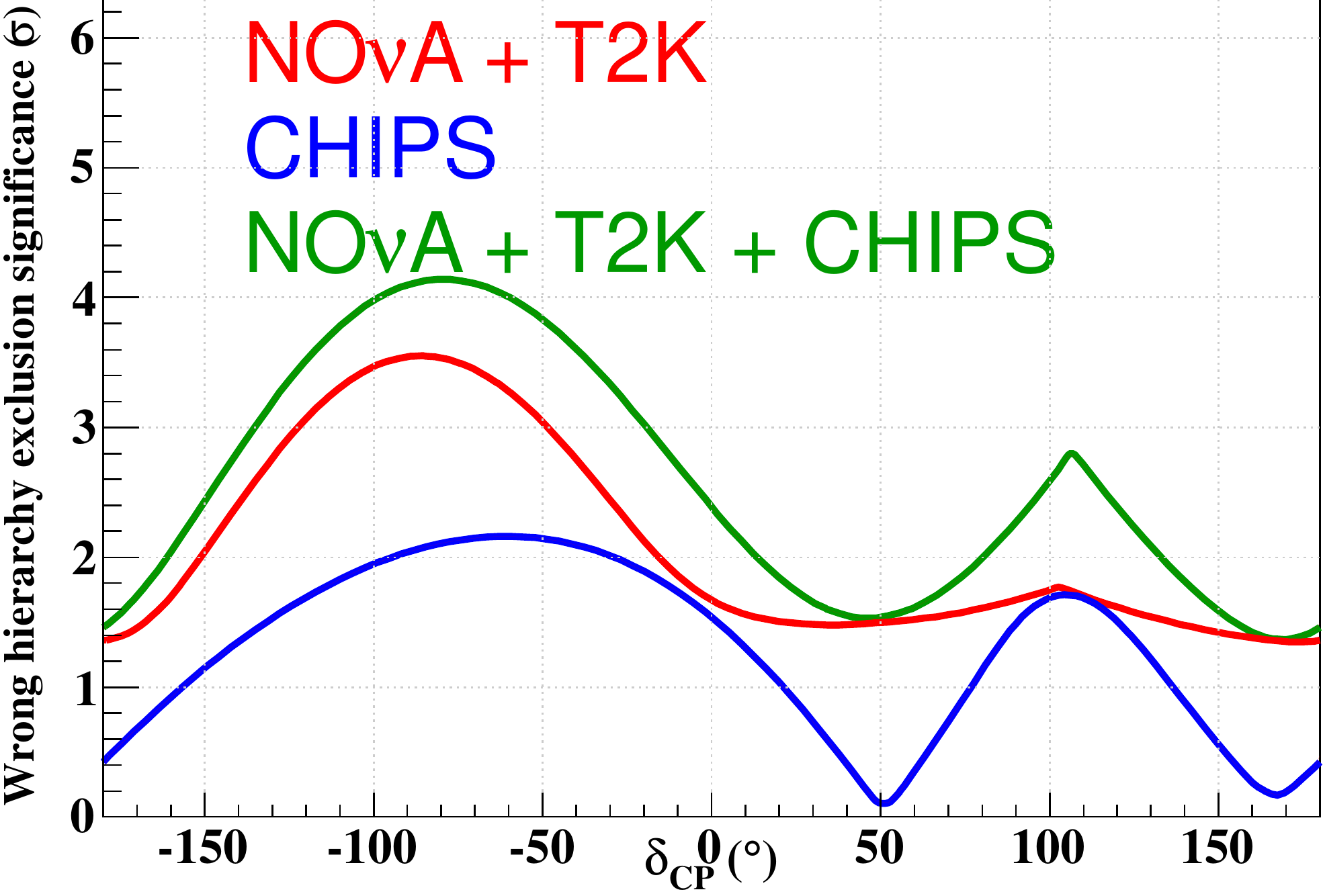}
\end{minipage}
\begin{minipage}{0.49\textwidth}
\includegraphics[width=\textwidth]{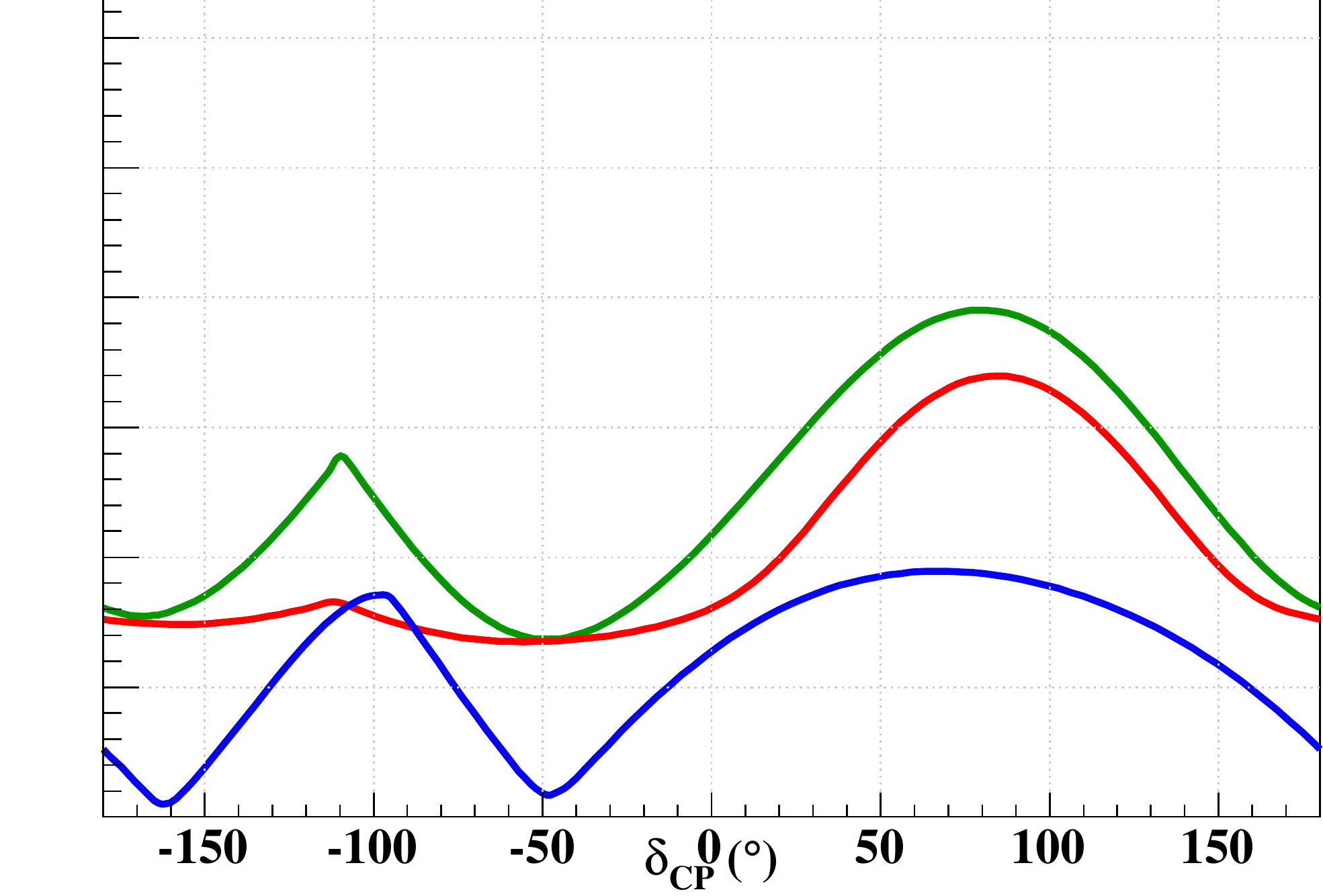}
\end{minipage}\\ 
\begin{minipage}{0.49\textwidth}
\includegraphics[width=\textwidth]{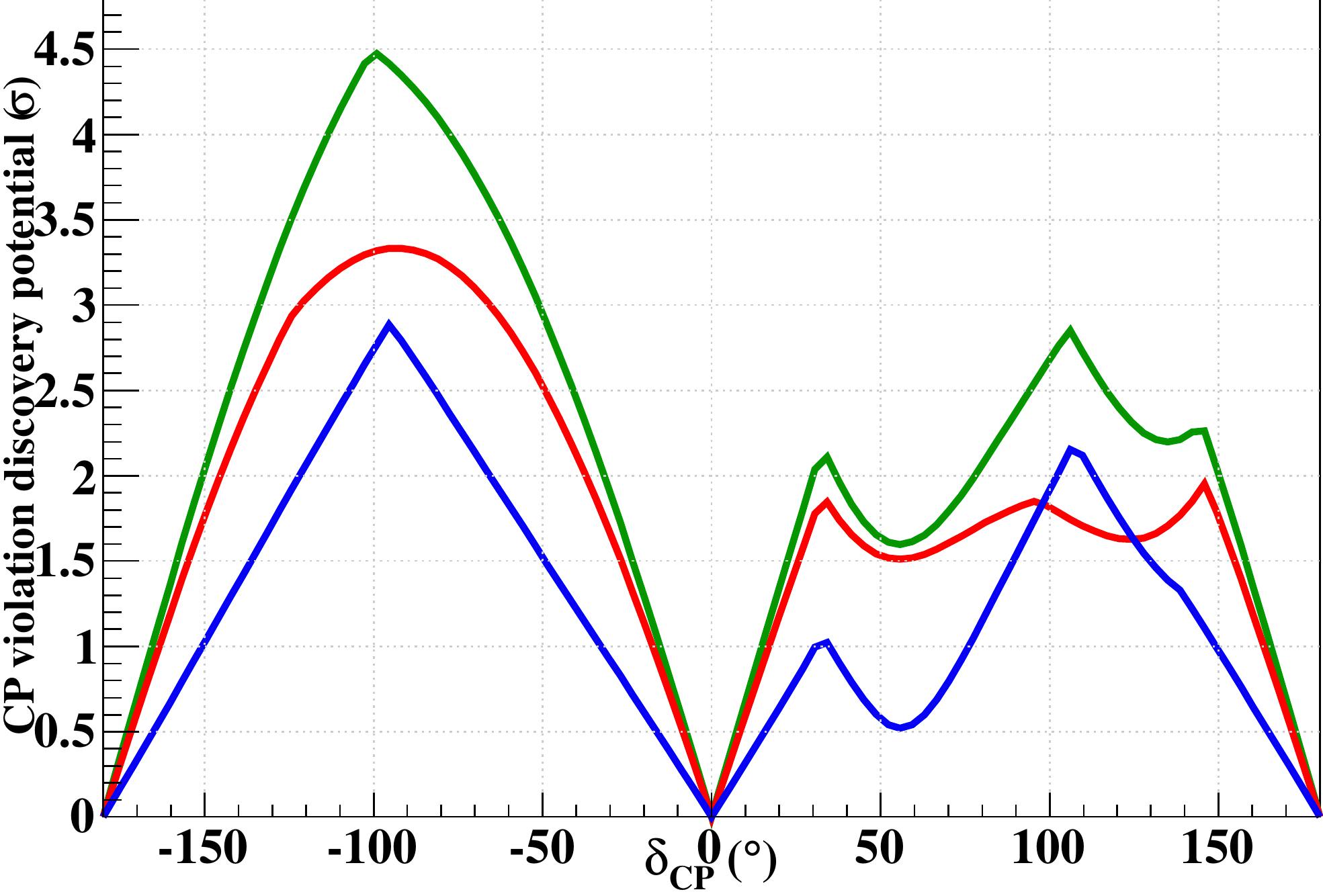}
\end{minipage}
\begin{minipage}{0.49\textwidth}
\includegraphics[width=\textwidth]{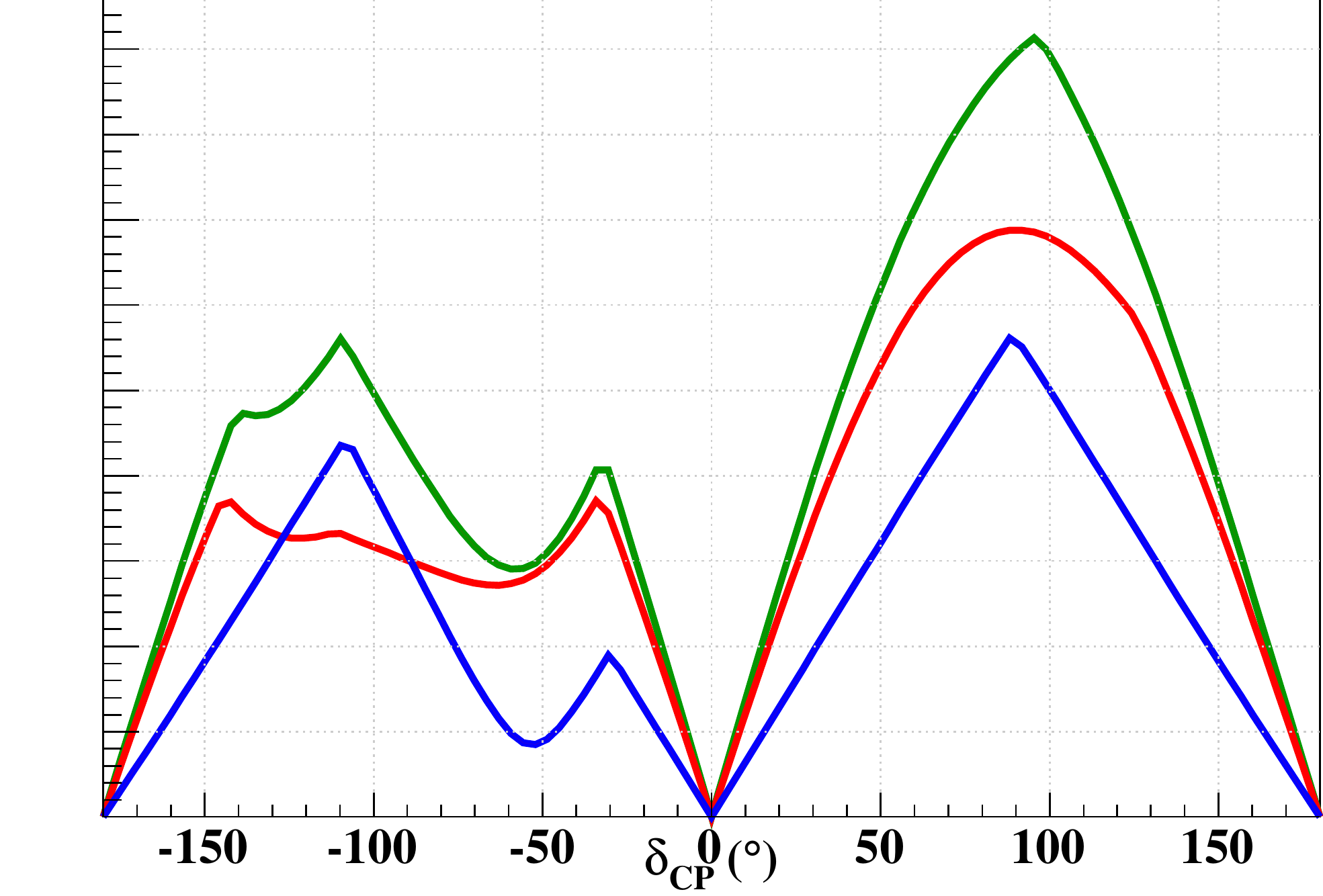}
\end{minipage}
\caption{\chips{} physics reach in the Normal Hierarchy (left) and Inverted Hierarchy (right), for \nova{} (5+5y) and T2K(\unit[8.8e21]{POT}), and \chips{}(3+3y).  (Top) \dcp{} resolutions.  (Middle) The significance of excluding the wrong hierarchy.  (Bottom) Significance of discovering CP violation.  The red line is \nova{} and T2K, the blue line is \chips{} and the green is the combination.}
\label{fig:reso}
\end{center}
\end{figure}

Figure~\ref{fig:reso} shows the resolution on \dcp{} when a \unit[100]{kton} fiducial mass \chips{} starts taking data four years after \nova{} starts. This resolution assumes that the mass hierarchy is known, and all other degeneracies are resolved. The \chips{} information is also combined with \nova{} and T2K in a simultaneous fit. The resolution ranges from around 15\dg{} to around 24\dg{}, across the whole range of $\delta_{CP}$. It can be seen from these figures that the information from \chips{} is complementary to \nova{}+T2K owing to the wider beam spectrum.  At large \dcp{} the \dcp{} resolution is much better than \nova{}, while at small \dcp{} it is worse. The wrong-hierarchy exclusion significance for the same configuration is shown in Figure~\ref{fig:reso} (middle).  The best combined exclusion in the ME tune reaches a 4$\sigma$ significance.  The potential for discovering CP violation (i.e. excluding \dcp~=~0\dg or 180\dg) is shown in Figure~\ref{fig:reso} (bottom). The features for one half of \dcp{} space (positive \dcp{} with NH, negative \dcp{} with IH) are due to the ambiguity in the hierarchy. If the hierarchy is determined, then the curves look symmetric.  While \chips{} can achieve lower errors on larger values of $\delta_{CP}$, the shape of the $\chi^{2}$ curve gives NOvA plus T2K more power to exclude CP conservation.  However, the combination of \chips{}, NOvA, and T2K can find evidence for CP violation (at above 3 sigma) in around 25\% of \dcp{} space, doubling to 50\% if the hierarchy is known.

\begin{figure}[htbp]
\begin{center}
\begin{minipage}{0.49\textwidth}
\includegraphics[width=\textwidth]{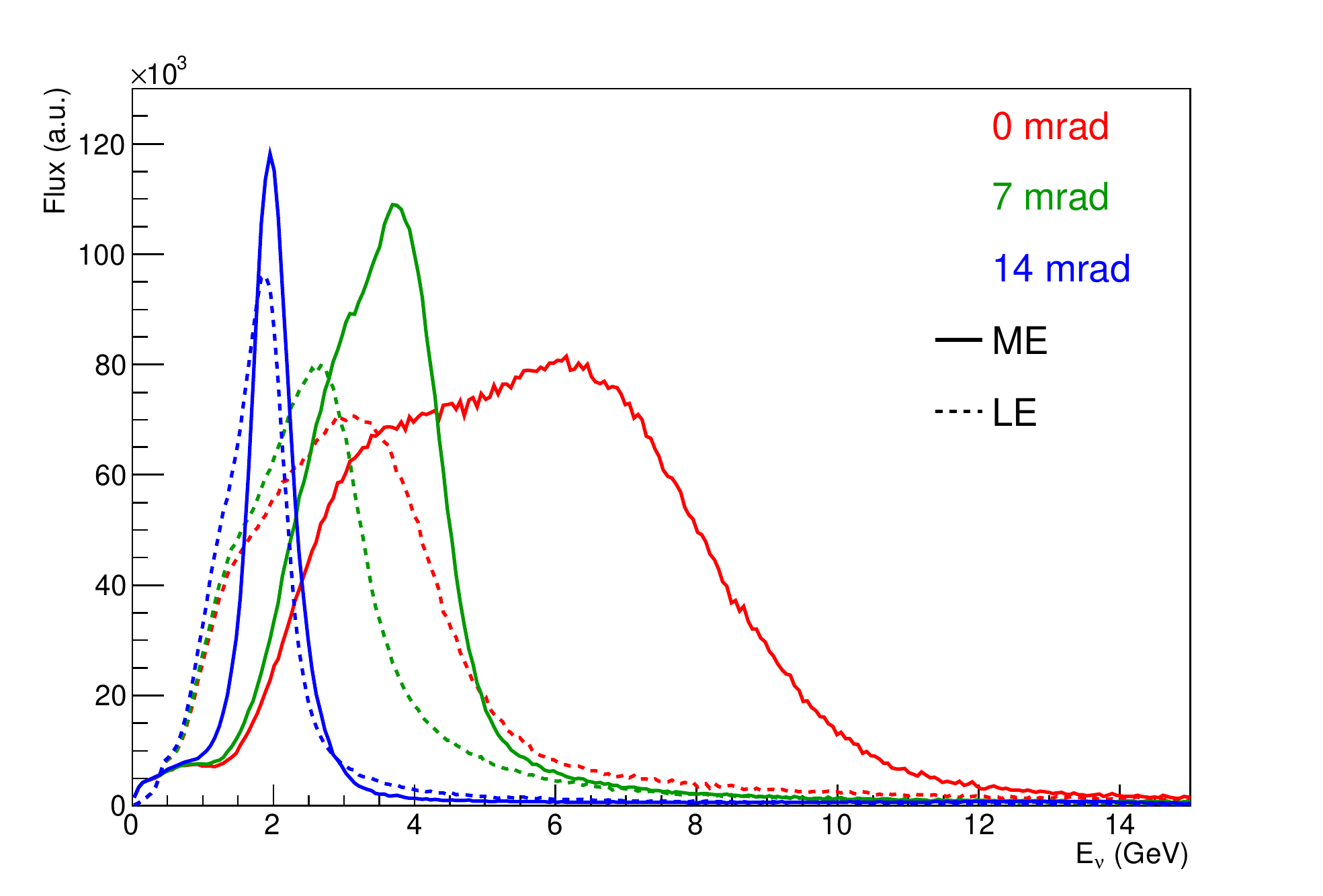}
\end{minipage}
\begin{minipage}{0.49\textwidth}
\includegraphics[width=\textwidth]{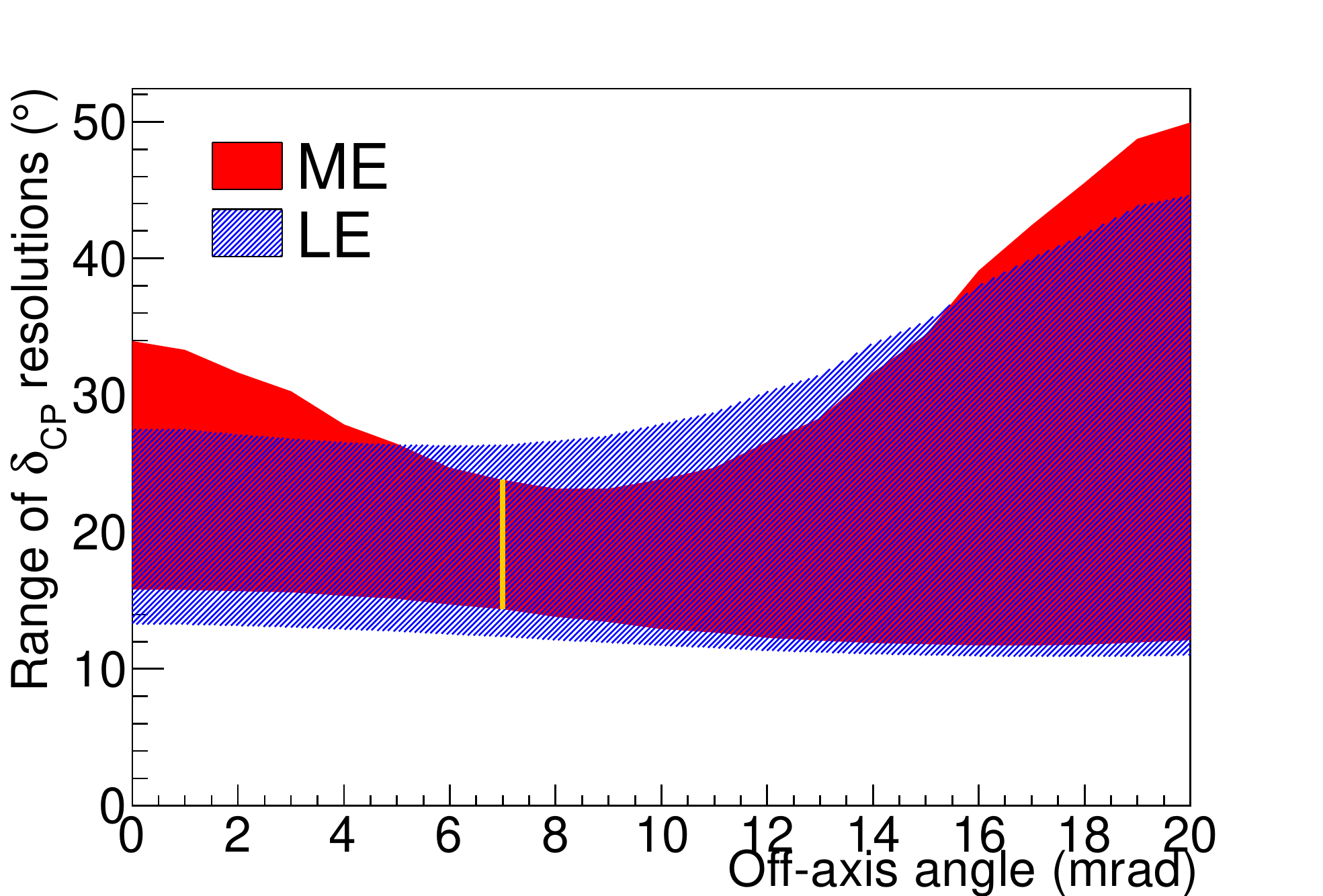}
\end{minipage}
\caption{(Left) $\nu_{\mu}$ flux (in arbitrary units) seen at 0, 7 and 14 mrad off-axis, in the Medium Energy (ME) and Low Energy (LE) beam configuration.  (Right) \dcp{} resolution band for off-axis angles from 0 to 20 mrad, for the ME and LE beams. The orange line at an angle of 7 mrad corresponds to the position of the Wentworth pit.}
\label{fig:angle-reso}
\end{center}
\end{figure}

The sensitivity of \chips{} in the lower energy beam tune was also explored.  An increase in the low energy beam flux can be achieved by moving the hadron production target closer to the magnetic focusing horns.  The standard NuMI low energy configuration is achieved by partially inserting the target into the neck of the first horn.  This configuration is harder to achieve, in terms of reconfiguring the beamline at Fermilab, now that the beam line has been upgraded for the \nova{} running.  For comparison, the event rates associated with the ME fluxes are shown in Figure~\ref{fig:angle-reso}.  The same figure also shows the band of \dcp{} resolutions (minimum to maximum across all values of \dcp{} in both hierarchies) against the off-axis angle of the detector. The choice of 7 mrad is the preferred location in both the LE and ME beams in terms of $\delta_{CP}$ resolution, owing to the combination of a high event rate with a low background for either beam tune.

\subsection{Staged NuMI Reach}
Owing to financial constraints, it may not be possible to construct a \unit[100]{kton} detector in the four years after starting \nova{}, and so the possibility of building \chips{} in a phased approach has been investigated. This would involve increasing the fiducial mass over multiple years, exploiting the experience to accelerate the expansion.

\begin{figure}[htbp]
\begin{center}
\includegraphics[width=0.5\textwidth]{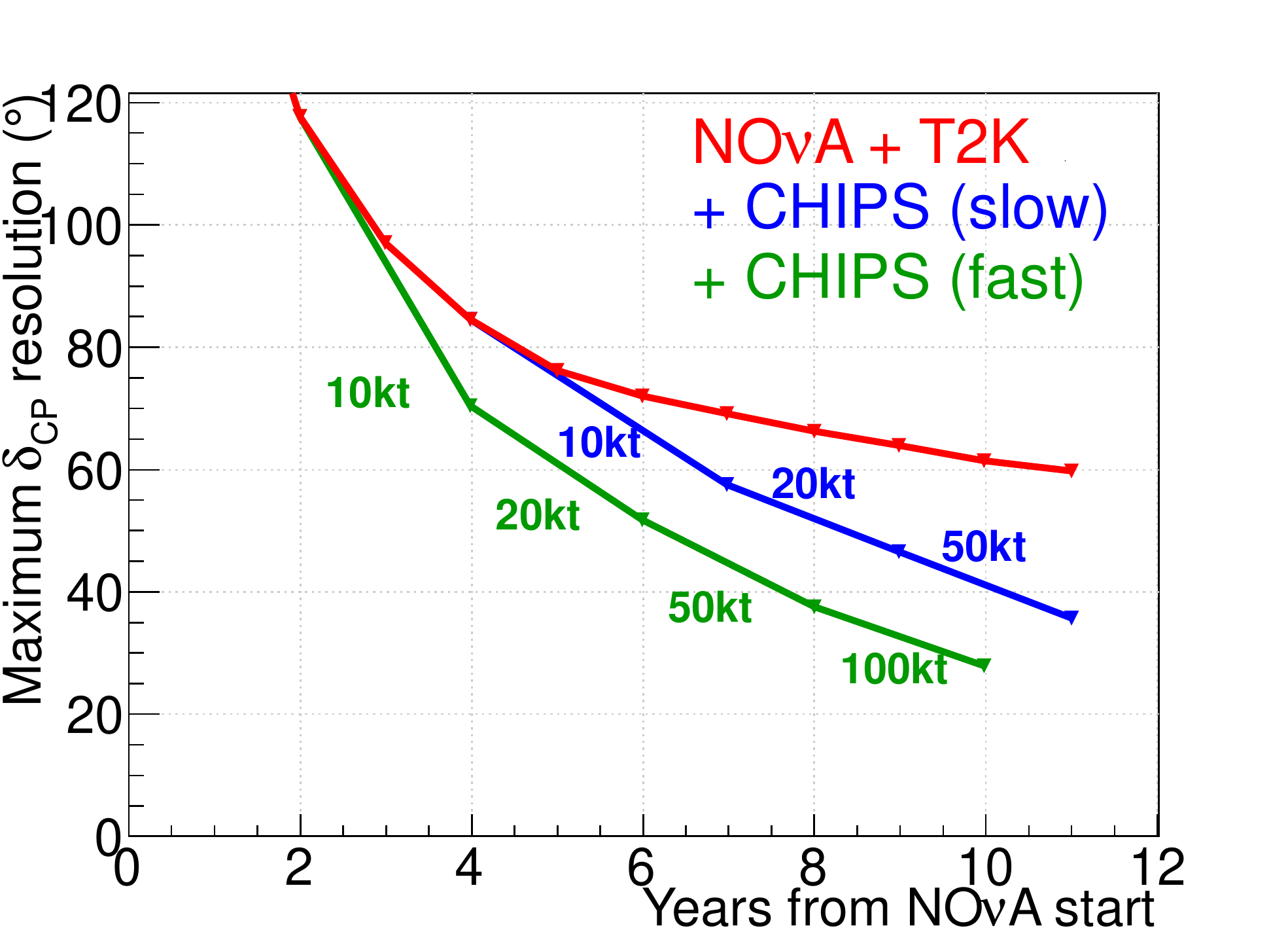}
\caption{Impact of a phased \chips{} program on \dcp{} resolution.}
\label{fig:phased-reso}
\end{center}
\end{figure}

Figure~\ref{fig:phased-reso} shows how adding a phased-\chips{} detector in the NuMI beamline improves the \dcp{} resolution over the default configuration of \nova{} and T2K only. Two approaches are shown; a fast track approach of building a \unit[10]{kton} detector two years after \nova{} starts data taking and increasing this to 20, 50 and \unit[100]{kton} every subsequent two years. The other is a slower-track approach, where \unit[10]{kton} is instrumented four years after the \nova{} turn on, and increased to 20 and \unit[50]{kton} after seven and nine years respectively.

\subsection{\chips{} in LBNE}

When the LBNE beam is completed, the \chips{} detector will be redeployed in that beam. The construction procedure will allow for the PMTs and electronics to be salvaged and reused. The question of where best to position \chips{} for the best complementarity to the LAr detector has been studied. As a first consideration, the off-axis angle was varied and the resolution on \dcp{} was studied.

\begin{figure}[htbp]
\begin{center}
\includegraphics[width=0.45\textwidth]{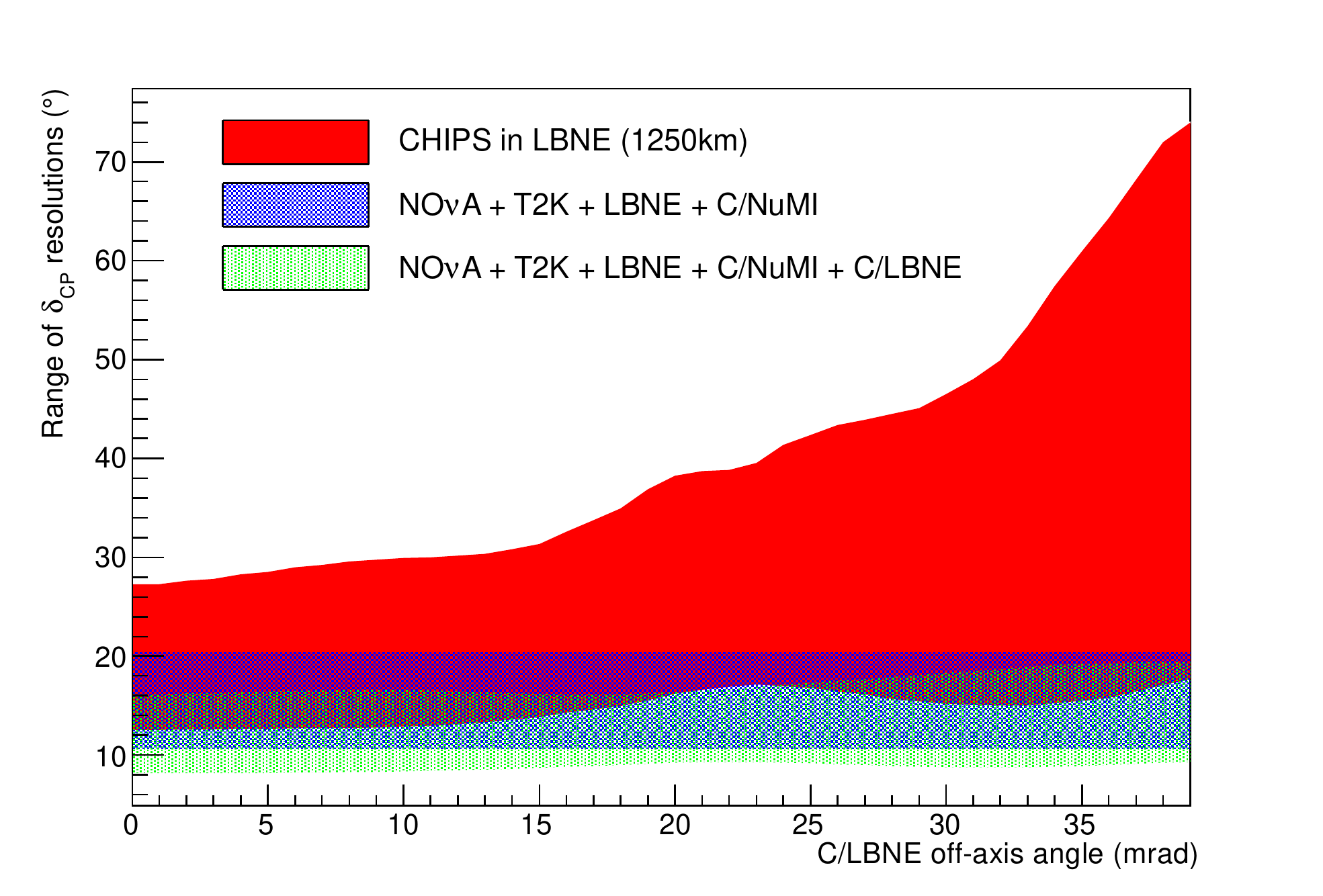}
\includegraphics[width=0.45\textwidth]{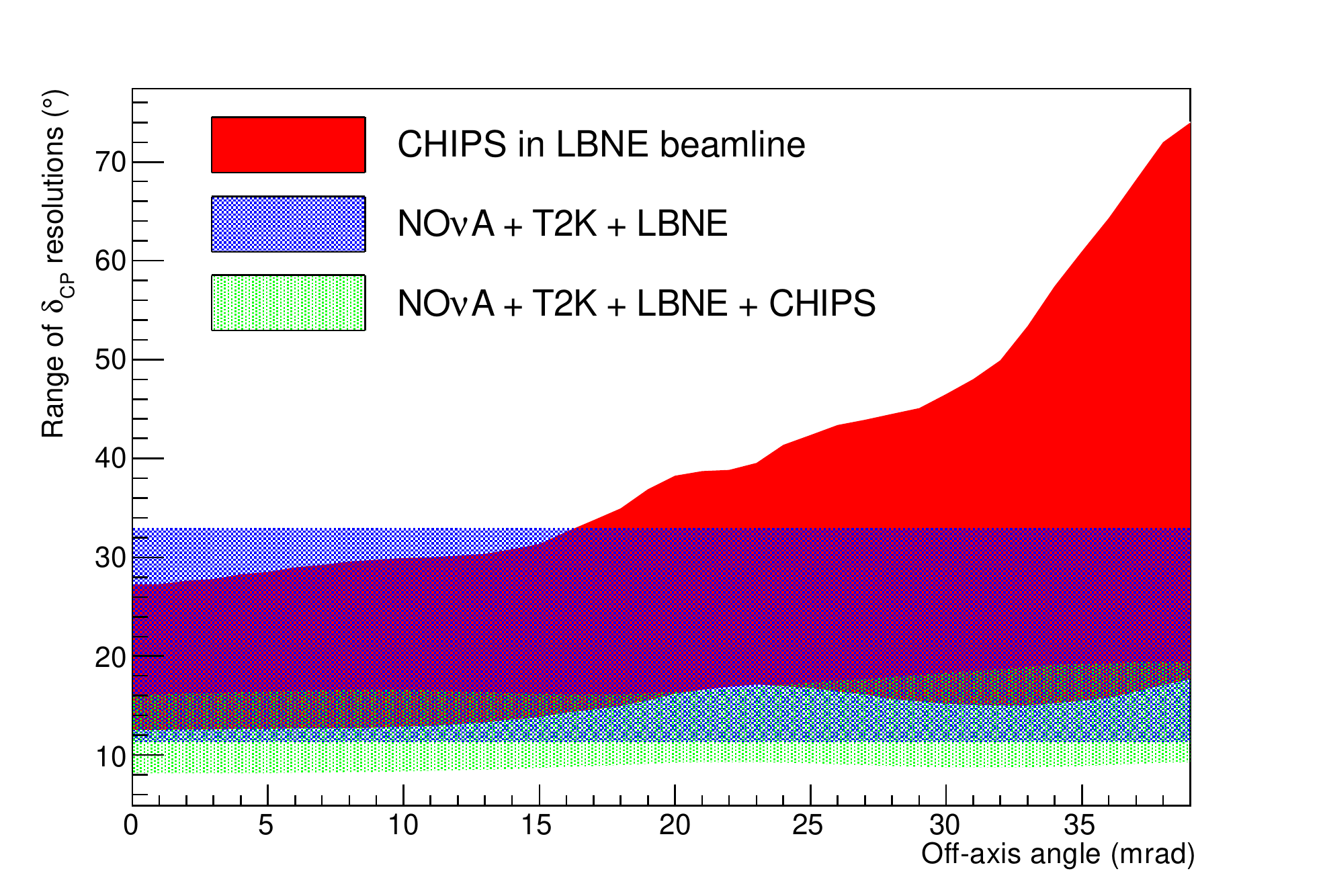}
\caption{ \dcp{} resolution bands for off-axis angles from 0 to 20 mrad, in the LBNE beam.  (Left) Assuming the \chips{} detector has already run in the NuMI beam.  (Right) Assuming \chips{} runs only in the LBNE beam. Only the \chips{} detector in the LBNE beam has been calculated at different off-axis angles; other detector positions are not varied.}
\label{fig:LBNE-off}
\end{center}
\end{figure}

\noindent Figure~\ref{fig:LBNE-off} shows the band of \dcp{} resolutions (minimum to maximum across all values of \dcp{} in both hierarchies) against the off-axis angle of the detector in the LBNE beam. The left plot shows the combined reach (in green) if the \chips{} detector has been already constructed in the NuMI beam. The right plot shows the \dcp~reach if \chips{} is only available in the LBNE beam. In either case, the \chips{} detector contributes a large weight to the resolution.

\begin{figure}[tbp]
\begin{center}
\begin{minipage}{0.49\textwidth}
\includegraphics[width=\textwidth]{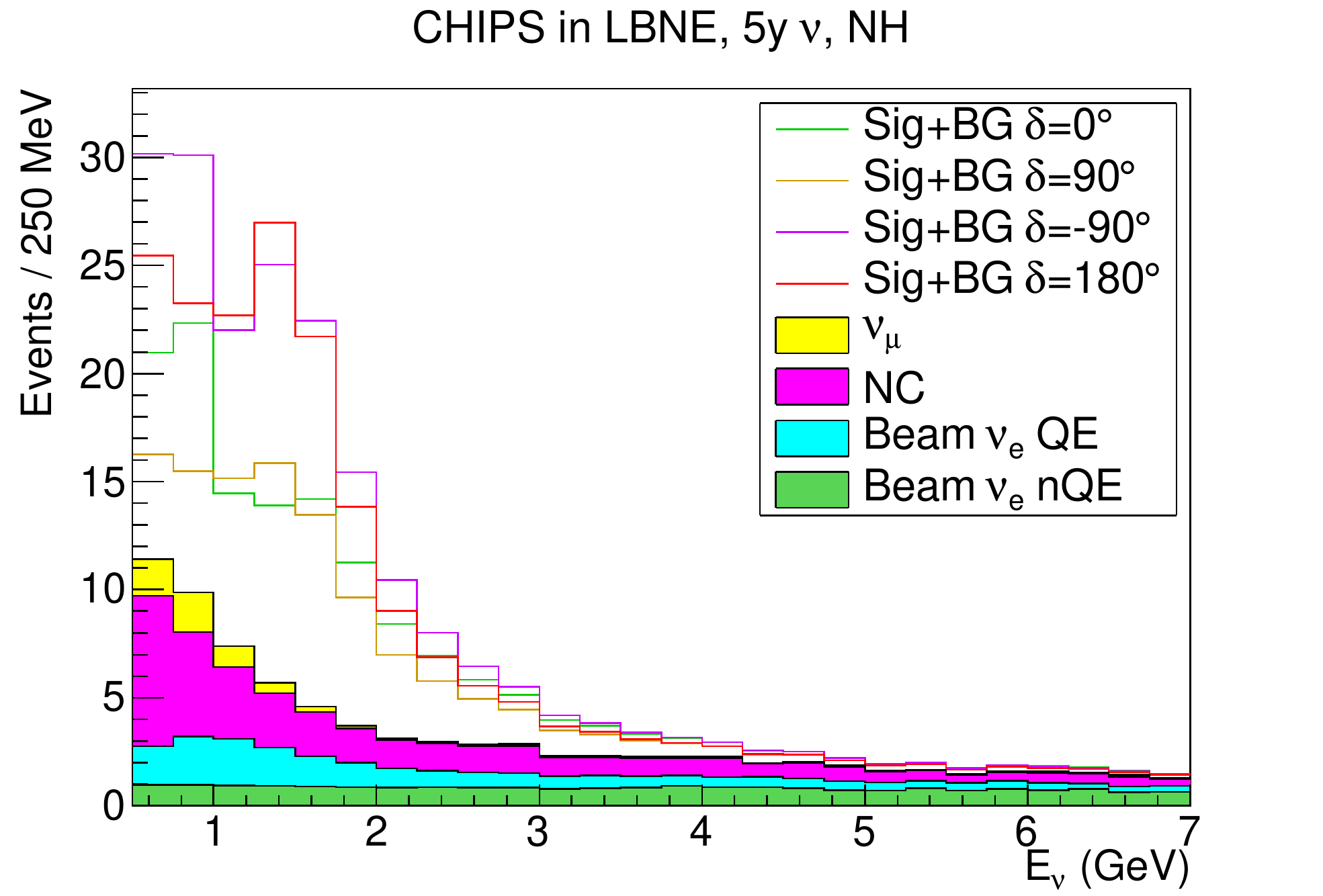}
\end{minipage}
\begin{minipage}{0.49\textwidth}
\includegraphics[width=\textwidth]{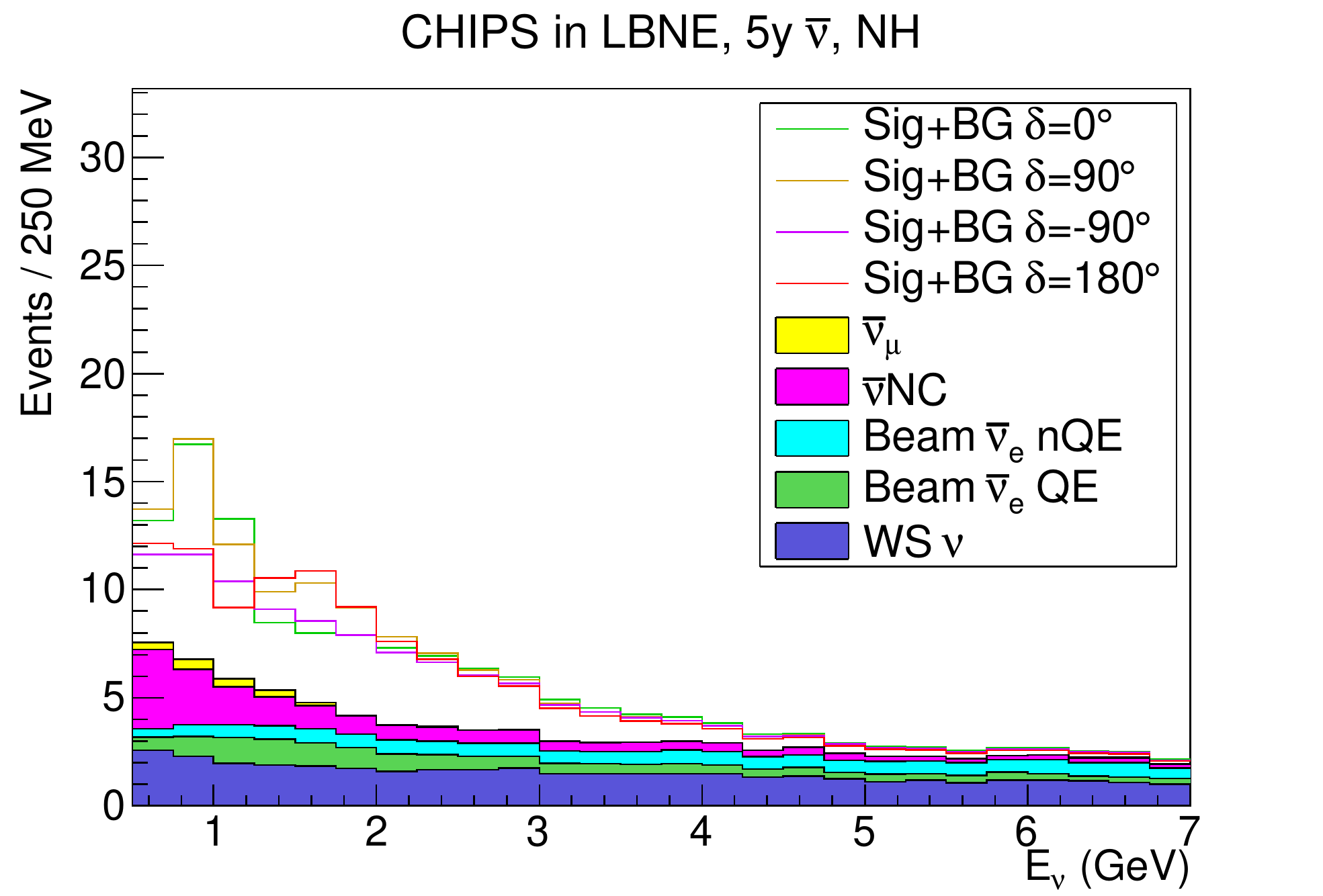}
\end{minipage}\\
\begin{minipage}{0.49\textwidth}
\includegraphics[width=\textwidth]{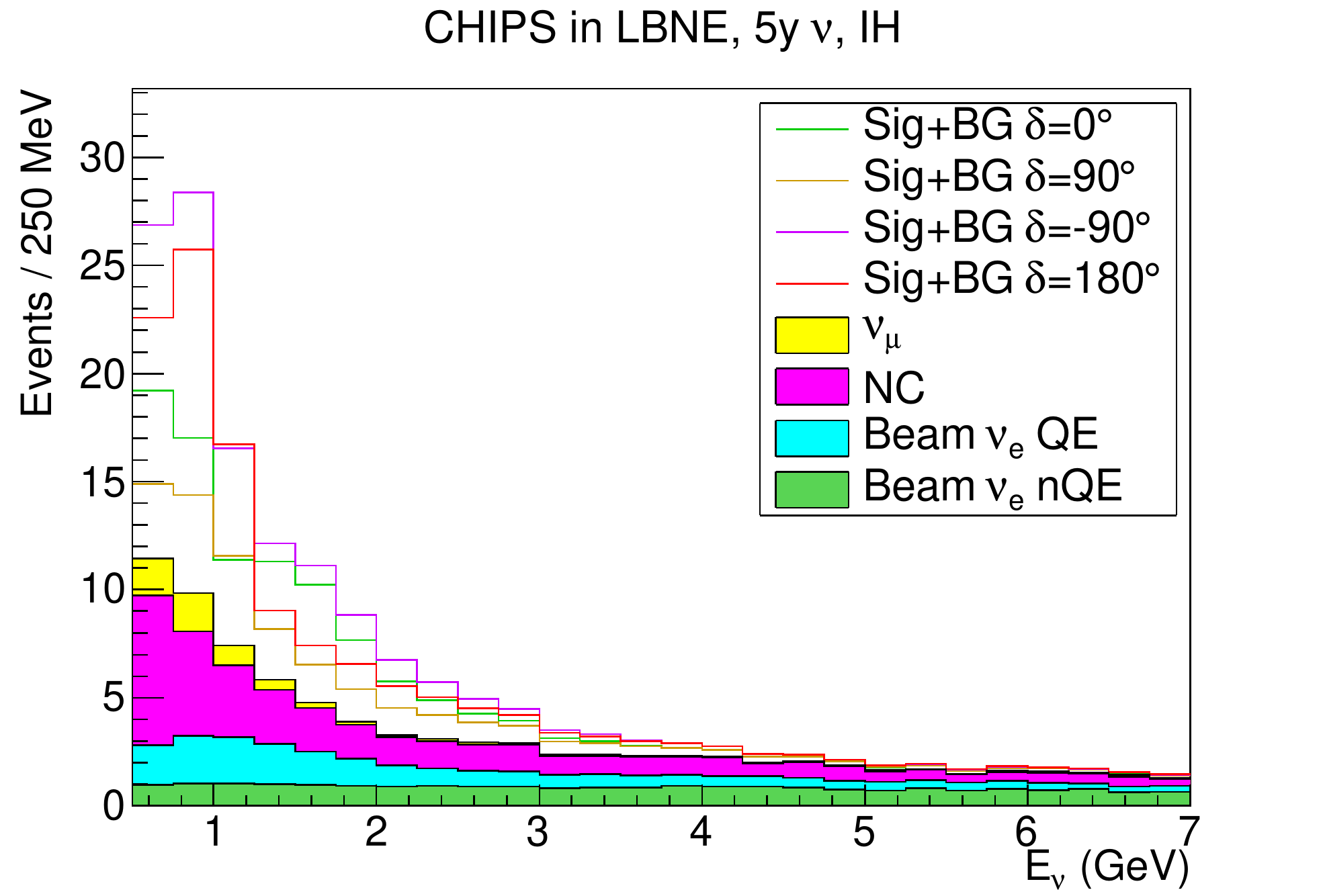}
\end{minipage}
\begin{minipage}{0.49\textwidth}
\includegraphics[width=\textwidth]{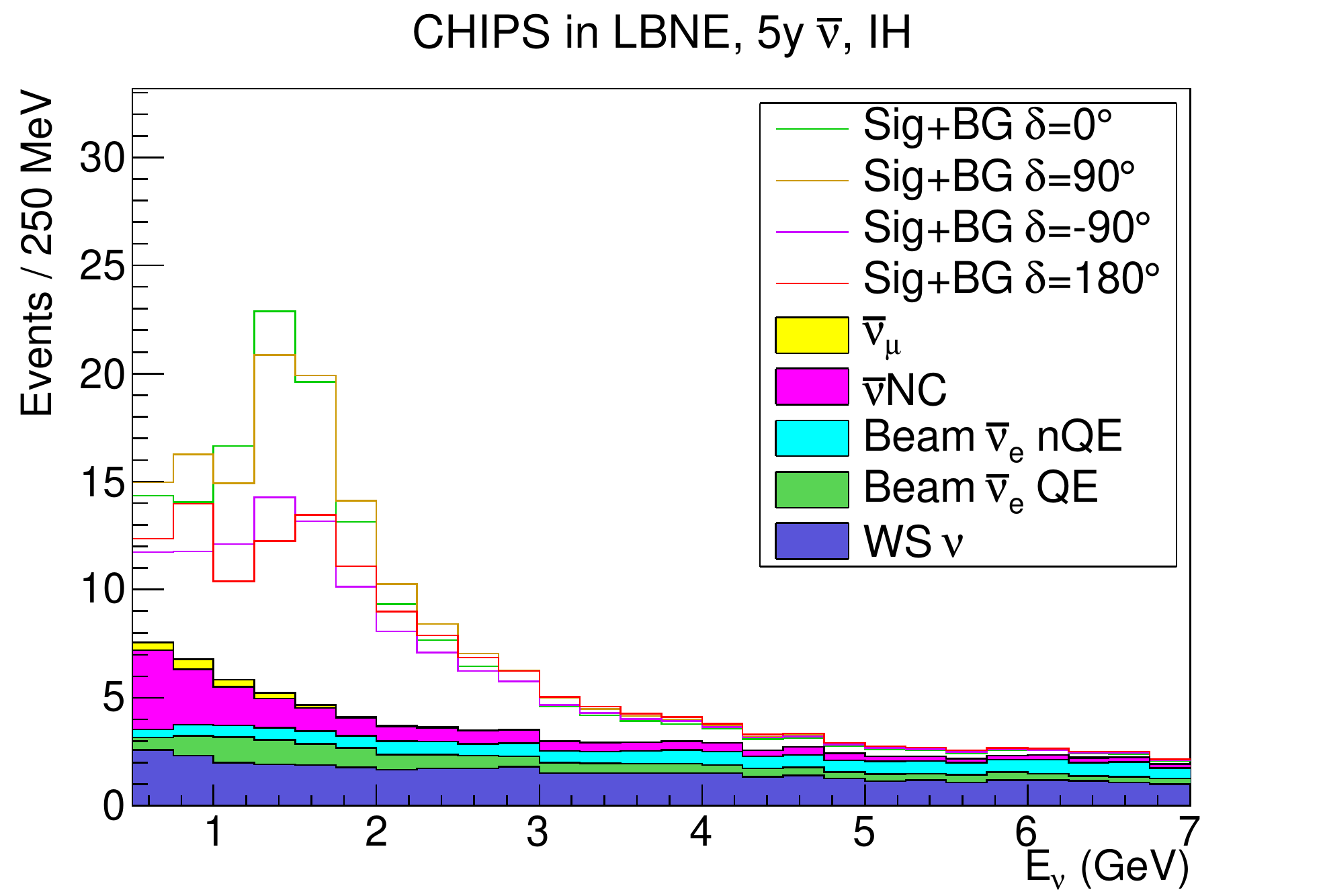}
\end{minipage}
\caption{{\bf } The expected event rates for a \unit[100]{kton} \chips{} detector \unit[20]{mrad} off-axis at the Pactola Reservoir in South Dakota, a hypothetical target for deployment of the \chips{} detector(s) in the LBNE beam. Beam \nue{} events are divided into quasielastic (QE) and non-quasielastic (nQE) samples.  The wrong sign (WS) neutrino sample is shown separately in the antineutrino beam plots.}
\label{fig:lbne-evrates}
\end{center}
\end{figure}

It would be preferable to place the redeployed \chips{} in a position for maximum complementarity, while taking into account the geographical considerations. There is a reservoir at a baseline of \unit[1250]{km} and at an angle \unit[20]{mrad} off-axis in the LBNE beam which could potentially house the \chips{} detector. In this case, the second maximum of the oscillation could be studied, which would be complementary to the on-axis LBNE detector.  When the \unit[100]{kton} fiducial mass \chips{} detector is placed in the LBNE beamline, with a baseline of \unit[1250]{km} and at \unit[20]{mrad} off-axis, the neutrino spectra produced are shown in Figure~\ref{fig:lbne-evrates}.  

\begin{figure}[p]
\begin{center}
\begin{minipage}{0.49\textwidth}
\includegraphics[width=\textwidth]{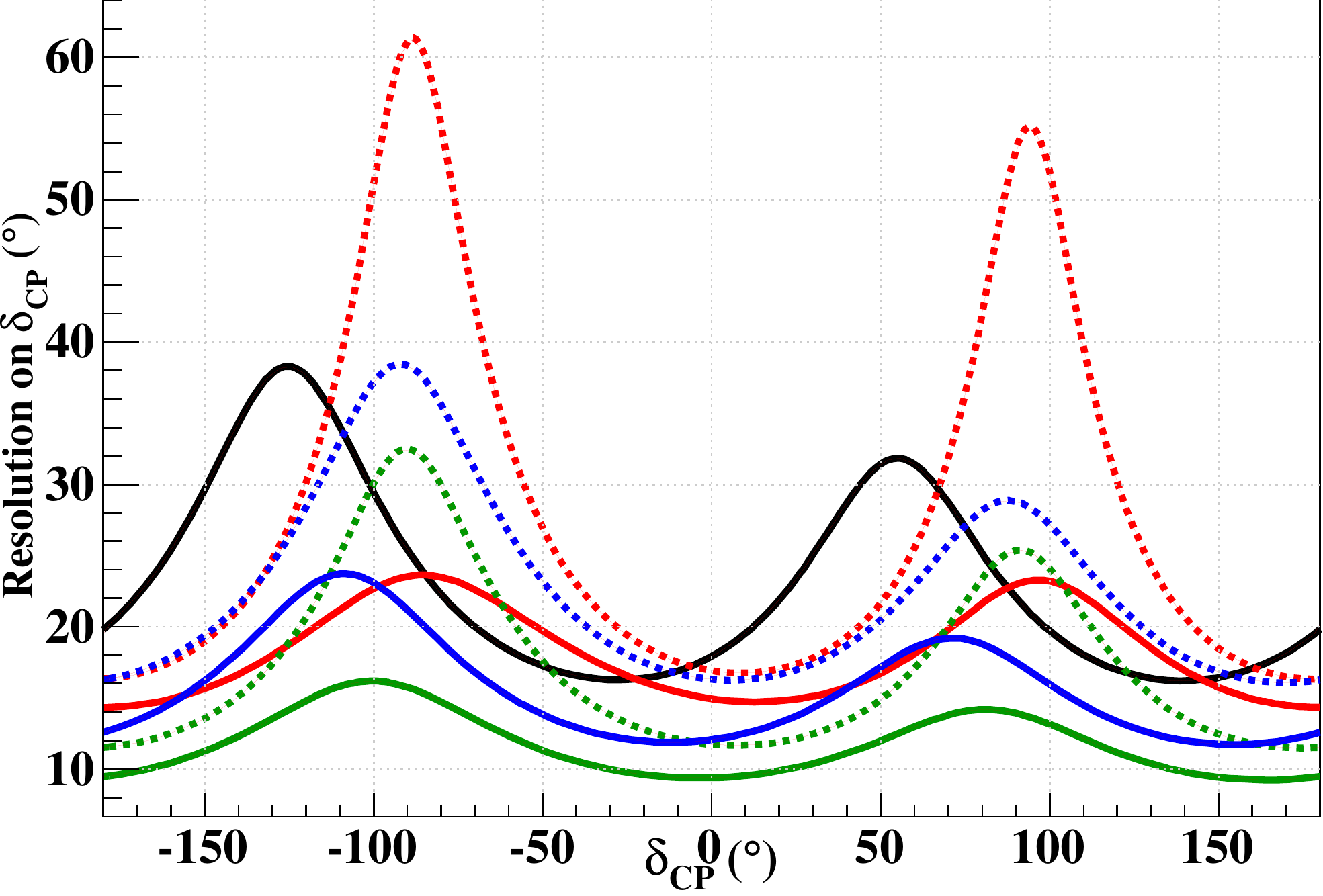}
\end{minipage}
\begin{minipage}{0.49\textwidth}
\includegraphics[width=\textwidth]{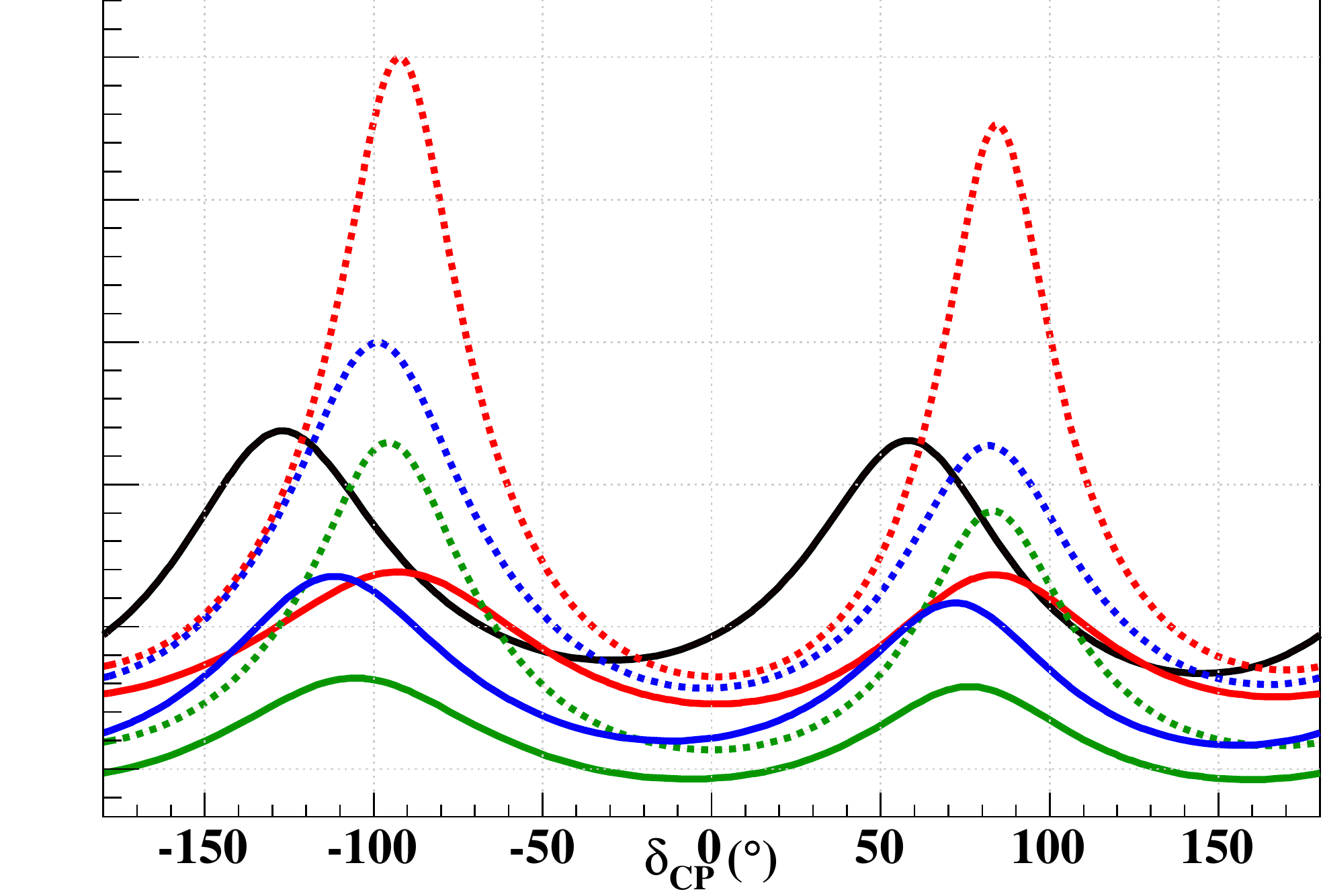}
\end{minipage}
\begin{minipage}{0.49\textwidth}
\includegraphics[width=\textwidth]{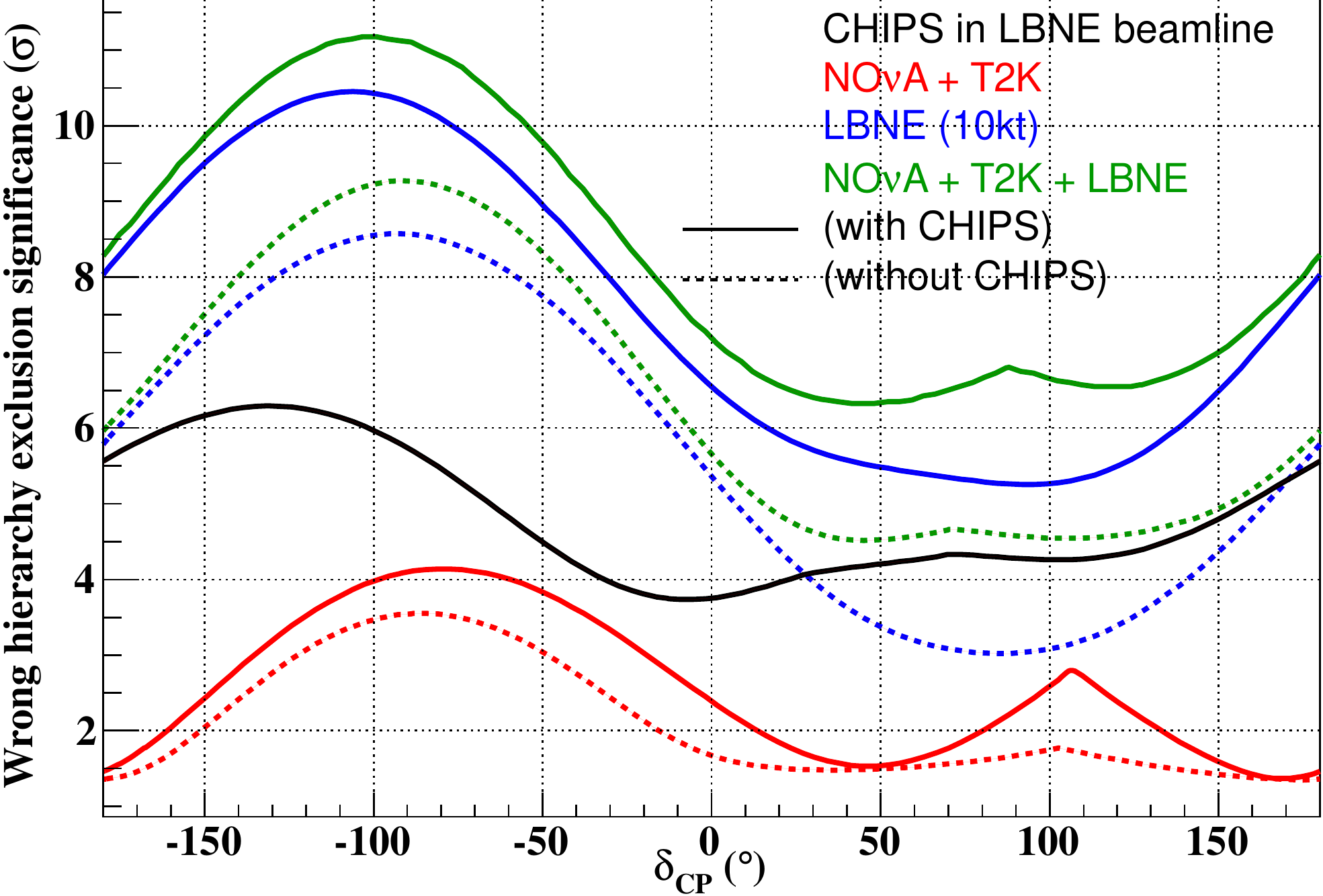}
\end{minipage}
\begin{minipage}{0.49\textwidth}
\includegraphics[width=\textwidth]{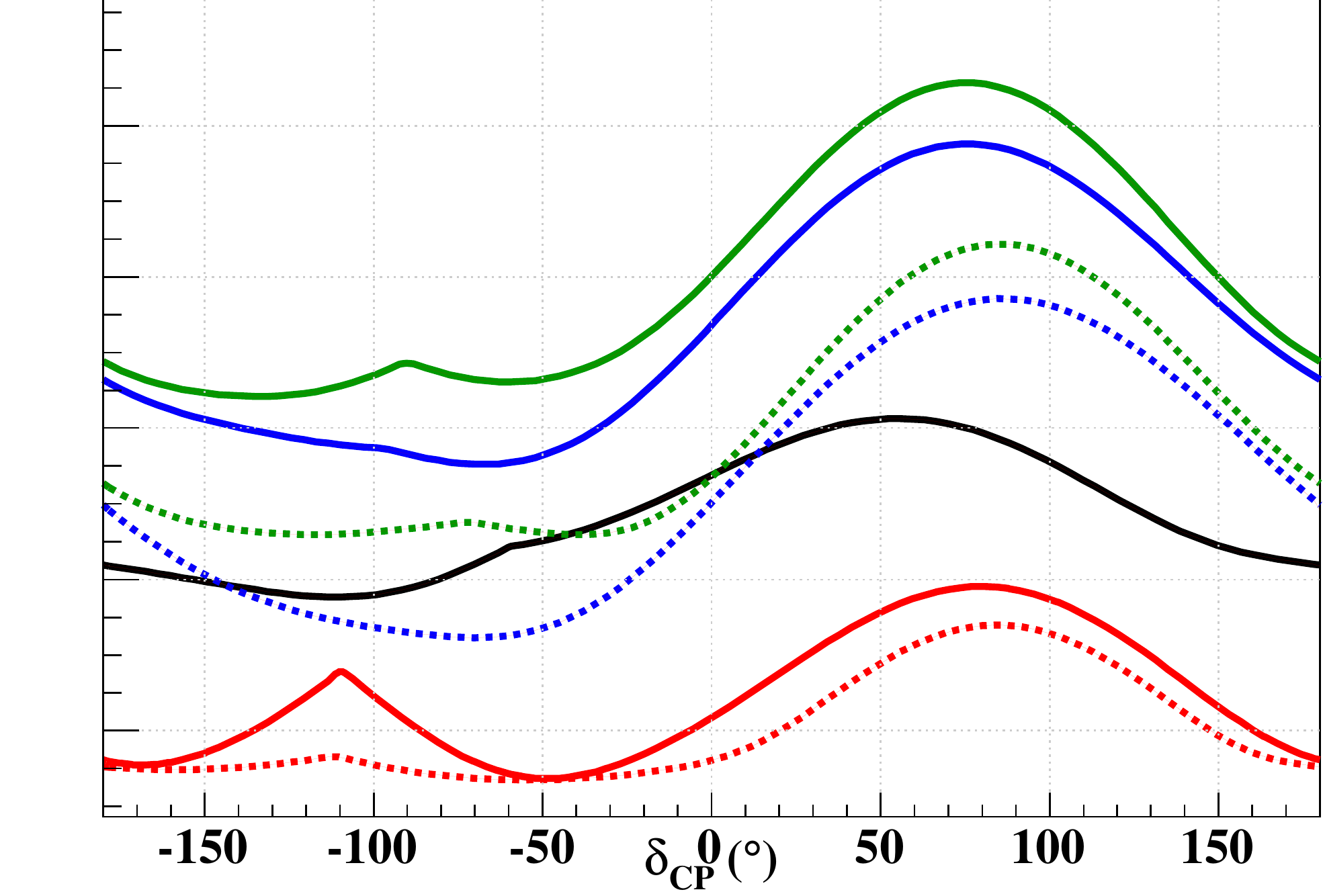}
\end{minipage}
\begin{minipage}{0.49\textwidth}
\includegraphics[width=\textwidth]{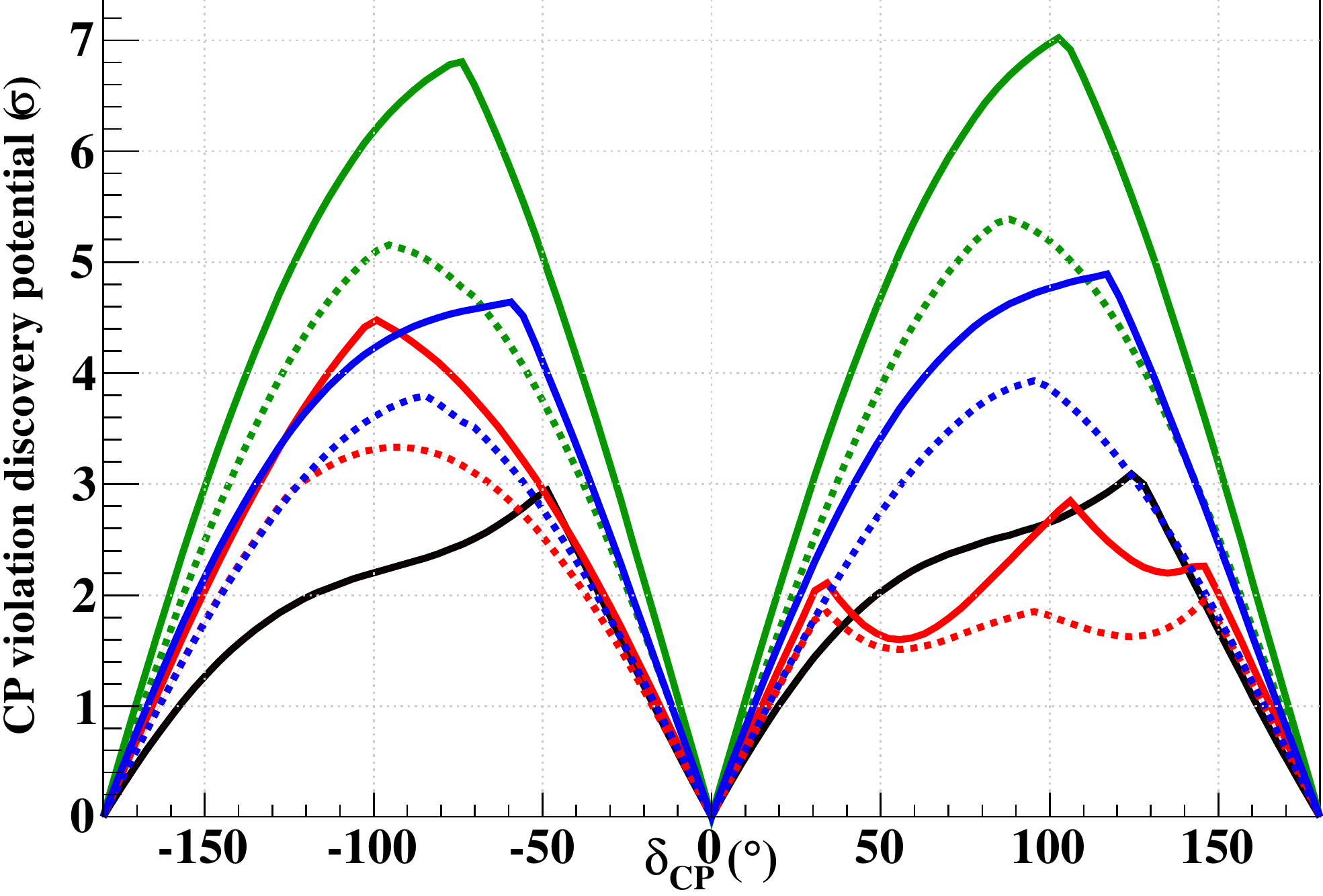}
\end{minipage}
\begin{minipage}{0.49\textwidth}
\includegraphics[width=\textwidth]{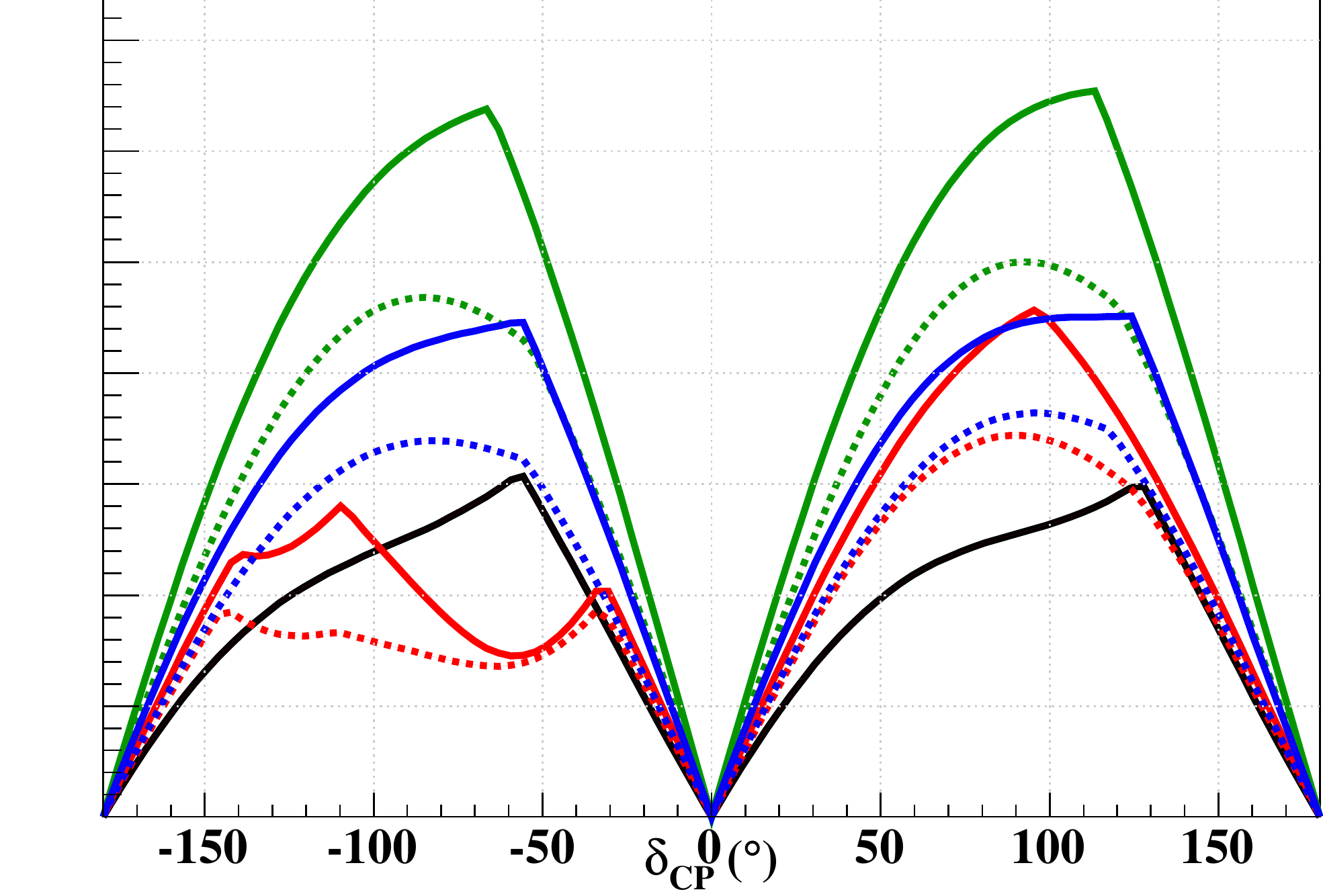}
\end{minipage}
\caption{Physics reach in the Normal Hierarchy (left) and Inverted Hierarchy (right), for \nova{}+T2K, \unit[10]{kton} LAr LBNE, and \chips{} in the LBNE beam at \unit[20]{mrad}.  (Top) \dcp{} resolutions.  (Middle) The significance of excluding the wrong hierarchy.  (Bottom) Significance of discovering CP violation.  The red line is \nova{} and T2K, the blue line is a \unit[10]{kton} LAr detector on-axis in the LBNE beam, and the green is the combination of those experiments.  Solid black line is for \chips{}, from both a NuMI and LBNE run.  Dotted lines show each experiment (or combination of experiments) without a \chips{} run.  Solid lines show the effect of adding \chips{} to the results.} 
\label{fig:lbne-cpv}
\end{center}
\end{figure} 

The hierarchy exclusion significance, resolution on \dcp{}, and CP violation discovery potential are shown in Figure~\ref{fig:lbne-cpv} for different combinations of the currently foreseen long-baseline neutrino experiments. A $5\sigma$ exclusion of the wrong hierarchy can be made for the whole phase space of \dcp{} only with the help of \chips{} (with a \unit[100]{kton} fiducial mass, in 6 years of NuMI beam and 10 years of LBNE beam). Likewise, a CP violation discovery potential of $5\sigma$ for a much larger range of \dcp{} can be achieved if \chips{} is included.  Finally, depending on the success of the \chips{} program in terms of reducing significantly the cost per kton for a water Cherenkov neutrino detector, a second \unit[100]{kton} module could be constructed in the first 3 years of running of the LBNE beam, thereby speeding up the collection of data and delivery of meaningful results.

\section{Proposed Location}
\subsection{The Mine Pit}
The proposed location for the \chips{} detector is in the Wentworth Mine Pit 2W.  The Wentworth Pit is on a disused surface iron mine property owned by Cliffs Natural Resources.  The site is a secure site and has the advantages of existing heavy industry infrastructure such as power and roads.  Space would be leased from Cliffs to do this experiment.  The center of the abandoned mine pit is located at a latitude of \latitude{} and longitude of \longitude{}.  It is \unit[7]{mrad} off the central axis of the NuMI beam at a baseline of \unit[712]{km}.  

\begin{figure}[htbp]
\begin{center}
\begin{minipage}{0.49\textwidth}
\includegraphics[width=\textwidth,clip=true,trim=0cm 5cm 0cm 5cm]{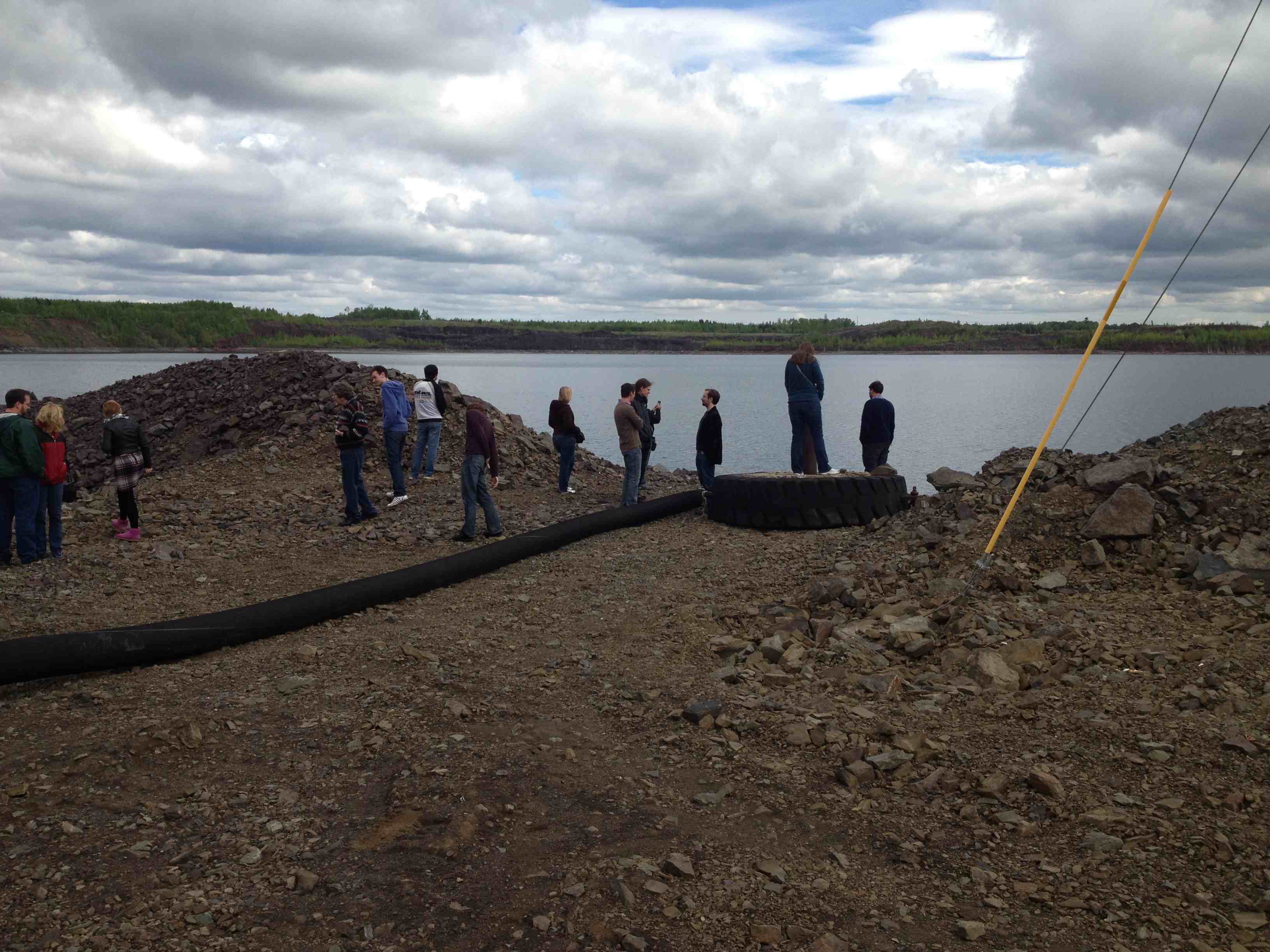}
\end{minipage}
\begin{minipage}{0.49\textwidth}
\includegraphics[width=\textwidth]{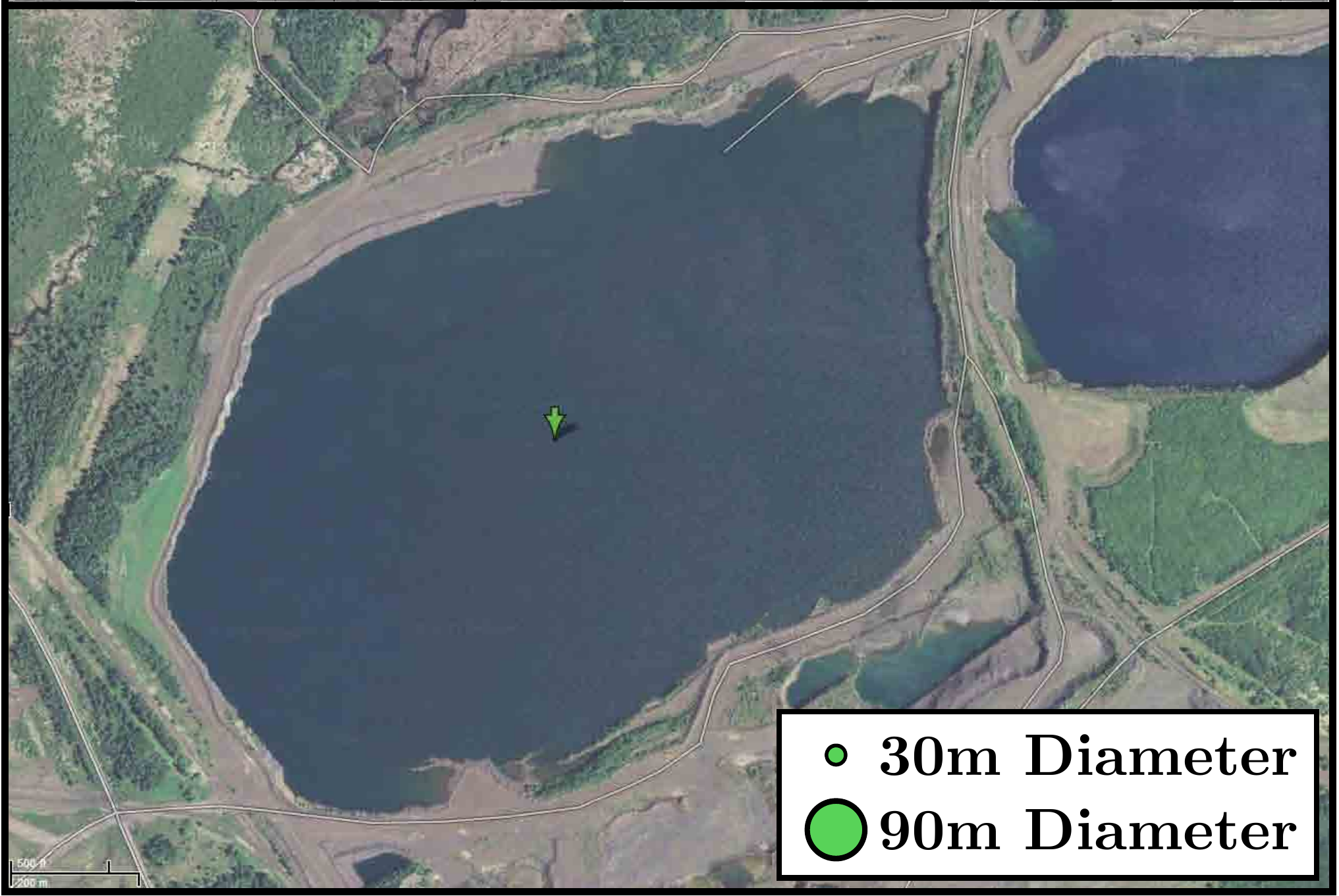}
\end{minipage}
\caption{(Left) A photograph of the Wentworth Mine Pit.  (Right) a composite satellite image of the region including roads.  A cross section marker of \unit[30]{m} and \unit[90]{m} diameter is given for scale.  Photograph courtesy J. Meier, satellite image from Google Earth.}
\label{fig:prettypics}
\end{center}
\end{figure}

A photograph of the site and a composite satellite image are shown in Figure~\ref{fig:prettypics}.  The aerial image details access to the mine location via railroad tracks and roads.  

\begin{landscape}
\begin{figure}[p]
\begin{center}
\includegraphics[height=0.95\textheight]{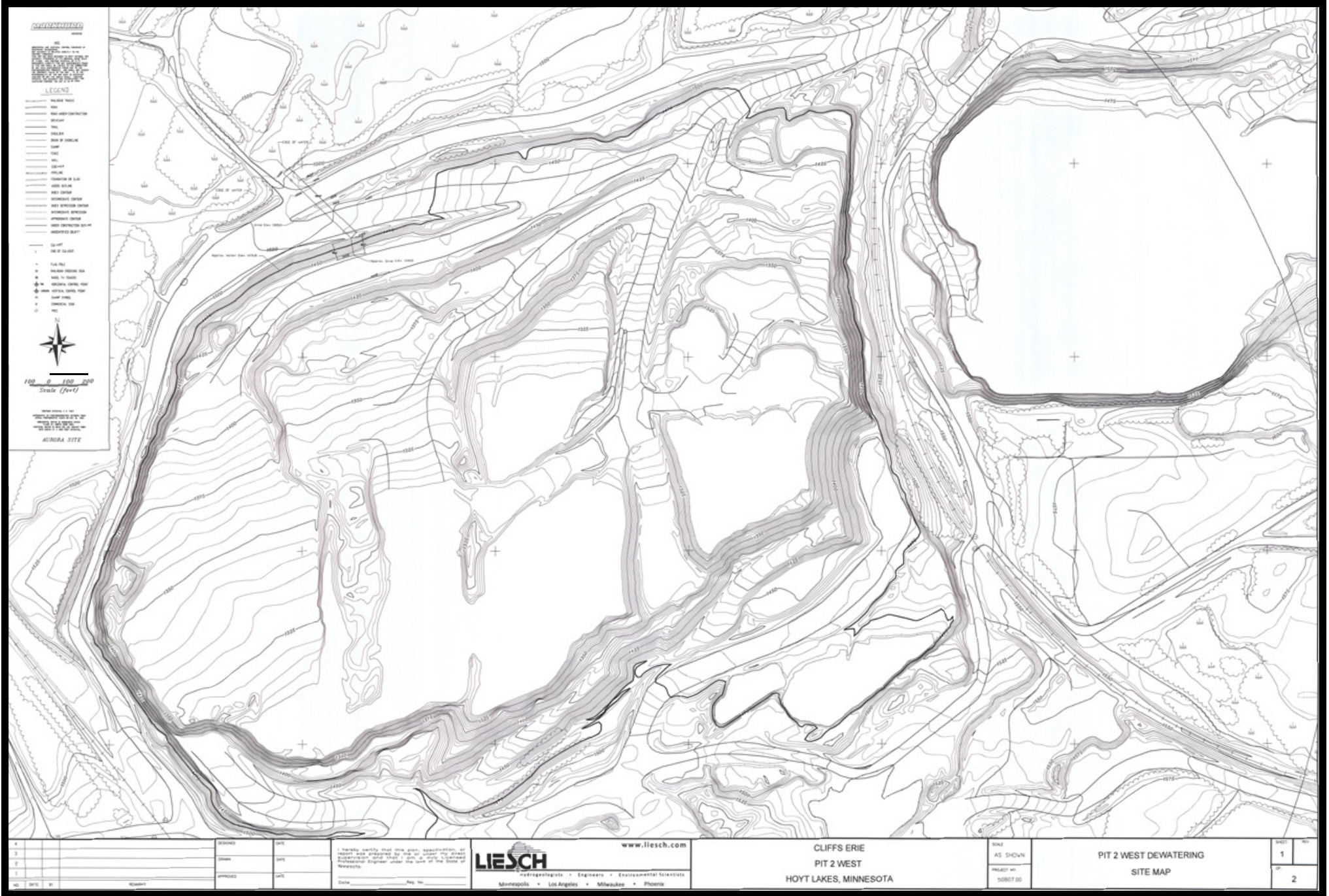}
\caption{An elevation map of the Wentworth Pit.  The intervals for the contour plot are 5 feet.}
\label{fig:location}
\end{center}
\end{figure}
\end{landscape}

\noindent Figure~\ref{fig:location} depicts a topographical map of the proposed location with elevation contours derived using photogrammetric methods from aerial photographs taken in May, 2001. From the contour map, the lowest elevation is \unit[1305]{ft} above sea level, but the contour map was made from a photo taken when the pit had about \unit[10]{m} of water in it.  The current water level is at \unit[1471]{ft}, making the local depth of the water about \unit[60]{m}.  This estimate of the water level is consistent with recent measurements taken with a commercial depth finder.  Water is drained from the pit in the spring to ensure the pit does not overflow during the summer rainy season.  Maximum fluctuations of the water level are estimated to be on the order of $\pm$\unit[10]{ft.}

The water in the Wentworth Pit was surveyed from January 2010 to September 2012 to characterize the quality.   Two separate types of testing were conducted. The first type consisted of monthly tests of surface water for standard mine water contaminants.  The temperature of the water was measured to range between  $0^\circ$ and \unit[$20^\circ$]{C} due to seasonal weather fluctuations. The turbidity of the water, a measure of the clarity, was measured to be $0.7 \pm 0.5$ NTU (Nephelometric Turbidity Units), implying that the water is quite transparent.  The pH at the surface was measured to be $8.3\pm0.3$.  A further set of profile measurements was also taken in September of 2011, from the surface to a depth of 123 feet. These profiles showed that the temperature varied from $20^\circ$ C at the surface to $5^\circ$ C at 123 feet. The pH was also observed to drop from 8.4 at the surface to 7.2 at 123 feet deep.  Detailed results of these tests are summarized in the appendix.

\subsection{Depth Considerations}
The water of the mine pit not only provides structural support for the detector, but also serves as an overburden to shield the detector from cosmic rays.  The depth of the Wentworth pit allows a relatively shallow overburden of a few tens of meters of water, implying a high rate of cosmic ray (CR) muons entering the detector.  To determine the feasibility of the shallow overburden, expected cosmic rates were computed as a function of detector depth, and detector dead time due to those rates was considered.

To first order, the energy-averaged intensity of muons at sea level, $I_{S}$, has a characteristic angular dependence proportional to $\cos^2\theta$, where $\theta$ is the zenith angle~\cite{PDG}: 

\begin{align}
I_S(\theta) =
\begin{cases}
I_{SV} \cdot \cos^2\theta &   \text{if } 0<\theta<\pi/2; \\
0, & \text{if } \pi/2 < \theta < \pi .
\end{cases}
\end{align}

\begin{figure}[h]
\centerline{ 
\includegraphics[keepaspectratio, width=.49\textwidth]{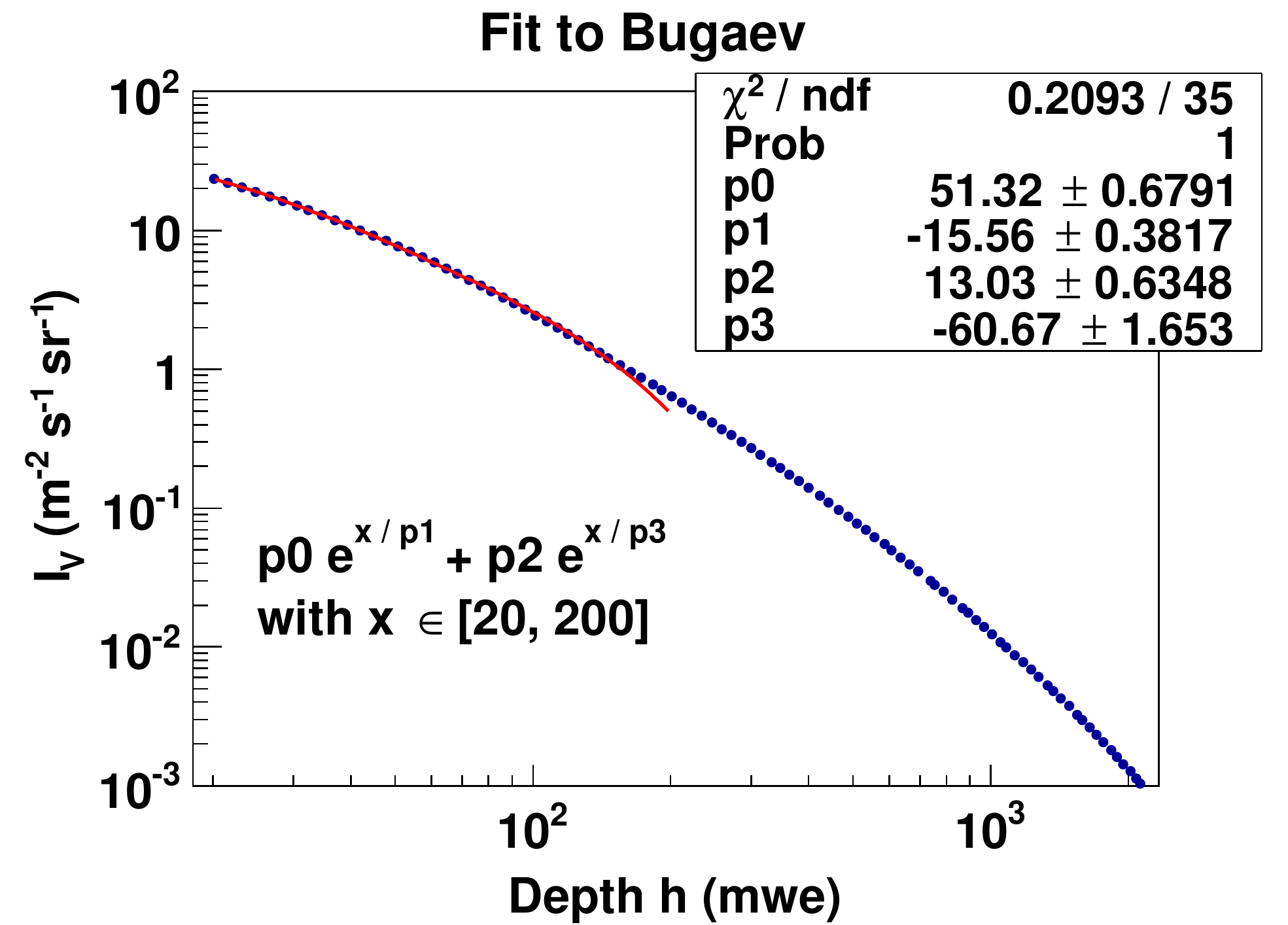}
\hskip0.150in 
\includegraphics[keepaspectratio, width=.49\textwidth]{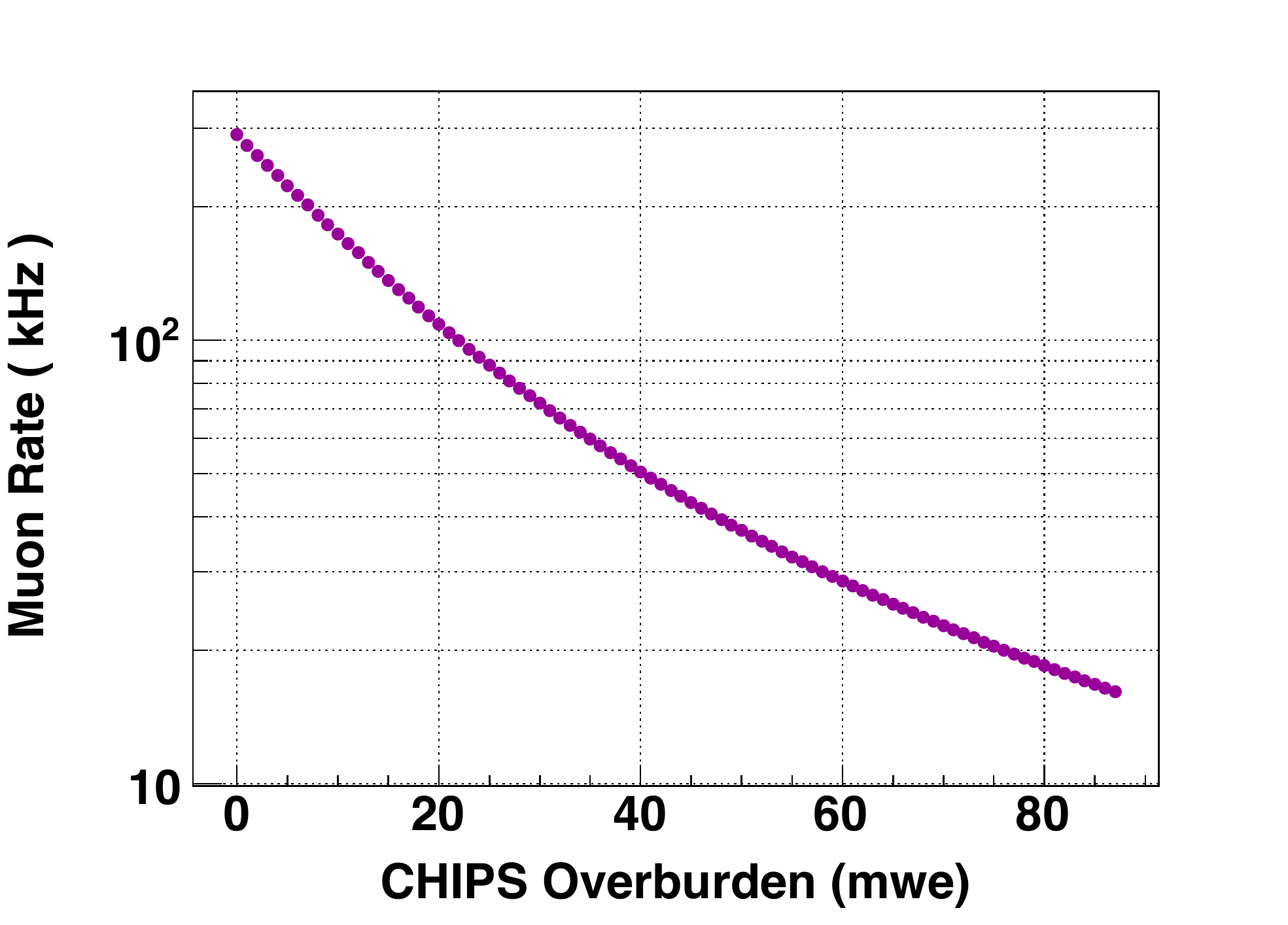}
}
\caption{(Left) The muon intensity from Bugaev~{\it et~al.}~\cite{Bugaev-1998}. From \unit[20]{m.w.e.} to \unit[200]{m.w.e}, a double exponential fits the calculation well.  
(Right) The \chips{} muon rate (using fits to Bugaev~{\it et~al.}~\cite{Bugaev-1998}) as a function of depth.}
\label{fig:Bugaev-1998}
\end{figure}

\noindent The vertical cosmic ray flux above \unit[1]{GeV} at sea level, $I_{SV}$, is \unit[70]{${\rm m^{-2}s^{-1}sr^{-1}}$}.  At the proposed detector depth, muon energy loss is primarily by ionization; radiative energy losses are completely negligible~\cite{Groom-Mokhov-Striganov}.  Using calculations from Bugaev {\it et al.}~\cite{Bugaev-1998} and Bogdanova~{\it et~al.}~\cite{Bogdanova-2006} we have estimated the rate of cosmic rays as a function of detector depth and detector geometry~\cite{CHIPS-CR-rates}.  Figure~\ref{fig:Bugaev-1998} shows rates as a function of depth.  With a \unit[40]{m.w.e} overburden, the cosmic rate is expected to be \unit[50]{kHz} in a cylindrical detector \unit[50]{m} in diameter and \unit[20]{m} high.  The impact of the cosmic rates on the NuMI beam events is mitigated by the short NuMI spill of about 10\,$\mu$s. The in-spill signal occupancy due to CR muons is a product of (CR muons rate)$\times$(spill-length), or \unit[50]{kHz}$\times$\unit[10$\times 10^{-6}$]{$\mu$s} = $0.5$ cosmic events per spill.

\begin{figure}[htbp]
\centering
\includegraphics[width=0.8\textwidth]{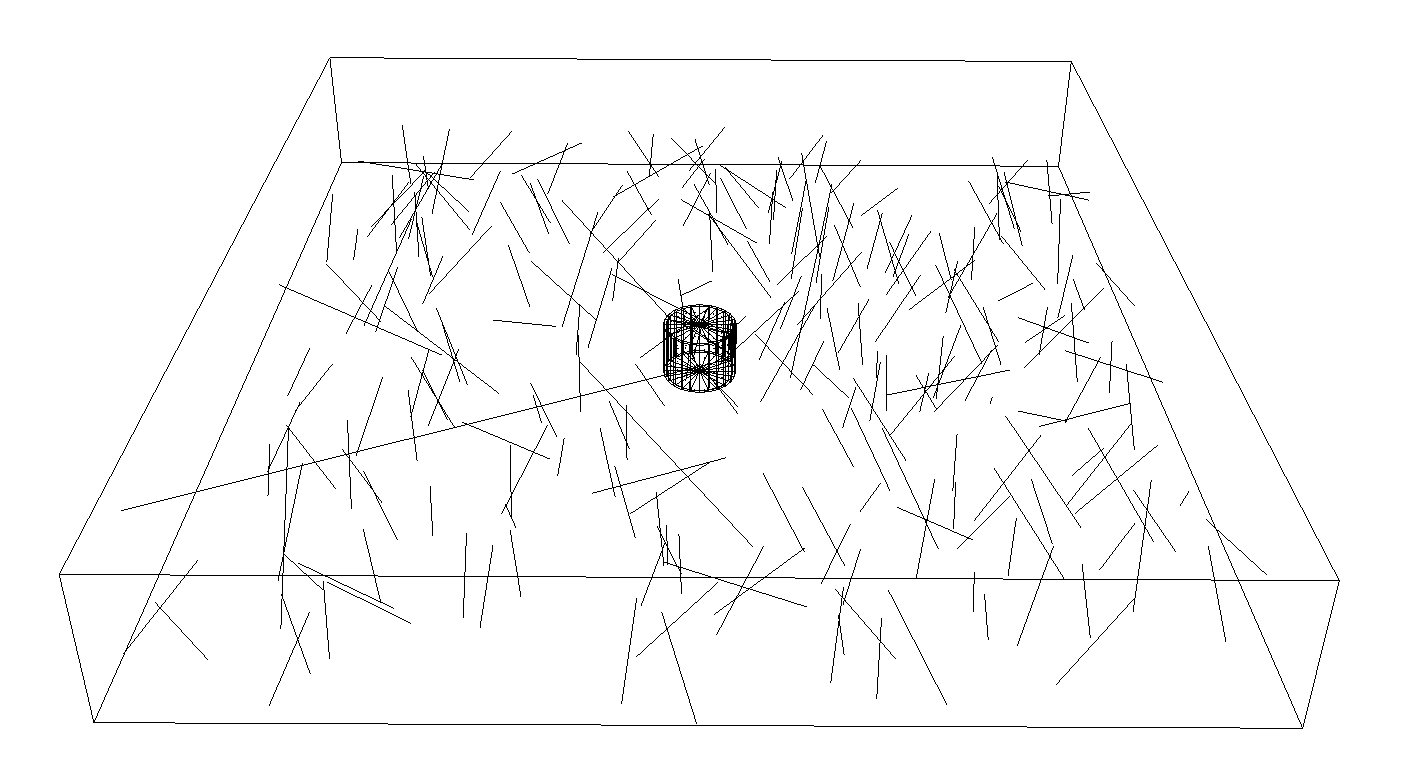}\\
\begin{minipage}{0.49\textwidth}
\includegraphics[width=\textwidth]{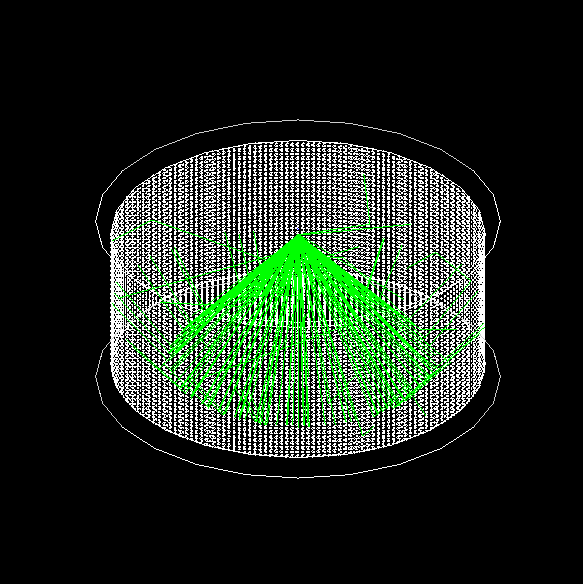}
\end{minipage}
\begin{minipage}{0.49\textwidth}
\centering
\includegraphics[width=\textwidth]{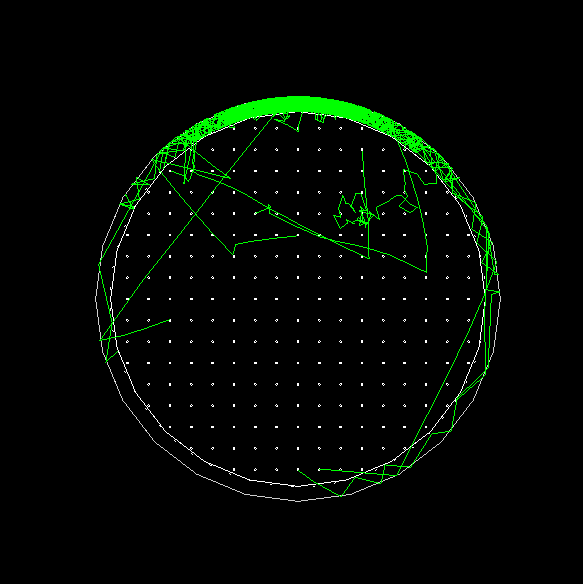}
\end{minipage}
\caption{(Top) Rain of cosmic rays around the \chips{} detector.  (Bottom Left) A 1\,GeV $\mu^-$ entering the Inner Detector from the top center and producing a Cherenkov cone. For a better view, most of the photons are not shown and the veto is disabled. (Bottom Right) A 1\,GeV $\mu^-$ entering from one side of the detector producing Cherenkov light in the veto. The inner detector is disabled for a better view. The white dots represent veto PMTs and the green lines represent photons.  The photons are trapped in the veto until they are absorbed or detected.}
\label{vetoevt}
\label{fig:Event-displays}
\end{figure}

To further understand the impact of these cosmic ray events on the \chips{} detector, we used the GEANT4 framework~\cite{GEANT4} and the cosmic ray flux available through the CRY package~\cite{CRY} to study the efficiency of photon detection and the effect of the event time span on the overall deadtime caused by the 0.5 cosmic ray events per spill.  In the simulations we have assumed a detector comprising two concentric cylinders: an Inner Detector (ID) surrounded by and optically separated from a larger Veto Detector (VD). The veto volume extends \unit[2]{m} outward from the inner detector boundary.  The walls of the ID volume are assumed to absorb light, while the walls of the VD volume are reflective. Figure~\ref{fig:Event-displays} shows event displays from this simulation package.  

\begin{figure}[h]
\centering
\begin{minipage}[b]{0.49\textwidth}
\includegraphics[width=\textwidth]{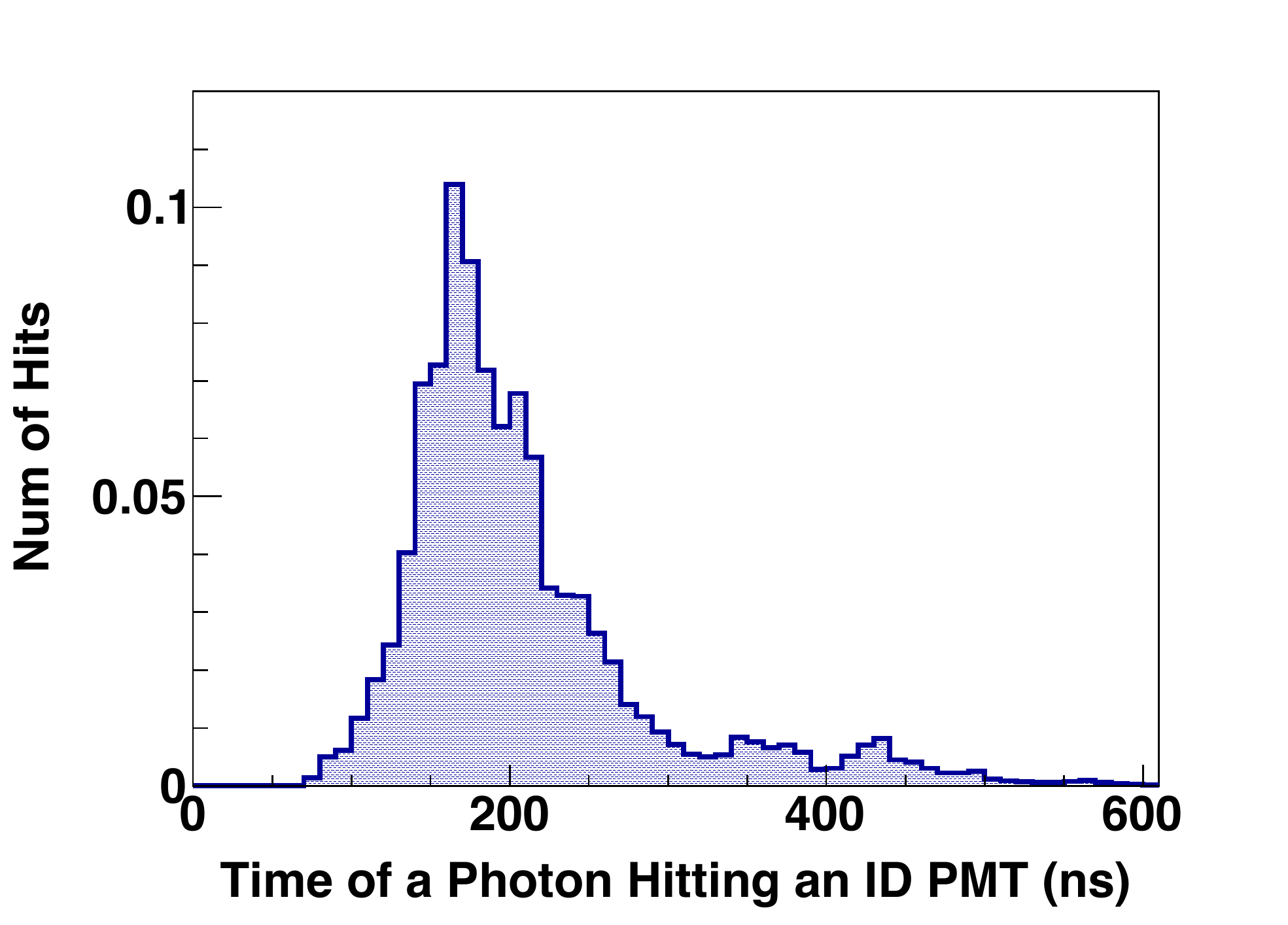}
\end{minipage}
\begin{minipage}[b]{0.49\textwidth}
\includegraphics[width=\textwidth]{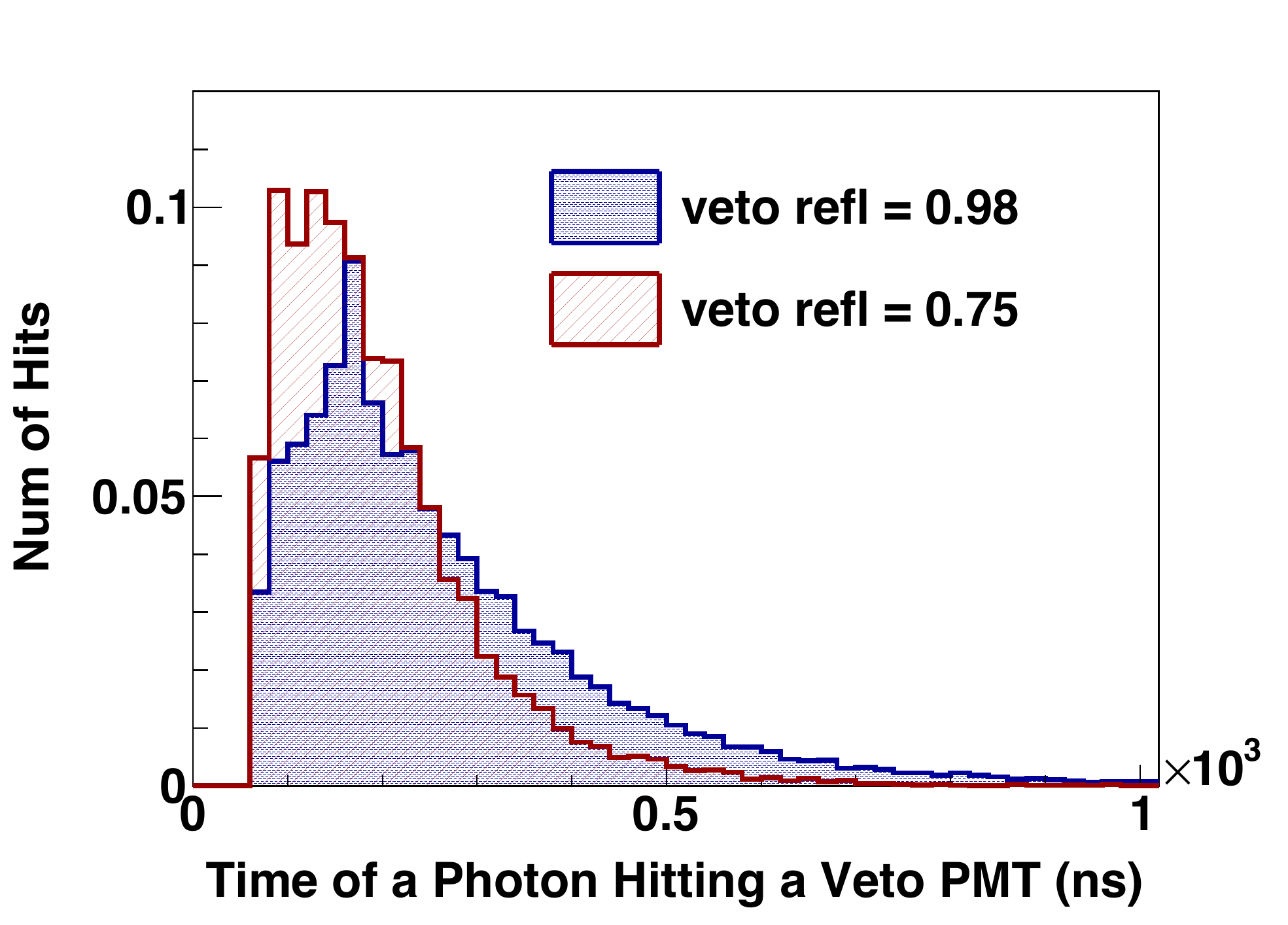}
\end{minipage}
\caption{(Left) Distribution of the event time span of CR muons in the inner detector volume. (Right) Distribution of the event time span of CR muons in the veto volume for different values of the veto wall reflectivity.
}
\label{fig:event-time-span}
\end{figure}

The distribution of cosmic ray event duration is shown in Figure~\ref{fig:event-time-span}.   Average dead time during the spill due to CR muons is (rate)$\times$(event time span), which results in a conservative estimate of the average dead time per spill of \unit[250]{ns}~\cite{CHIPS-veto}.  This is 2.5\% of the beam spill.  For contained CR muon events, the dead time window could be enlarged, perhaps to \unit[1-2]{$\mu$s}, to minimize the impact of muon decay Michel electrons on the beam events.  

\section{Detector Design}
\label{standarddet}
\subsection{Detector Concept}
\begin{figure}[htbp]
\begin{center}
\includegraphics[width=0.7\textwidth]{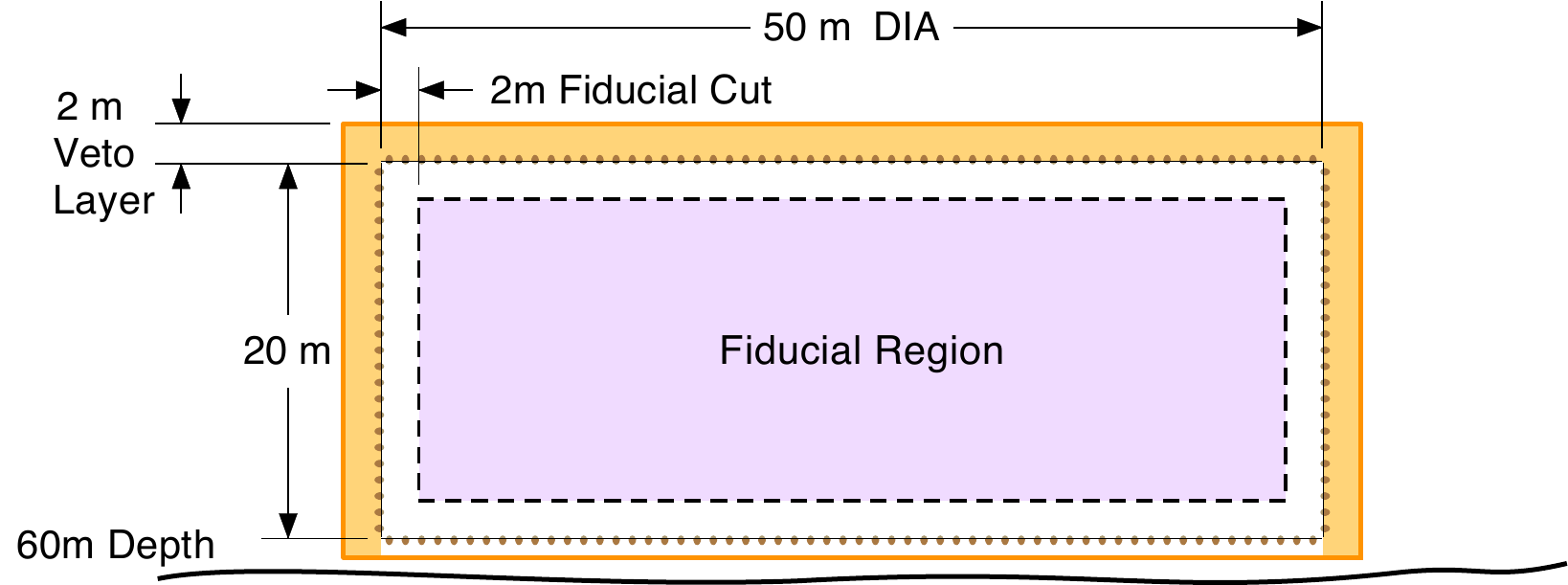}
\end{center} 
\caption{Illustration of the basic module geometry and dimensions}
\label{detscheme}
\end{figure}

Due to practical considerations deploying very large detectors, we propose to build up the needed detector mass in independent, cylindrical units.  Each unit will sit at the bottom of the mine pit.  The detector height is constrained by the depth of the water and the overburden requirements.  Detector dimensions are further limited by the attenuation length of light in the water.   To respect these constraints, each unit will comprise a cylinder of photodetectors surrounding a water volume \unit[20]{m} high and \unit[50]{m} diameter.  Excluding interactions closer than \unit[2]{m} to the photodetector surface, the proposed dimensions yield a fiducial mass of \unit[27]{kton}.  The water enclosed in each unit will be kept dark and isolated from the outside lake water by a reinforced polymer membrane.  Additionally, some photodetectors will be arranged to point outwards into a \unit[2]{m}-thick veto volume along the top and side of the cylinder.  This volume also provides room for the support framework, and is optically separated from the active volume by an opaque plastic sheet between the photodetectors.  Figure~\ref{detscheme} illustrates the module geometry, while Table~\ref{dettab} tabulates the detector parameters.

\begin {table}[h]
\begin {center}
\begin {tabular} {| l r |}
\hline
Feature&\\
\hline
Inner Detector mass&\unit[39.3]{kton}\\		
\hline
Detector geometry&cylinder\\		
Detector dim. (D x H)&\unit[50]{m} x \unit[20]{m}\\		
Inner surface area&\unit[7,069]{${\rm m^2}$}\\
PMT diameter&\unit[10]{in}\\
Photocathode Coverage&10\%\\
No. of PMTs&13,854\\
\hline
Overburden&\unit[40]{m.w.e}\\
CR muon rate&\unit[50.5]{kHz}\\
In-spill CR occupancy&0.51\\	
Event dead time&\unit[500]{ns}\\
\hline
Veto Detector dim. (D x H)&\unit[54]{m} x \unit[22]{m}\\
Veto medium&water\\
Veto photocathode Coverage&0.5\%\\
No.  of veto PMTs&626\\
Veto PMT diameter&\unit[10]{in}\\
\hline
\end {tabular}
\end {center}
\caption{Summary of cosmic rate and key features of the basic \chips{} module.}
\label{dettab}
\end {table}

\subsection{Photodetectors} 
\begin{figure}[h]
\begin{center}
\begin{tabular}{cccc}
\includegraphics[width=.3\textwidth]{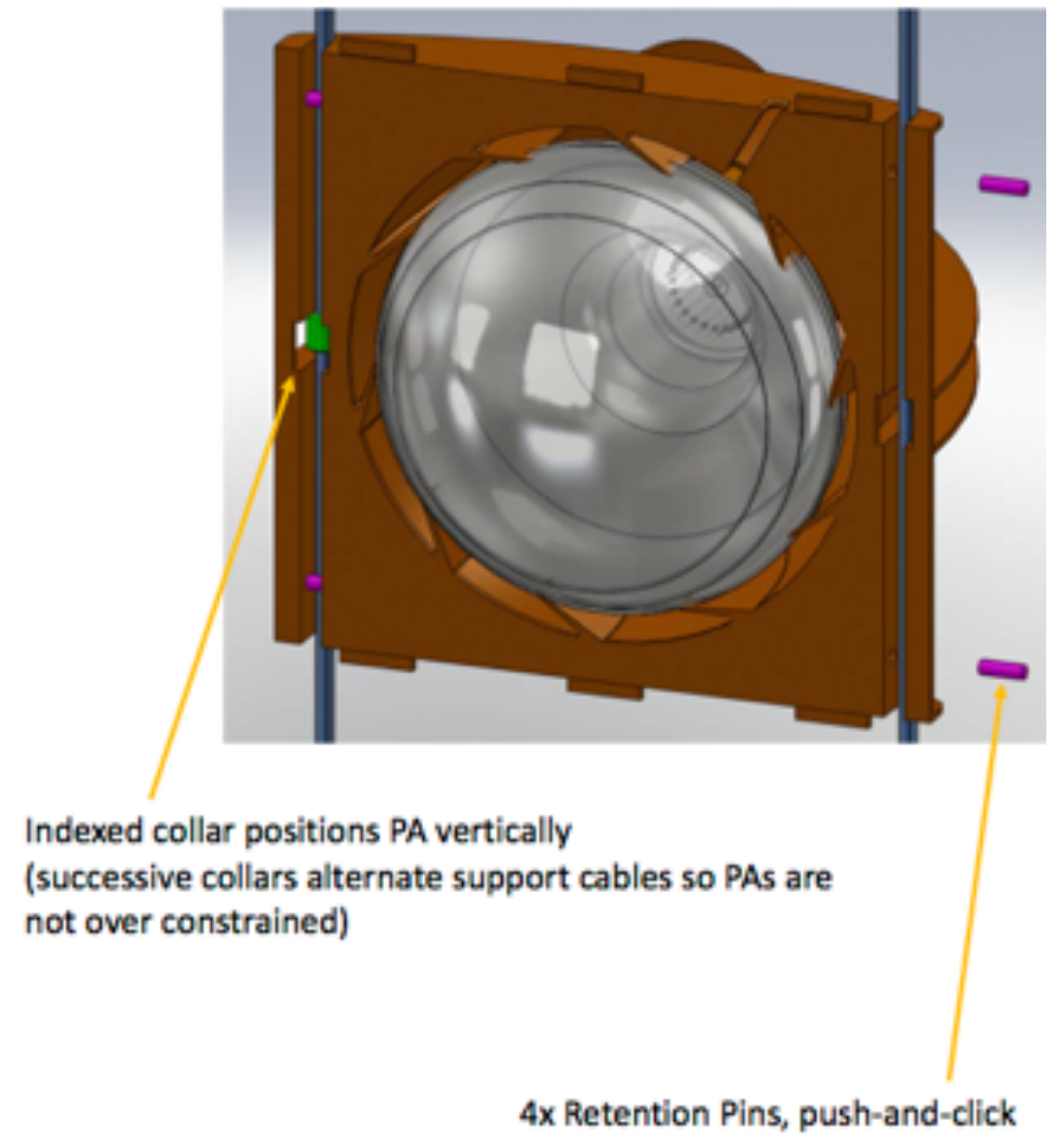}&\includegraphics[width=.1\textwidth]{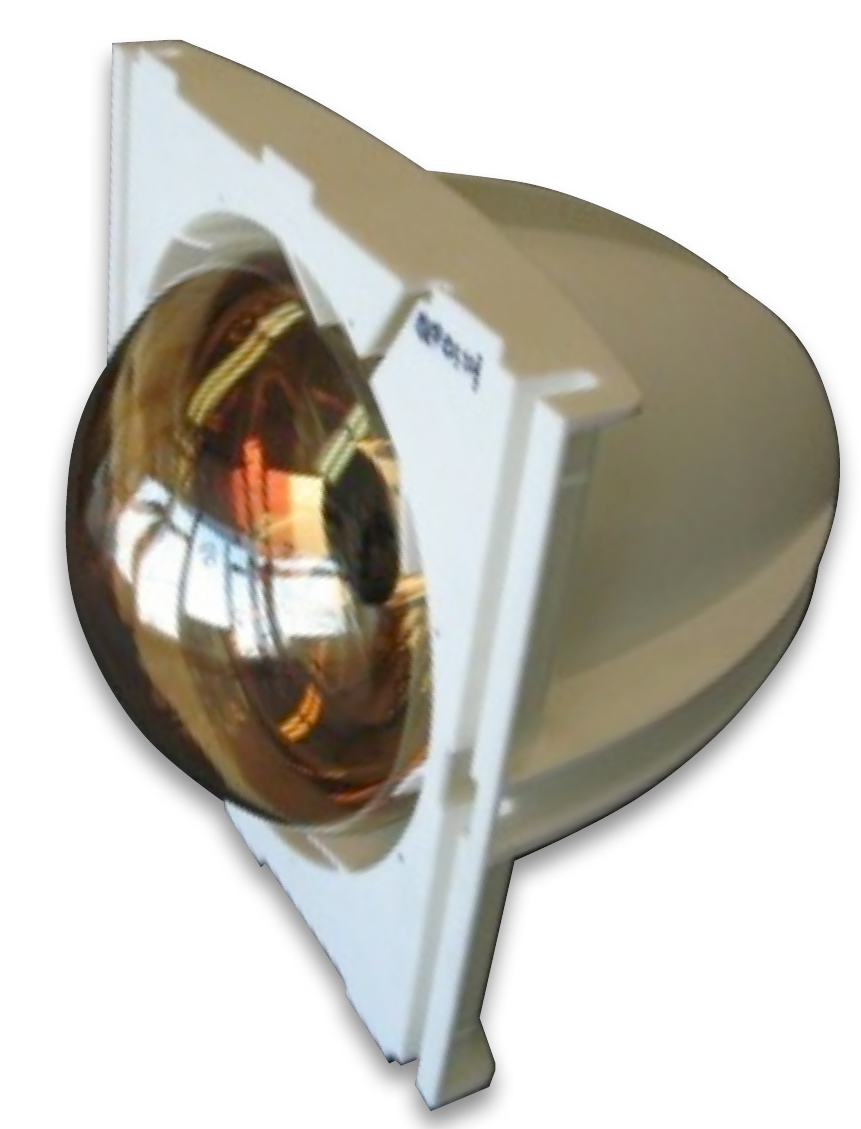}&
\includegraphics[height=.2\textheight]{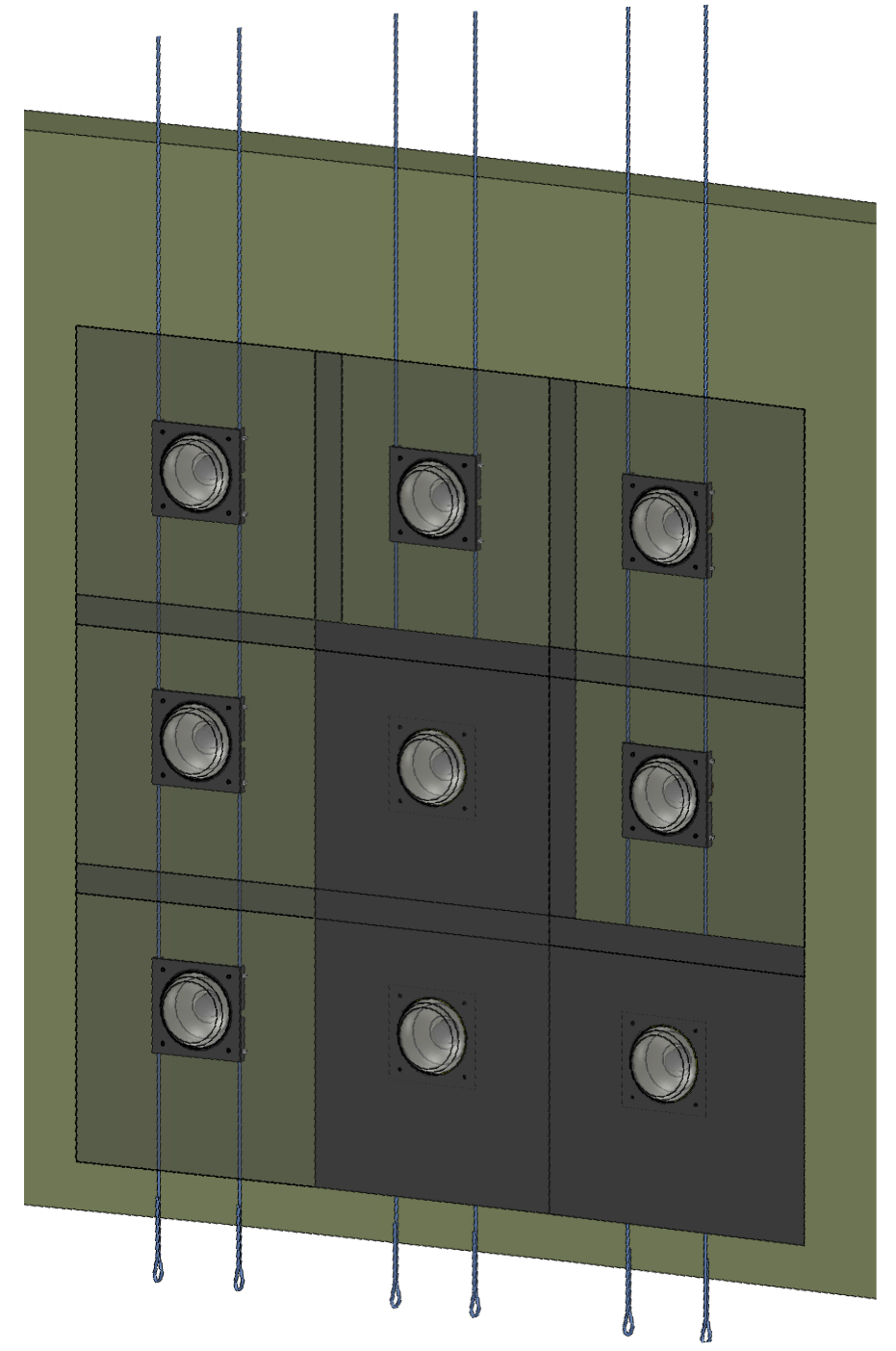}&\includegraphics[height=.3\textheight]{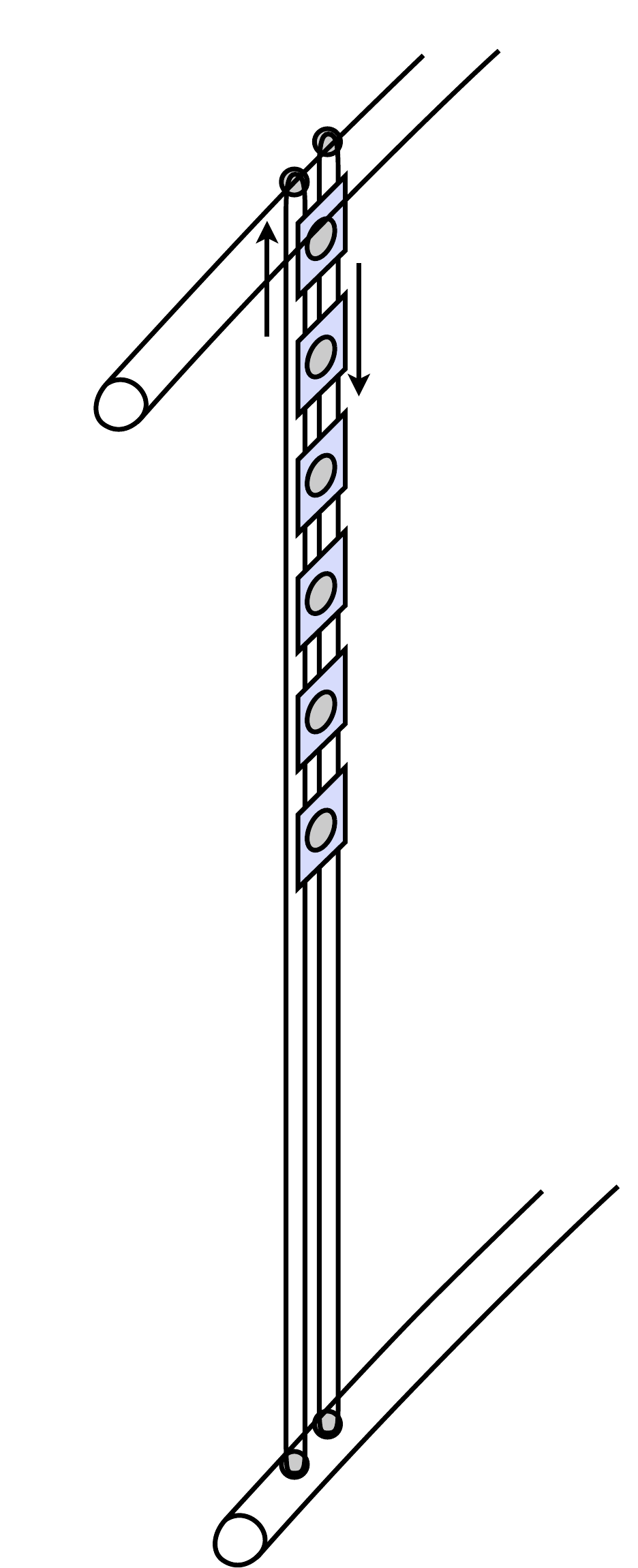}\\
\end{tabular} 
\end{center}
\caption{Depictions of the PMT housings.  (Far Left) The collar of the PMT housing.  (Middle Left) Side view of the PMT housing and collar.  (Middle Right) PMT housings in the framing mounts.  (Far Right) PMT housings on the cable mounts.  Figures from Ref~\cite{LBNECDR}.}
\label{pmtdeployfig}
\end{figure}

The nominal design calls for high quantum efficiency (HQE) \unit[10]{\inch} photomultiplier tubes from Hamamatsu~\cite{Hamamatsu}.  As in the LBNE water Cherenkov design~\cite{LBNECDR}, we assume the HQE tubes can achieve the same efficiency with 10\% photosensor coverage as Super-K achieved with 20\% coverage using lower QE tubes.  A simulation and reconstruction program is under development to determine the optimal coverage and placement of the tubes, as is described in Section~\ref{simsec}. The tubes will need to withstand substantial pressure.  The \unit[10]{\inch} tubes have been shown in tests done by LBNE~\cite{LBNECDR} to survive down to \unit[60]{m}, but this is at the edge of the comfort zone.  The \unit[12]{\inch} tubes from Hamamatsu do withstand more than \unit[60]{m} hydrostatic pressure~\cite{LBNECDR, Hamamatsu, ADITPMT}, and would be an appropriate replacement should the \unit[10]{\inch} tubes not suffice.    The comparatively low cost of the deployment and support system is a strong motivation for use of cheaper photodetectors currently in development~\cite{HybridPMT} (see Section~\ref{randdsection} for further discussion on photodetector strategy).  Individual photodetectors on the bottom and top surface of the cylinder will be mounted on a lightweight truss framework, or ''space frame'', extending \unit[1-2]{m} perpendicular to the instrumented plane; frame components will incorporate only enough mass to approximately cancel the buoyancy of the photodetectors, so that the assembly remains neutrally buoyant and spans the \unit[50]{m} diameter without significant distortion.  Molded plastic housings will be used to gently hold photodetectors while providing a secure mounting system, similar to those designed and tested for the LBNE-WCD option~\cite{LBNECDR}.  Photodetectors on the cylinder walls can be secured to vertical steel support cables~\cite{LBNECDR,IMB} or to a framework similar to the bottom and top planes.  The framing and/or support cables are in turn secured between large stiff rings, defining an overall \unit[20]{m}-high cylinder of instrumentation that can be raised and lowered as needed.  Illustrations of the PMT housings developed at Physical Sciences Laboratory (University of Wisconsin) are shown in Figure~\ref{pmtdeployfig}, which are compatible with either framing or cable supports as shown.  Cost estimates for the PMT assemblies are given in Table~\ref{pmtcosttable}.

\begin{table}[h]
\begin{center}
\begin{tabular}{|lr|}
\hline 
Item&Cost per Channel\\
\hline
PMT, 12'' HQE&\$1,800\\
Frame Housing&\$34\\
Base Encapsulation&\$93\\
HV Base&\$34\\
HV Supplies&\$45\\
Front End,trigger,DAQ&\$80\\
Cables&\$150\\
PIU (support framing)&\$200\\
\hline
Total per channel&\$2,436\\
Total 13K channels&\$33,750,000\\
Engineering Cost&\$3,000,000\\
\hline
\end{tabular}
\end{center}
\caption{Cost of the PMT assemblies}
\label{pmtcosttable} 
\end{table}

\subsection{Detector Vessel}
Surrounding each \unit[27]{kton} detector unit is a reinforced polymer membrane (liner) that blocks outside light and isolates the pure water inside the modules from the pit water.  Many commercially available liner material options exist and are regularly used in the geomembrane and roofing industries for blocking water over large areas~\cite{Scheirs}.  For \chips{}, the liner will be maintained in a cylindrical shape by a framework and cables connecting two large stiff rings.  Such rings and associated mooring lines are routinely used for construction of net cages in the aquaculture industry~\cite{Sunde, Moe}.  Rings up to \unit[64]{m} diameter have been deployed in open sea conditions~\cite{Aqualine}.  

The polymer liner may be contracted as a design-build project.  There are several relevant examples in the literature to guide the design of the \chips{} liner.  A baseline material is Hypalon, which was the proposed material in the GRANDE detector design~\cite{Grande}. Hypalon is a chlorosulfonated polyethylene (CSPE) synthetic rubber (CSM) that was previously manufactured by DuPont. In 2010, DuPont ceased manufacturing Hypalon, but several other manufacturers are still operational and there are other viable options on the market~\cite{HypalonDone}. Some alternative liner materials have been investigated, including XR-5 manufactured by Layfield~\cite{XR-5}, polypropylene, and polyethylene materials. These materials may provide more economical alternatives to the mainstream Hypalon/CSPE.

Because of forces acting on the liner surface, the design is expected to require additional support to relieve stresses in the liner and maintain its cylindrical shape.  The main challenge is posed by differences in water density that may occur between the inside and outside of the liner volume.  Lakes exhibit a time-dependent temperature vs. depth profile, chemical concentration profile, and density profile~\cite{Boehrer}.  If the temperature profile inside the liner lags that outside by \unit[5]{\degree C}, the density effect causes differential pressures up to \unit[100]{N/${\rm m}^2$} across the liner surface; for a surface with curvature radius \unit[25]{m}, the resulting tension approaches the tearing strength of available liner materials.  Supporting such a pressure difference on either the top or bottom flat of the cylinder is even more impractical than on a curved side.  In addition, while motion of water in small lakes is modest compared to open seas, storm driven flows (seiches) reach up to \unit[20]{cm/sec} well below the surface~\cite{Boegman}.  Such flows create a dynamic pressure on the cylinder wall, which can be regarded as a ``bluff body'' in the turbulent flow regime~\cite{King}.  While corresponding forces are only of order \unit[1]{N/${\rm m^2}$}, they are asymmetric and act over large almost-flat areas; the forces can also be magnified by vortex-excited oscillations~\cite{Bearman}.

\begin{figure}[h]
\begin{center}
\includegraphics[width=\textwidth]{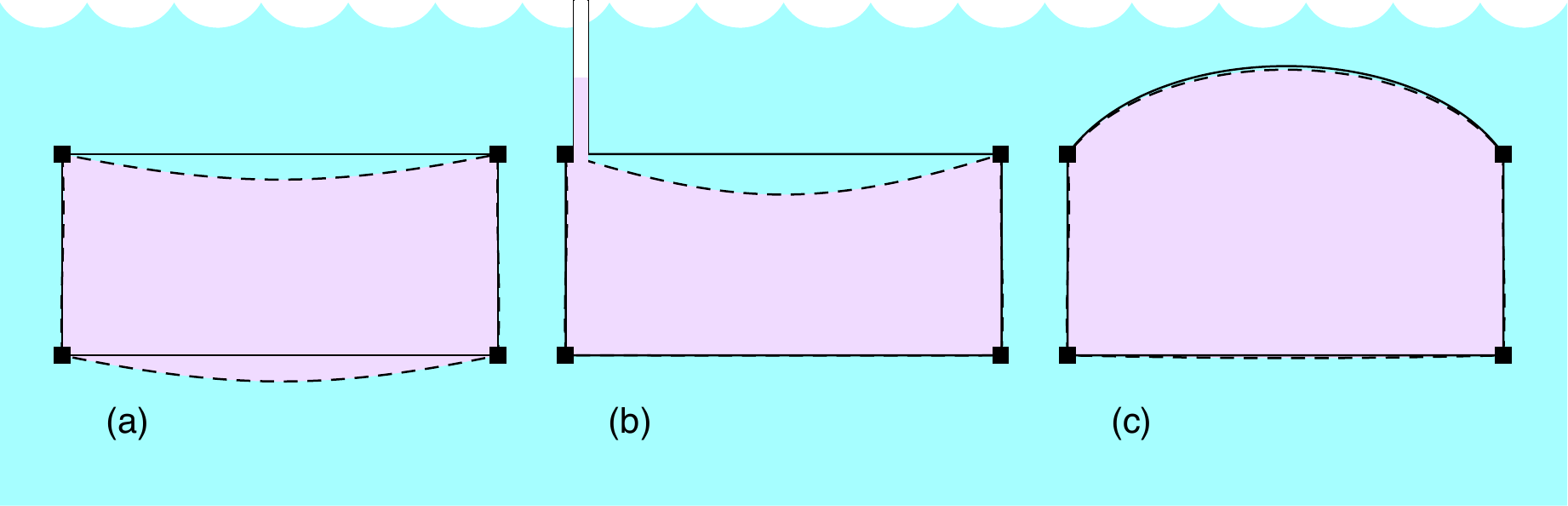}
\end{center}
\caption{Side view of submerged thin-walled cylinders filled with liquid denser than surroundings.  (a) Sealed cylinder with flat top and bottom.  (b) Cylinder with flat top and bottom, with riser tube allowing reduction of pressure until bottom pressure is in equilibrium.  (c) Sealed cylinder featuring domed top.  Solid squares show constraints assumed around perimeter to prevent the structures sinking.  Solid lines indicate nominal shape; dashed lines indicate deformation under load, which is greatest for large flat surfaces in (a) and (b).}
\label{structgeomfig}
\end{figure}

Analysis of the density profile issue reveals that a completely submerged cylinder tends to experience substantial differential pressure on the top, bottom or both flat surfaces, which are hard to restrain.  This follows from computing the pressure increase from top to bottom, which cannot match between inside and outside if the densities are different.  However, a single horizontal surface on a submerged volume naturally experiences low differential pressure as long as the remainder of the structure is much stiffer, for example by virtue of a support frame that maintains a curved shape.  For \chips{}, the bottom of each cylinder is chosen as the single horizontal surface, whereas the top will be covered by a structural dome, as illustrated in Figure~\ref{structgeomfig}.  The dome needs to be sealed with a liner similar to the side wall, but an additional membrane will isolate its volume from the pure detector water without compromising its structural function. The vertical side walls and the dome will still be subject to differential density pressure rising with distance above the bottom surface, but this can be reduced to around \unit[10]{N/${\rm m}^2$} by management of the thermal profile inside the liner.  A truss framework would provide effective reinforcement of the liner walls against remaining forces, and other options such as tension cables and rope netting will also be considered.  Costs associated with the detector structure are provided in Table~\ref{vesselcosttable}.

\begin{table}[h]
\begin{center}
\begin{tabular}{|lrr|}
\hline
Item&First \unit[50]{m}$\times$\unit[20]{m}&Additional \\
&module (DxH)&modules\\
\hline
Engineering&\$1,000,000&\$200,000\\
Marine Cage Superstructure, 3 rings&\$250,000&\$250,000\\
Steel framework for liner support&\$1,000,000&\$1,000,000\\
Liner (\unit[\$50]{${\rm m^{-2}}$})&\$500,000&\$500,000\\
Deploy PMT modules, 9 FTEs&\$900,000&\$600,000\\
Water Purification System&\$1,400,000&\$1,400,000\\
\hline
Total&\$5,050,000&\$3,950,000\\
\hline
\end{tabular}
\end{center}
\caption{Cost of a detector vessel module.}
\label{vesselcosttable}
\end{table}

\subsection{Construction and Deployment}
\begin{figure}[h]
\begin{center}
\includegraphics[width=\textwidth]{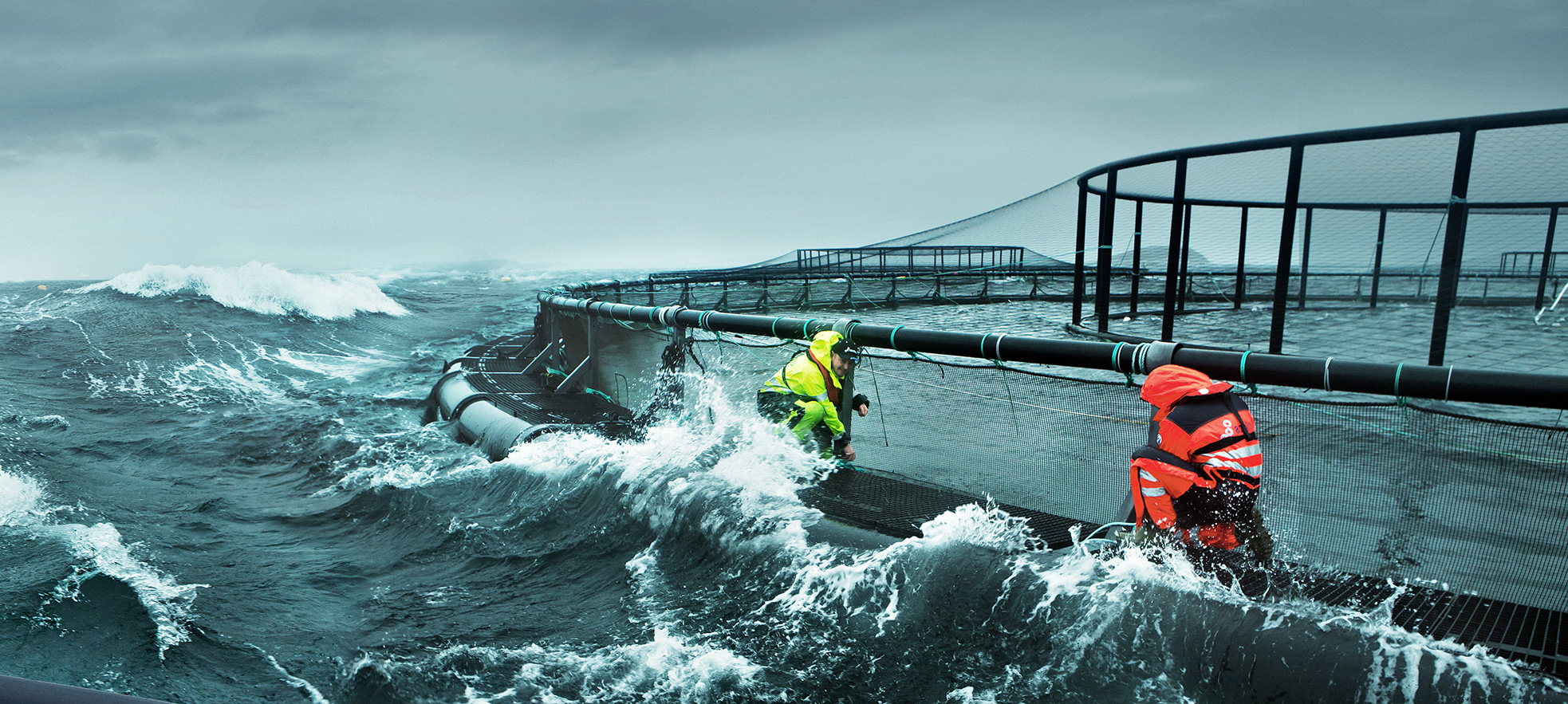}\\
\vspace{3mm}
\begin{minipage}{0.32\textwidth}
\includegraphics[width=\textwidth]{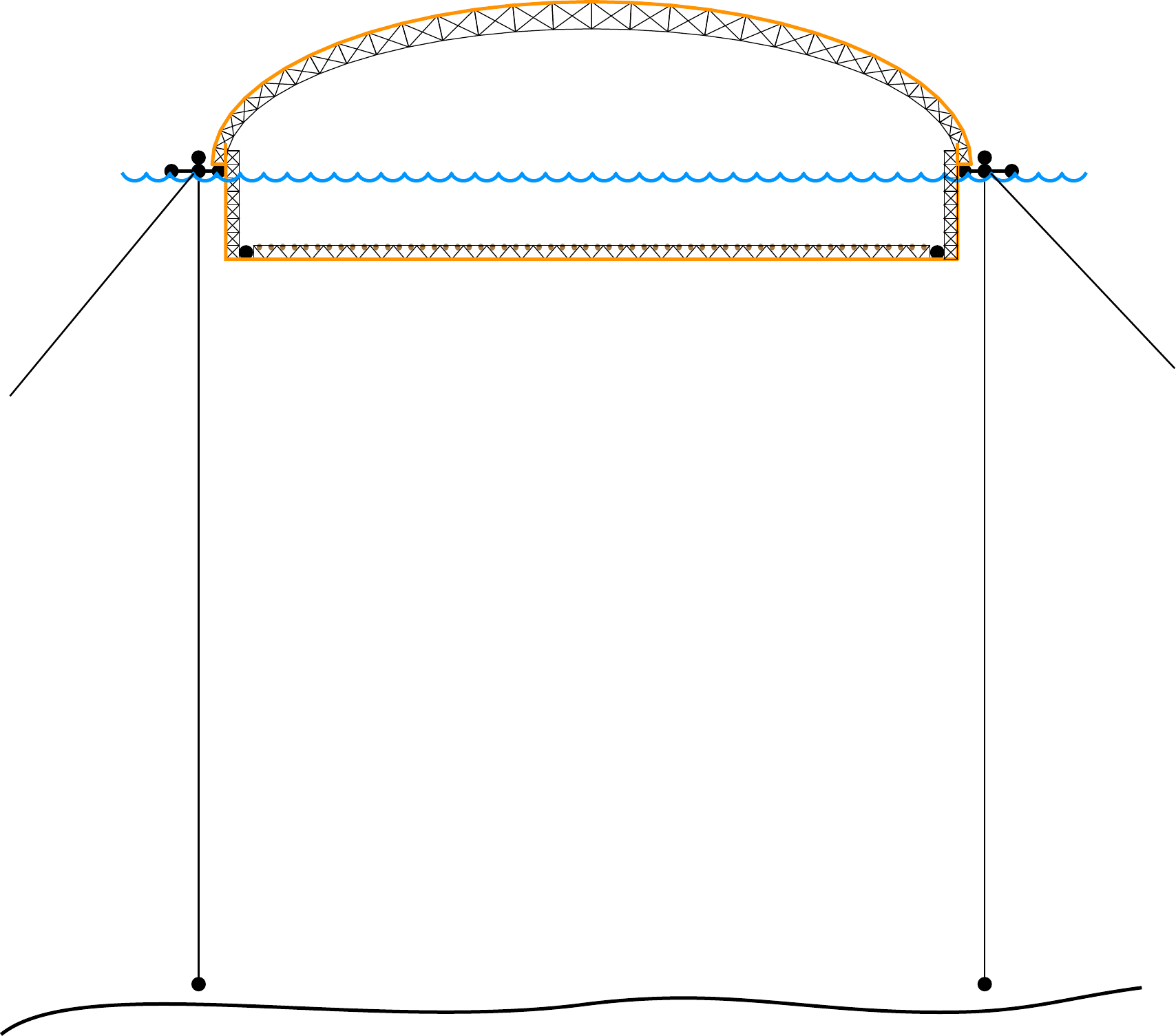}
\end{minipage}
\begin{minipage}{0.32\textwidth}
\includegraphics[width=\textwidth]{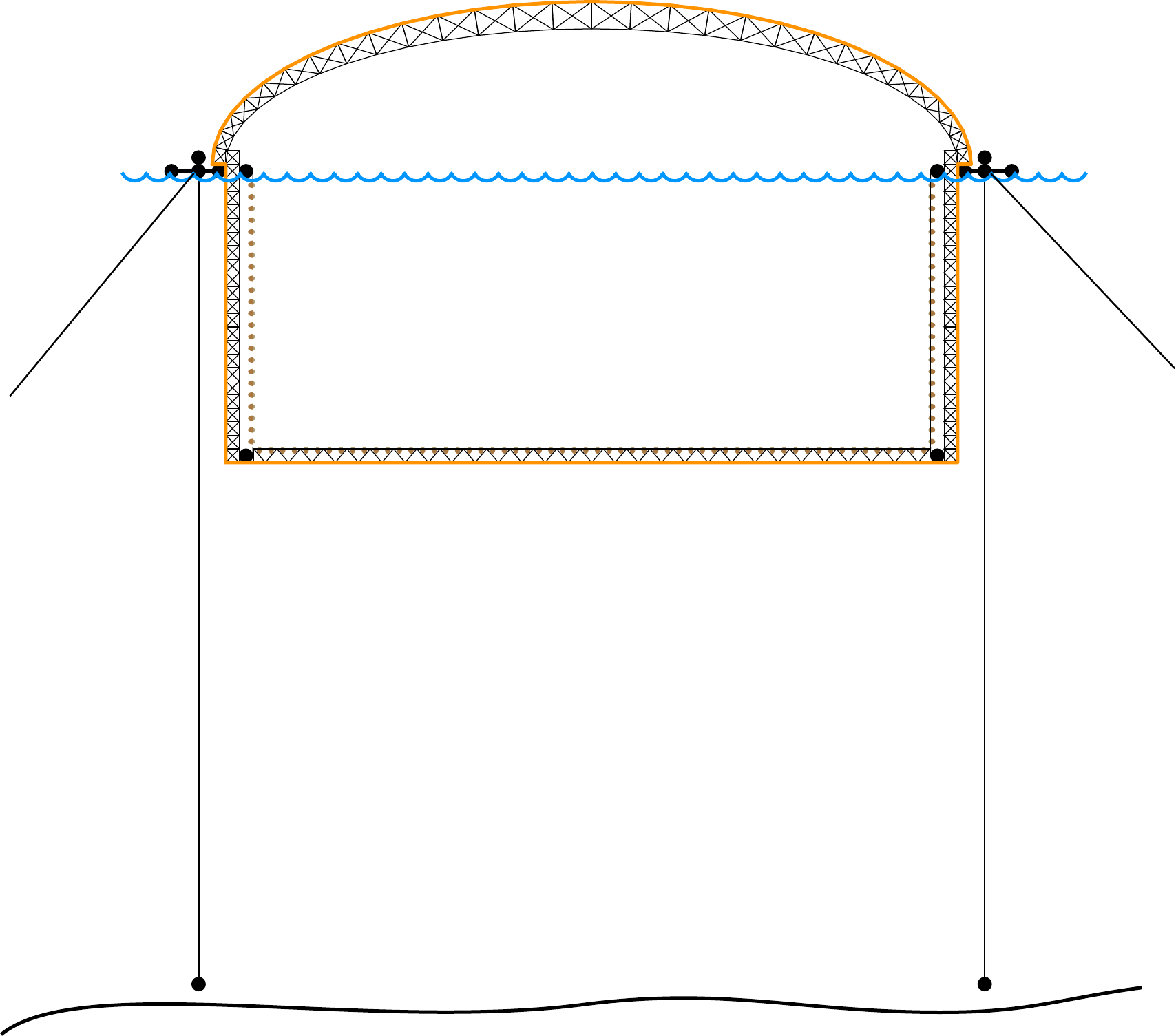}
\end{minipage}
\begin{minipage}{0.32\textwidth}
\includegraphics[width=\textwidth]{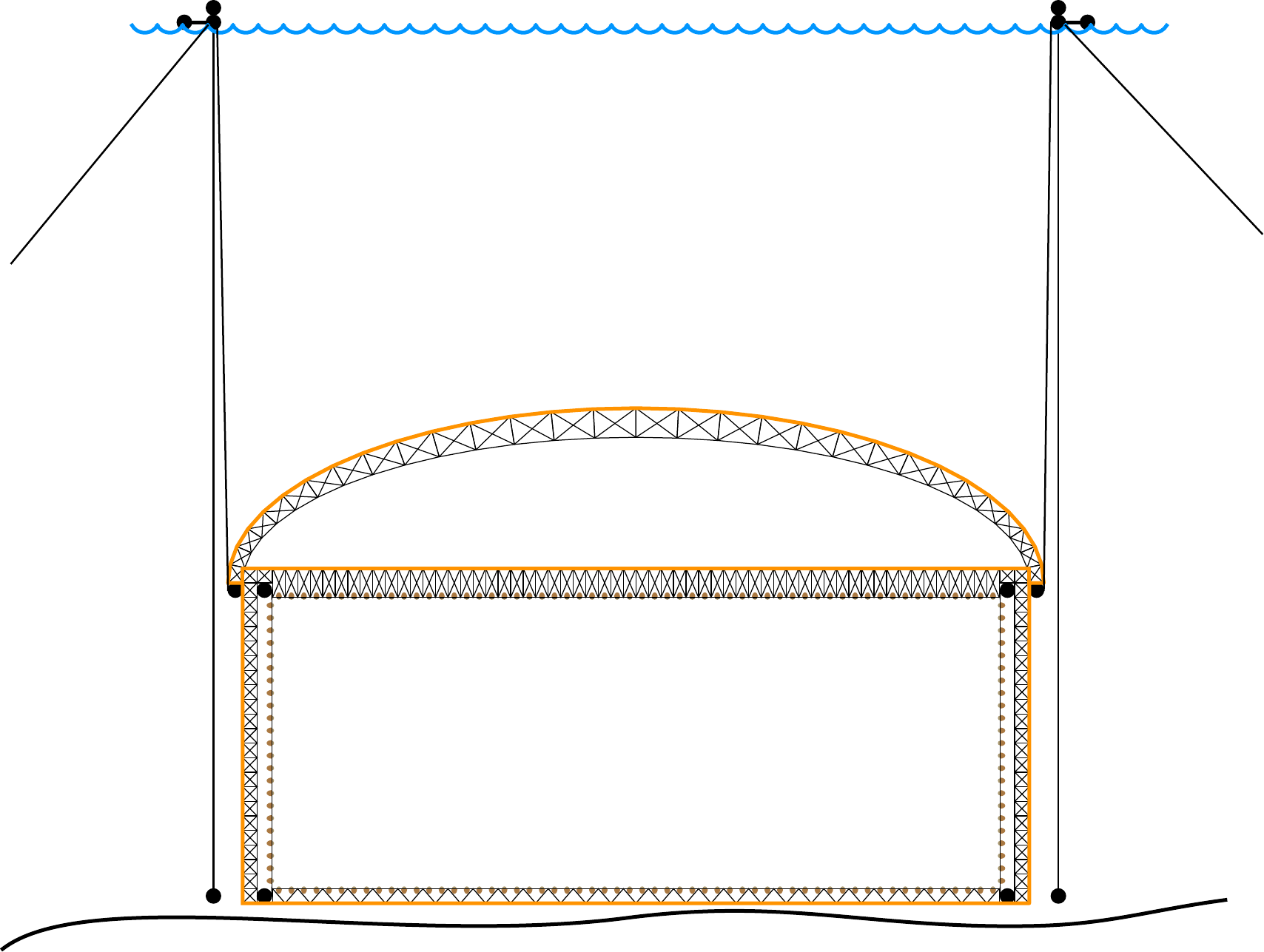}
\end{minipage}
\end{center}
\caption{(Top) Floating ring and platform used in aquaculture cages.  Note the Wentworth Pit is not expected to ever be as turbulent as the open sea.  (Bottom) Pictorial representation of detector construction then submersion.}
\label{scaryfig}
\end{figure}

A cable or net cage will be moored in the lake surrounding each intended detector location, supported at the surface by a large floating ring and held at the bottom by a large sinker ring, like those used for the aquaculture cages~\cite{Sunde, Moe, Aqualine} shown in Figure~\ref{scaryfig}.  As shown in the bottom panels of Figure~\ref{scaryfig}, the detector will be built incrementally downward, supported by the surface ring, gradually flooded and lowered (or raised) inside the outer cage by cables.  During construction activities, the top dome will be held above the lake surface (empty) and will serve as a sheltered work area preventing contamination of the interior purified water volume.  Once the cylinder is complete and capped by liner, the dome will be sealed to the cylinder top and then also flooded and sunk.  During the winter, standard marina equipment will be used to circulate water near the floating ring and keep it decoupled from the ice sheet around it.

Each photodetector is served by a single electrical cable which is routed along the support framework and emerges in bundles near the top of the cylinder.  The emerging bundles are sealed or surrounded by additional liner material and routed to on-shore power supply and data acquisition equipment.  Alternatively, floating trailer-sized enclosures at one side of each surface ring could house the first level electronics, with consolidated power and high-speed communication connections from there to shore.  Additional connections to shore will be necessary for the flow of purified water.  These connections can be sealed to suitable openings in the liner during deployment, and the corresponding umbilicals will then be routed to on-shore purification equipment.  Multiple connections at different depths will allow better matching of the lake's thermal profile if each umbilical is supported at approximately constant depth from a line of buoys.

\begin{figure}[h]
\begin{center}
\begin{minipage}{0.49\textwidth}
\includegraphics[width=\textwidth]{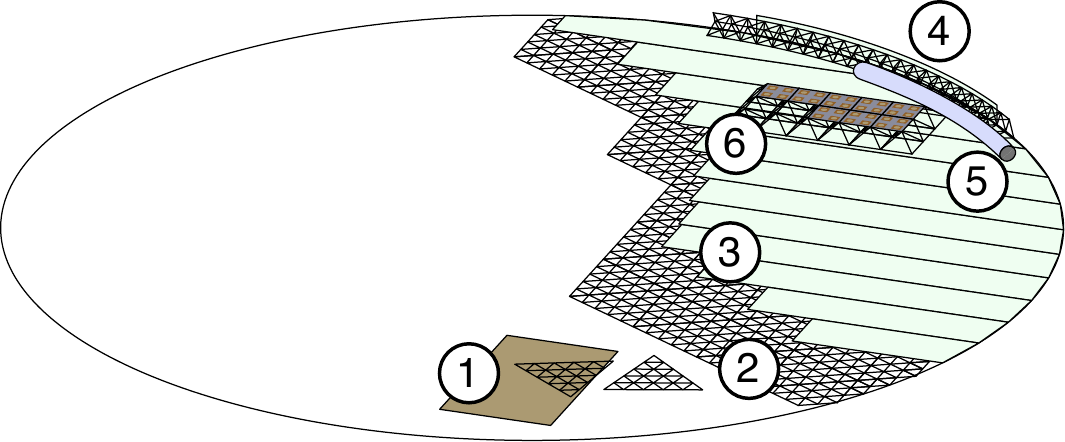}
\end{minipage}
\begin{minipage}{0.49\textwidth}
\includegraphics[width=\textwidth]{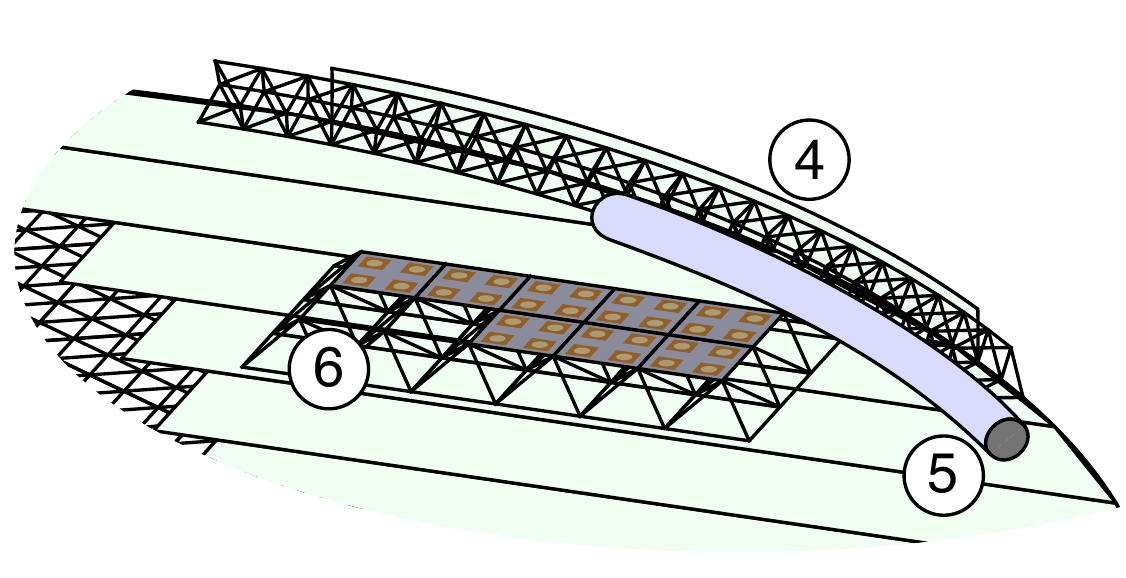}
\end{minipage}
\end{center}
\caption{Construction of cylinder bottom floating on water inside the dome.  1 - Assembly of raft on work platform.  2 - Joining of rafts.  3 - Unrolling and seaming of liner sections.  4 - Framing for beginning of vertical wall, with vertical liner strip attached.  5 - Bottom support ring for vertical cables.  6 - Bottom surface framing with panels of photodetectors partially installed.}
\label{platformfig}
\end{figure}

Construction of each cylinder will begin with assembly of the upper and lower rings defining the outer cage, together with an inner concentric ring on top of which the dome is built.  This work can be done near shore, then towed to the final location and moored.  Within the dome, additional temporary floating docks will enable workers to assemble sections of the large bottom panel and subsequently to join them, as shown in Figure~\ref{platformfig}.  Each section could start with a triangular raft \unit[5]{m} per side, assembled on the dock as a \unit[1]{m} lattice of PVC pipe lengths, then floated and finally secured to preceding raft sections.  With inner pipe diameter of \unit[4]{\inch}, each raft section can temporarily support \unit[360]{kg} for the additional work of unrolling and welding together sections of liner and building a \unit[1]{m}-high network of support framing on top of that.  Because of the raft support, plastic welding of liner strips or large prefabricated liner sections can be carried out above the water line, and standard field techniques~\cite{Scheirs} can be used.  Around the growing perimeter of the base section, the liner will be wrapped upwards along the support framing.  Photodetectors in housings will also be attached to the framework at this stage, and cables routed as needed.  Once complete, the raft pipes can be filled with water to eliminate their buoyancy, but still leaving the entire bottom panel temporarily in a floating state.  

An internal ring will also be assembled and joined to the framing inside the perimeter, which helps to stiffen the bottom circumference and provide an attachment point for vertical deployment cables that ultimately extend to a corresponding ring \unit[20]{m} above.  Again, as in Figure~\ref{scaryfig}, the remaining vertical wall of the liner will then be built up around the perimeter in \unit[1]{m} increments, while flooding the existing volume with enough purified water to maintain the top edge just above the water line where it is easily accessed by workers on floating dock sections. After completion of the walls, the second ring will be attached and serve as the top attachment point for the vertical deployment cables.  Similar to IMB~\cite{IMB} and the concept for LBNE-WCD, photodetectors in plastic frames will be attached in sequence and lowered along each pair of support cables, with the cables looped around pulleys at top and bottom to allow the necessary motion.

After completion of the walls, the top framework is built in from the perimeter, utilizing integral PVC pipe for temporary flotation.  Construction of the wall and top sections includes installation of light barrier material to define the veto volume.  Finally the cylinder top is sealed with liner and all temporary flotation devices are filled with water to allow submerging the detector. This includes the dome which can be flooded with pit water. In principle, the assembly process can be stopped at any time, allowing a partially complete detector to be lowered and later retrieved for further work.

\subsection{Cosmic Veto}
\label{cosmicvetosection}
At \unit[40]{m.w.e} in such a large detector, the cosmic ray rate will not be negligible;  these background muons must be tagged for removal.  To this end, a fraction of the photodetectors will be arranged to point outwards into a \unit[2]{m}-thick veto volume along the top and side of the cylinder, where they will reliably detect Cherenkov light from background muons.  This veto volume also provides room for the support framework, and is optically separated from the active volume by opaque plastic sheets between photodetectors.   The PMTs are the same type as those of the inner detector and are mounted on the same wall. 

\begin{figure}[h]
\centering
\begin{minipage}[b]{0.49\textwidth}
\includegraphics[width=\textwidth]{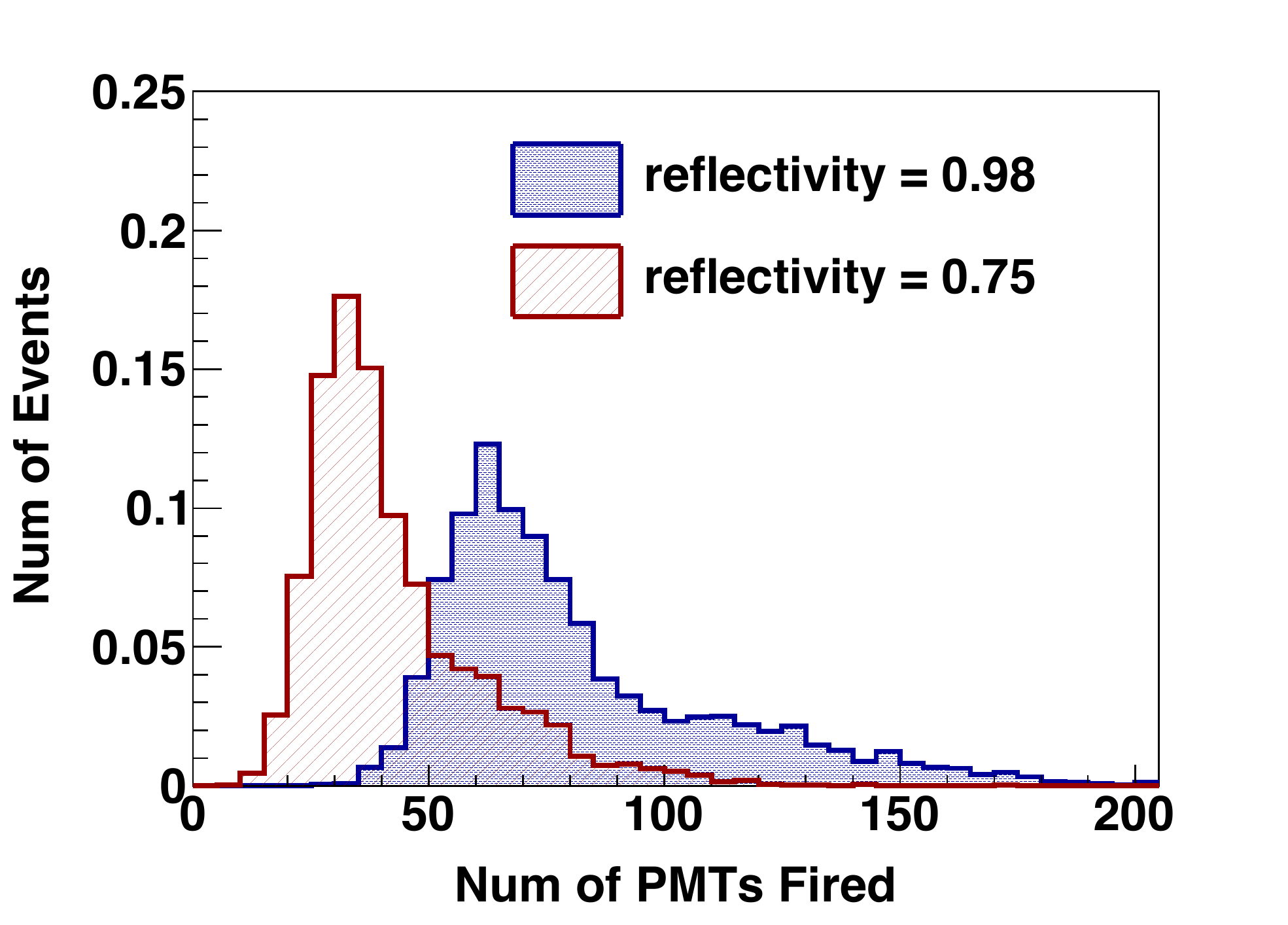}
\end{minipage}
\begin{minipage}[b]{0.49\textwidth}
\includegraphics[width=\textwidth]{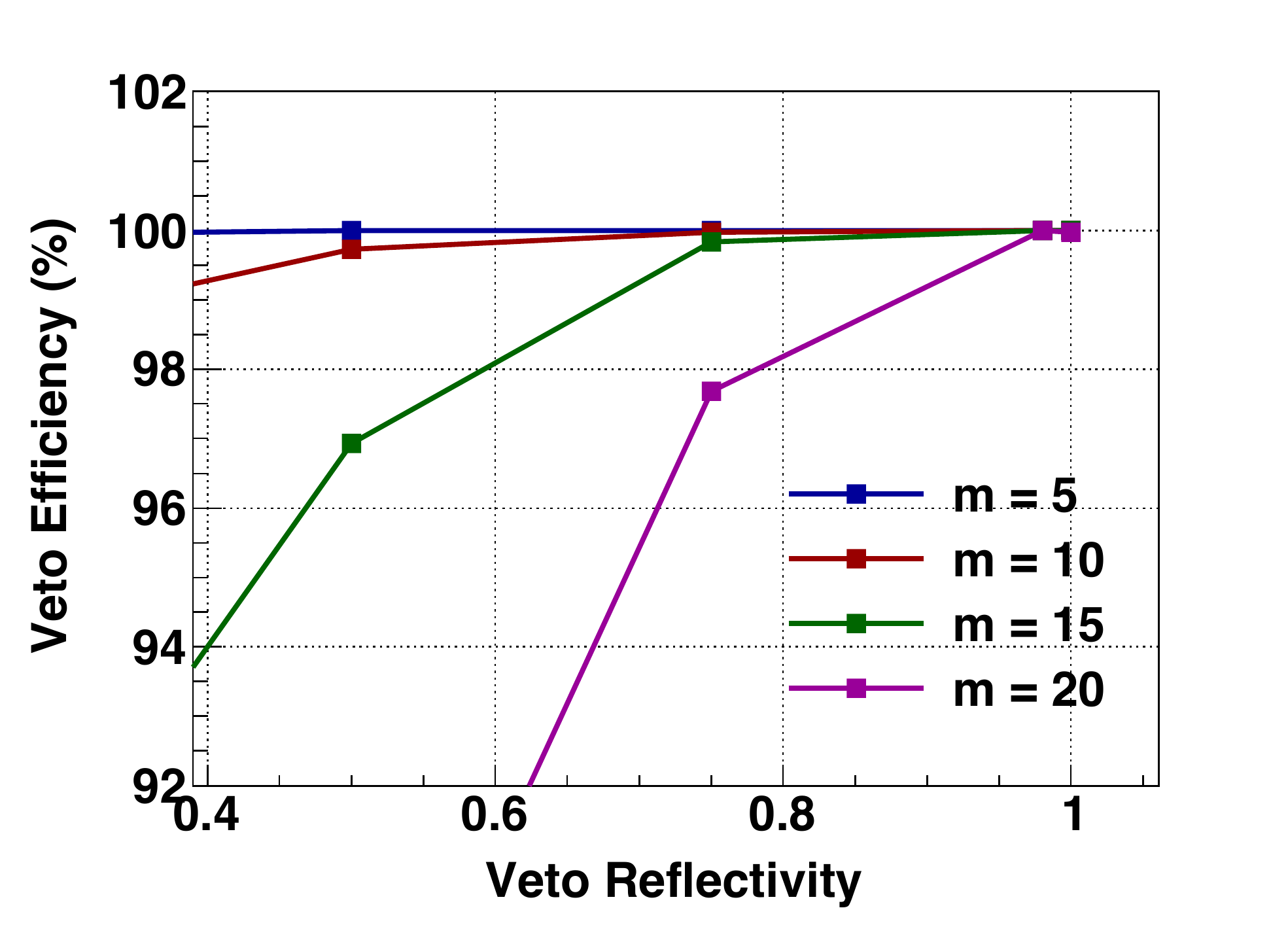}
\end{minipage}
\caption{(Left) The dependence of number of PMTs fired in the VD per event on the assumed reflectivity of the veto detector wall.  Histograms are area normalized to 1.  (Right) The number of cosmic ray muon events vetoed divided by the total number of cosmic ray muon events vs. reflectivity.  The different colors are for different requirements on the minimum number of fired PMTs to tag a muon (m).  PMT spacing is fixed at \unit[284.7]{cm}.}
\label{reflpmt}
\end{figure}
\begin{figure}[h]
\centering
\begin{minipage}[b]{0.49\textwidth}
\includegraphics[width=\textwidth]{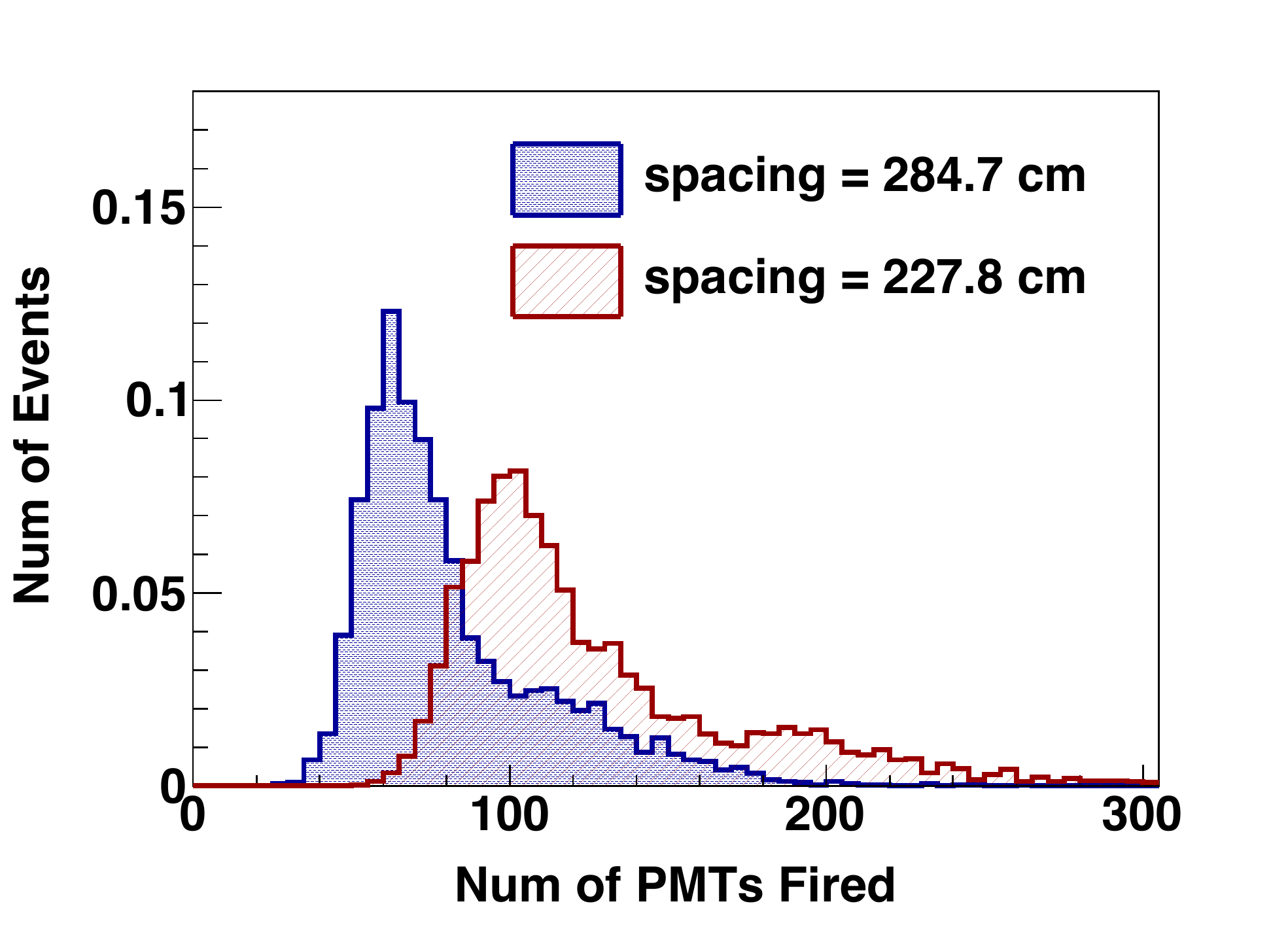}
\end{minipage}
\begin{minipage}[b]{0.49\textwidth}
\includegraphics[width=\textwidth]{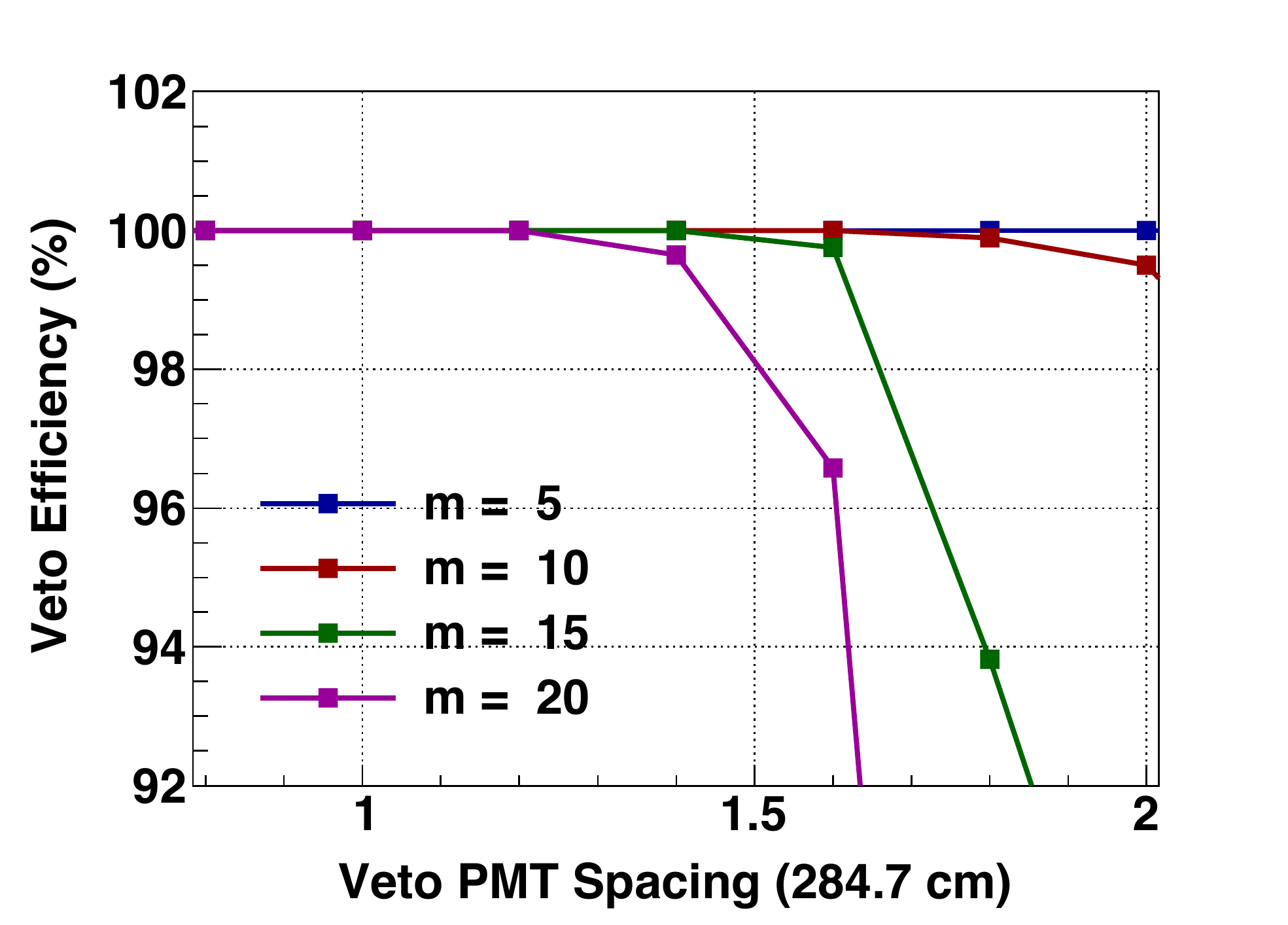}
\end{minipage}
\caption{
(Left) The number of PMTs fired in the VD as a function of the PMT spacing. 
(Right) Veto efficiency as a function of the PMT spacing.  In the right plot, 1 represents \unit[284.7]{cm} which is the spacing for one veto PMT to cover a  $4\times 4$ ID PMT array. Veto detector wall reflectivity is set to 0.98.
}
\label{pmtdnum}
\end{figure}

The role of the veto is to efficiently tag and measure time of CR muons entering the veto and possibly penetrating into the inner detector. The efficiency of different configurations was studied using the cosmic ray simulation.  The PMT readout threshold is set to \unit[1]{PE}. Taking into account the QE and the threshold, a PMT ``fires'' if 10 photons hit its surface.  A veto is defined as a coincidence of $m$ or more veto detector PMTs firing, where $m$ is an integer allowed to vary in the tests.  Figure~\ref{reflpmt} shows the number of veto PMTs that fire and the veto efficiency for different assumed liner reflectivity values.  Figure~\ref{pmtdnum} shows the number of veto PMTs that fire and the efficiency for different veto PMT spacing.  It is found that it is sufficient if each veto PMT covers a $4\times 4$ ID PMT array.  Assuming a 10\% photocathode coverage, a total of 626 PMTs are needed for this configuration.  

\section{Water Purification}
A long attenuation length of light in the water is critical for the Cherenkov radiation to reach the PMTs.  Furthermore, knowledge of the attenuation length is critical for accurate modeling of the detector.  Though remarkably clear, the Wentworth pit water is not clear enough for the detector volume, which requires a light attenuation length of $\sim$\attenlength.  The detector volume water will need to be purified to attain and maintain water clarity.  Water purification is a standard technology that will likely be implemented through a design-build or a design-build-operate contractor. 

\begin{figure}[p]
\begin{center}
\begin{minipage}{0.80\textwidth}
  \includegraphics[width=\textwidth]{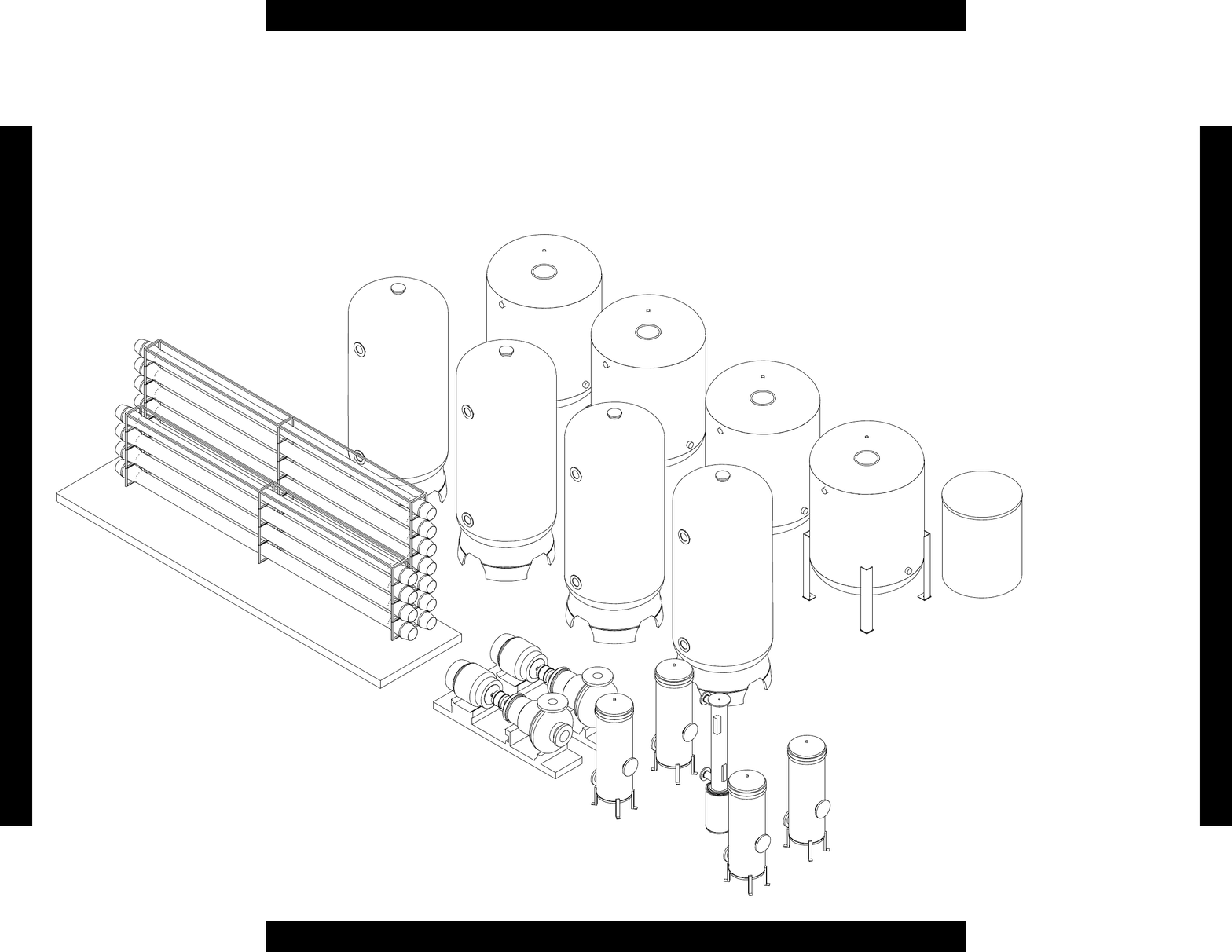}
\end{minipage}
\begin{minipage}{0.80\textwidth}
\includegraphics[width=\textwidth]{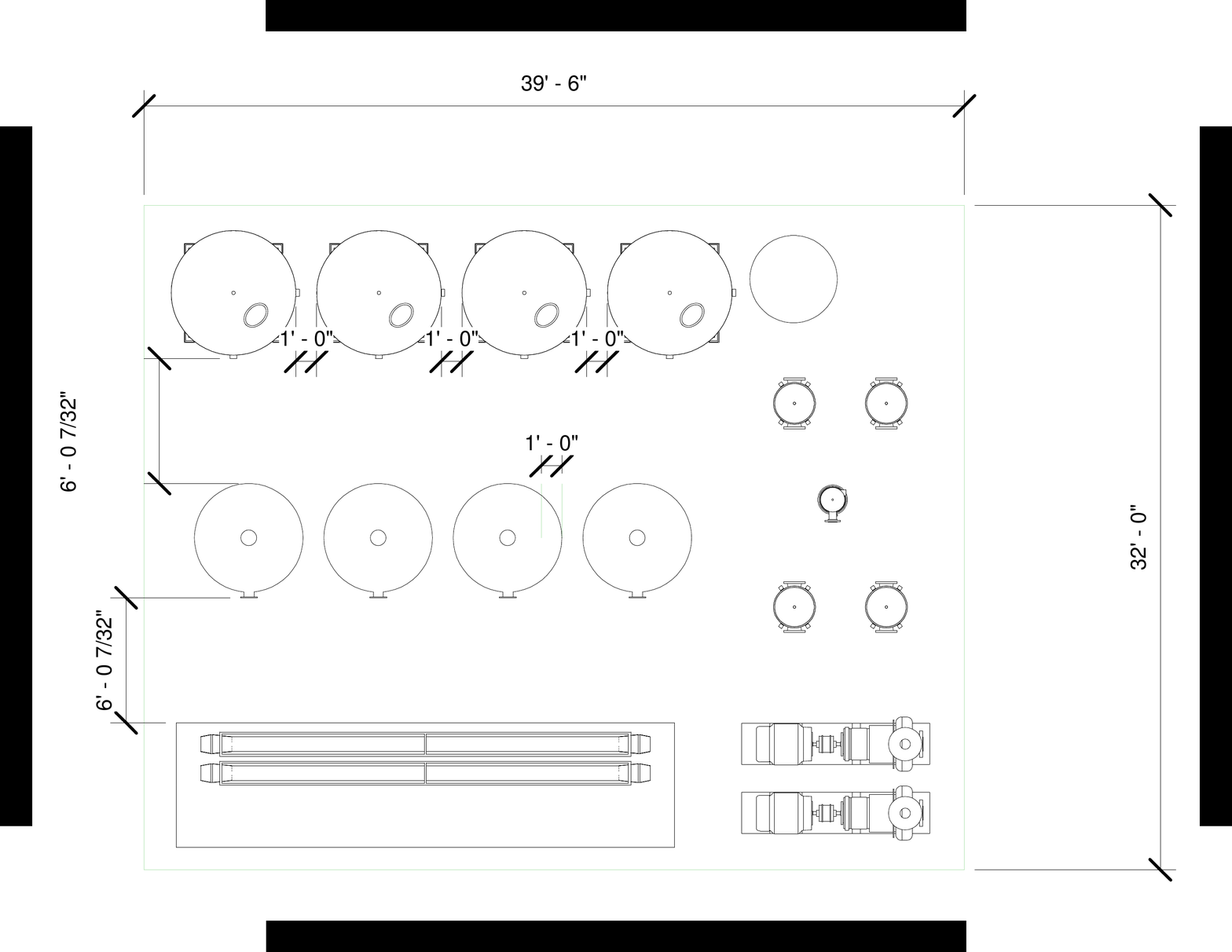}
\end{minipage}
\end{center}
\caption{Physical layout of a water purification system containing pretreatment equipment (carbon filters, water softeners, micron filtering), reverse osmosis unit and post treatment (pumps, UV sterilizer, sub-micron filters, deionization vessels).}
\label{hankfilter}
\end{figure}

Given a fiducial mass of \unit[27]{kton}, and a total mass of \unit[39]{kton}, we can scale the water system requirements from past detectors using total volume and surface area considerations.  The scaling results in a required recirculation rate of about \unit[300]{gal/min}.  The \unit[300]{gal/min} system would fill and recirculate the \unit[39]{kton} in 24 days. Figure~\ref{hankfilter} shows an outline of a \unit[200]{gal/min} filtration system that could be scaled for this application.  To reduce the cost of civil construction, the system could be mounted in three, \unit[40]{foot}-long shipping containers using modified tanks.  Such a project has been implemented by the U.S. Navy.  

\begin{figure}[h]
  \begin{center}
    \begin{minipage}{0.60\textwidth}
      \includegraphics[width=\textwidth]{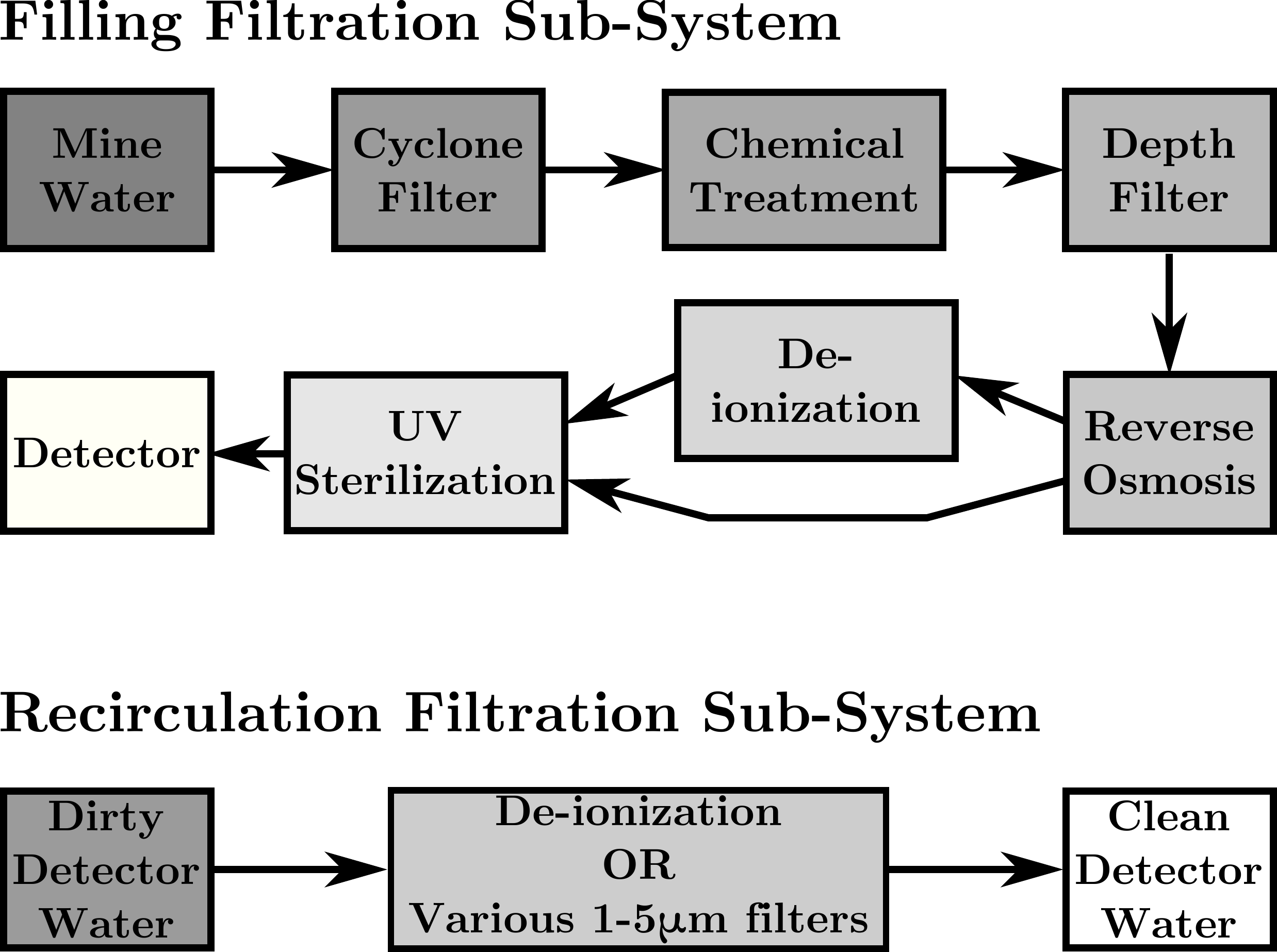}
      \caption{A sketch of the proposed water filtration systems that will fill and maintain the \chips{} detector.}
      \label{fig:filtration}
    \end{minipage}
  \end{center}
\end{figure}

The water system, shown schematically in Figure~\ref{fig:filtration}, will be used to both fill and recirculate the detector water.  The system will be used to fill the volume enclosed by the polymer liner initially, and then will be used to provide additional pure water as the detector leaks or evaporates over time.  Recirculation will be accomplished by bypassing the Reverse Osmosis stage of the system as described below.  Recirculation is necessary to maintain the high degree of purity while also eliminating ``dead zones'' in the detector volume where detrimental bacteria growth would normally take place. While bubbling of compressed air may be useful in deterring surface freezing in the winter, a water heating system may also need to be included during recirculation.  A preliminary and conservative budget, meant to be able to cover the cost for details that have not been included, is \$1.4M for the system, containers, and the internal and interconnect piping.

\paragraph{The Filling Filtration System}
As described for previous water Cherenkov detectors~\cite{SuperK,IMB,Grande}, we can expect that the water filtration system will include an initial depth and/or cyclone filter to remove contaminants down to a few $\mu {\rm m}$. Chemical treatment will then be implemented to remove undesired chemicals from the water. A reverse-osmosis (RO) stage will follow (possibly multiple stages - IMB used a 3-stage RO system), which will reduce the remaining particulate size down to less than \unit[0.01]{$\mu {\rm m}$}. A deionization (DI) filter may also be necessary following reverse-osmosis, but due to the high cost of deionization, this stage may be eliminated or only a portion of the water may pass through the DI filter. A UV sterilization stage will kill any bacteria. 

\paragraph{The Recirculation Filtration System}
The recirculation flow an bypass the reverse osmosis stage since the input water (coming from the detector) will already be quite pure. The remaining portion of the system will remove particulates down to \unit[0.2]{$\mu {\rm m}$} and remove substances that leach into the detector water from the detector materials themselves.  Cost may drive the final solution towards eliminating the deionization process, but this decision will be made later by the water purification contractor.

\section{\chips{} ND Concept}

\subsection{Physics Considerations}
The NuMI beam is not a pure \numu{} beam. It has a small inherent admixture of \nue{} that is an irreducible background to the \numue{} oscillation signal. In addition, neutral current \numu{} events, particularly those with a $\pi^{0}$ in the hadronic recoil system, can mimic the \numue{} oscillation signal.  \chips{} requires a Near Detector to study all neutrino interaction types before they have had a chance to oscillate and to provide understanding of the initial composition of the beam.  A Near Detector would study the neutrino-nucleus interactions in a high statistics environment close to the beam target and could also monitor the neutrino beam's performance. Beam monitoring is not a requirement given the array of detectors that already exist that monitor the NuMI beam.

With these goals in mind, a key issue of the Near Detector design is that it should be as similar as possible to the Far Detector design and material. Sufficiently similar Near and Far Detectors would allow one to use the same event reconstruction and particle identification in both detectors. This would minimize the systematic uncertainties in the predicted background at the Far site. A water \CER{} Near Detector design is preferred to maximize the benefits.  It provides the same neutrino-interaction target, ensuring that the efficiencies for signal and background events are similar in each detector.  Such a detector would be low cost per ton and utilize a well known technology; however, it should be noted that a water Cherenkov detector has never been proven in high intensity environments. Deployment in the NuMI beam would represent such a scenario.

\subsection{Shape and Size Considerations}
The challenges related to containment and multiplicity of the neutrino interactions drive the specifications of a water \CER{} Near Detector design. On one hand, the detector needs to span enough radiation lengths for a developing electromagnetic shower to be identified.  The design must also allow for the separation photon rings with the ring identification algorithms that are used for the event reconstruction in the Far Detector. On the other hand, if the detector is too large then event pile-up and high rock event overlap rates will incur high dead times, significantly reducing statistics. 

The current location being considered for the Near Detector placement is \unit[100]{m} underground on the Fermilab site upstream of the MINOS and MINERvA detectors. For reference, the MINOS Near Detector is situated \unit[500]{m} downstream of the end of the decay pipe. This is a compromise between the cost of digging a new cavern, finding available space in the heavily-congested cavern in front of the NuMI beam, and trying to achieve the necessary physics, driven by the flux requirements. 

The MINOS Near Detector was used to estimate the expected event overlap rate. It has a fiducial mass of approximately \unit[24]{tons}, with a total mass of about \unit[1]{kton}. In a \unit[10]{${\rm \mu s}$} beam spill, 5.6 events are expected in the fiducial volume compared to a total of 35 events in the whole detector. A simple study found that 14.8\% of neutrino interactions inside the fiducial volume had additional activity in the fiducial volume within a time window of \unit[50]{ns} from other neutrino interactions.  These other neutrino interactions occur either in the rock or in the non-fiducial part of the detector. This fairly large event overlap rate demonstrates the importance of keeping the CHIPS Near Detector as small as possible in order to minimize overlaps and cost.
 
\subsection{Design Possibilities} 
\begin{figure}[h]
  \centering
  \includegraphics[width=0.5\textwidth]{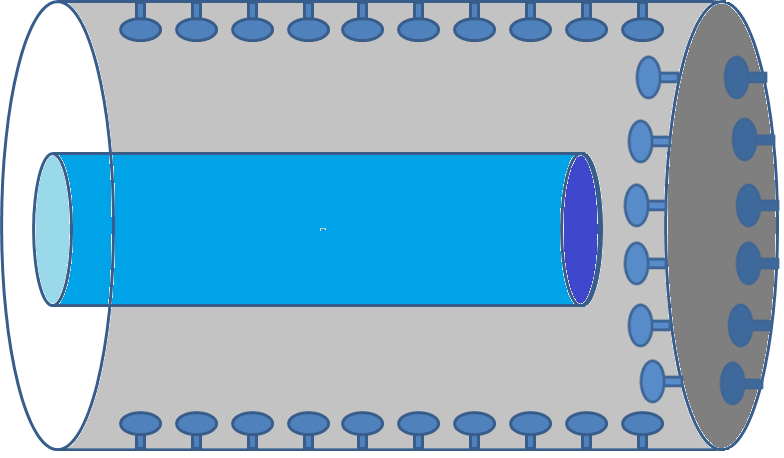}
  \caption{A conceptual design of the \chips{} Near Detector.}
\label{ndfig}
\end{figure}

Three ND options are being explored.  One design under consideration is a thin side-on inner cylinder (IC) of radius \unit[0.5-1]{m} and of length $\sim\,$\unit[4]{m}, which encloses a $\sim$\unit[12.5]{ton} volume. The IC would be filled with water to serve as the water target and would be contained and supported along the center of a larger side-on light-tight outer cylinder (OC). The readout PMTs would be instrumented along the sides and back face of the inside of the OC.  This option is illustrated in Figure~\ref{ndfig}.

Another possible design would entail a larger water detector using photodetectors with significantly better time resolution ($\sim$\unit[100]{ps}) and finer granularity ($\sim$\unit[1]{cm}) than the standard PMTs to be used in the Far Detector~\cite{pdwork,Sanchez2012525}. The better time sampling of the showers can help in both enhancing the particle identification in a small volume as well as overcoming high overlap rates. Furthermore, the finer granularity can help mitigate the deadtime issues as each channel could be read out independently. These enhancements would allow a single volume of water with the size driven by space and cost constraints. The sampled bins in time and space for each interaction can then be merged to simulate the geometry and granularity at the Far Detector. 

The third possible option sacrifices the idea of a common particle ID between the Near and Far Detectors.  Instead, it utilizes a combination of existing MINOS+ and MINERvA detectors with a water target to study neutrino-nucleus interactions on water.

\section{Data Acquisition}
The Data Acquisition (DAQ) for the \chips{} detector is designed with the aim of being flexible enough to be cost effective enough for the full detector but also simple enough to implement for an R\&D stage. These disparate aims can be accommodated within a modular solution based around the MicroTCA crate, which is rapidly becoming the industry standard~\cite{microtca}. The principle advantage of this modular approach is that in the early stages (low channel count), off-the-shelf components can be used, then be replaced with custom boards for the final (high channel count) detector.  The system follows a fairly typical design featuring front-end boards, digitizers, and FPGAs for triggering and digitizer readout, and fiber optic links to transfer the triggered data to an on-line CPU farm for full event building.  A summary of the requirements and channel count is provided in Table~\ref{tab:daq}.

\begin{table}[ht]
\centering
\begin{tabular}{|c c c|}
\hline
Item & Initial System & Large Detector Module \\
\hline
Number of photodetectors &   2-128 & 13,000 \\
Number of MicroTCA crates & 1 & 17 \\
Used slots per crate & 2 & 12 \\
Channels per ADC FMC  & 32 & 32 \\
Sampling speed & \unit[125-1024]{MHz} & \unit[125-1024]{MHz} \\
Bits per ADC & 8-12 & 8-12 \\
\hline
\end{tabular}
\caption{Summary of electronics, DAQ requirements, and channel counts.}
\label{tab:daq}
\end{table}

\subsection{Front End Electronics}
The front end electronics will handle the output signal of the photodetectors. The exact design of the front-end board will necessarily depend on the final choice of photodetector, although in general terms the board will contain a preamp to amplify the signal before digitization, a discriminator to provide a digital signal for triggering and timing, and a shaper (if required).  Several particle physics experiments have developed application-specific integrated circuits (ASIC) which perform all three of these tasks in a single package. The final choice of photodetectors will determine whether these existing ASICs could be used in the \chips{} front end electronics. The amplified and shaped signal is then sampled and digitized, either directly via a high speed analog-to-digital converter (ADC) or using a switched capacitor array (SCA) and lower speed ADC.

\subsection{Digitization}
During the early stages of the experiment when only a handful of PMTs will be read out, the digitizer will be an off-the-shelf solution. In the reference design, this is a multichannel FPGA mezzanine card sitting on a carrier board in the MicroTCA crate. Several vendors sell suitable crates, carrier boards, and ADC mezzanine cards, some of which are already in use at CERN, DESY and other laboratories.  Ultimately, the only way to cost effectively instrument a 10,000+ channel detector is to replace the off-the-shelf high speed ADC components with custom, in-house designed boards. There are several possible high speed sampling or digitizing ASICs on which the electronics could be based, including the IRS/TARGET family of chips from the University of Hawaii~\cite{Bechtol:2011tr} or the DRS family from PSI~\cite{Ritt:2004gm}. 

\subsection{Clock, Control and Triggering}
The global time references will be \unit[10]{MHz} and pulse per second signals distributed to the MicroTCA crates from a GPS receiver. Existing experiments have demonstrated channel synchronization is possible to better than the \unit[1]{ns} level.

There are two distinct levels of triggering in the reference design: the local trigger that determines when the signal from a given photodetector is digitized, and the global event trigger which determines when digital data from the digitization modules is transferred to the offline.  Simulation studies are currently ongoing to determine the optimal local (single channel vs. single string vs. logical channel group) and global trigger conditions. The local trigger will be implemented in the FPGA of the ADC carrier board, the global event trigger will either be implemented in the FPGA, or if rate permits, in the CPU farm.

\subsection{Event Readout and Storage}
Triggered event data and small amounts of housekeeping information will be transferred from the MicroTCA crates to a Linux CPU farm via standard optical Ethernet links.  High level software triggers can be run in the CPU farm to reduce the data rate further if necessary.  

\section{Calibration}
The concept for monitoring and calibration of the photodetectors for \chips{} is based on the light-injection system currently being deployed for SNO+, which in turn builds on systems used successfully for Double Chooz and MINOS~\cite{Adamson:2002ze}. The SNO+ system uses \unit[50]{m} long poly-methyl methacrylate (PMMA) and quartz fiber optic cables to route LED and laser light (respectively) into the detector from the deck above the detector. The detector-ends of each of the 92 fibers are mounted on the PMT support structure in SNO+ and the light shines all the way across the detector to illuminate the PMTs 18~m away on the opposite side. Controlled pulses of light with between 1000 and 1,000,000 photons are injected.  The SNO+ system is capable of providing data for PMT timing and gain calibration as well as measurements of scattering and attenuation monitoring. Adapting the SNO+ design and scaling the dimensions up to match those envisioned for \chips{} is expected to be straightforward.  A similar design was proposed for the LBNE-WCD option that also included a light-diffusing ball located near the center of the water volume\cite{LBNECDR}.  As proposed in the LBNE CDR, energy and vertex calibration can be performed using naturally occurring events in the detector such as cosmic muons or Michel electrons.

\section{Simulation and Reconstruction tools}
\label{simsec}
While the initial physics reach of \chips{} was established using \globes{}, a program is already underway to develop a full simulation of the beam and detector and a full ring reconstruction protocol.  This program draws from extensive work on simulation of WC detectors (WCSim) and will be leveraged to determine the optimal geometry and photodetector coverage for a massive, cost-effective water Cherenkov detector.  

\subsection{Beam Simulation}
\label{fluxsec}
\begin{figure}[h]
  \centering
  \includegraphics[width=0.6\textwidth]{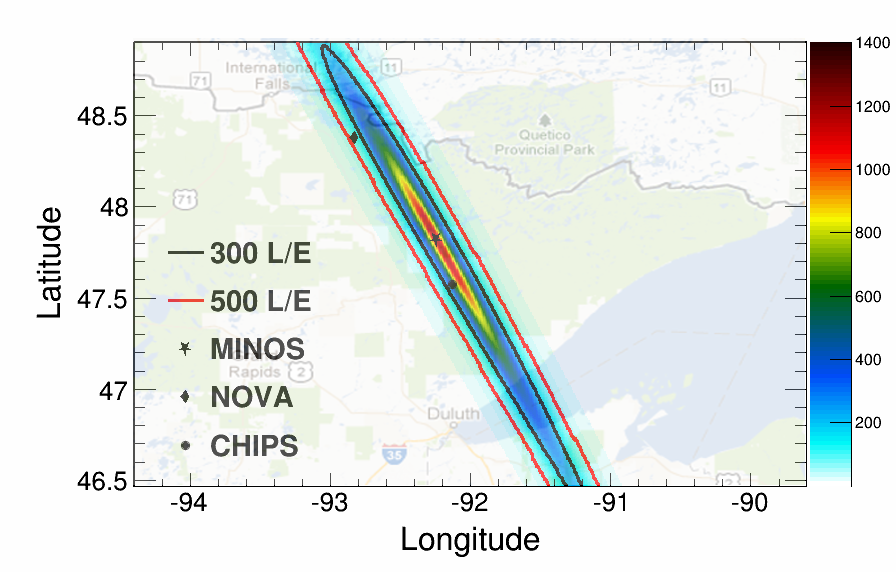}
  \caption{A map of potential neutrino event rates, assuming no oscillations, between \unit[0-30]{GeV} for an exposure of \unit[1]{kton-year}. Contours show lines of constant L/E where L is the distance from the hypothetical detector to the NuMI target and E is the peak energy of the reweighted neutrino spectrum}
\label{eventratemesabi}
\end{figure}

\begin{figure}[h]
\centering
\begin{minipage}{0.49\textwidth}
\includegraphics[width=\textwidth]{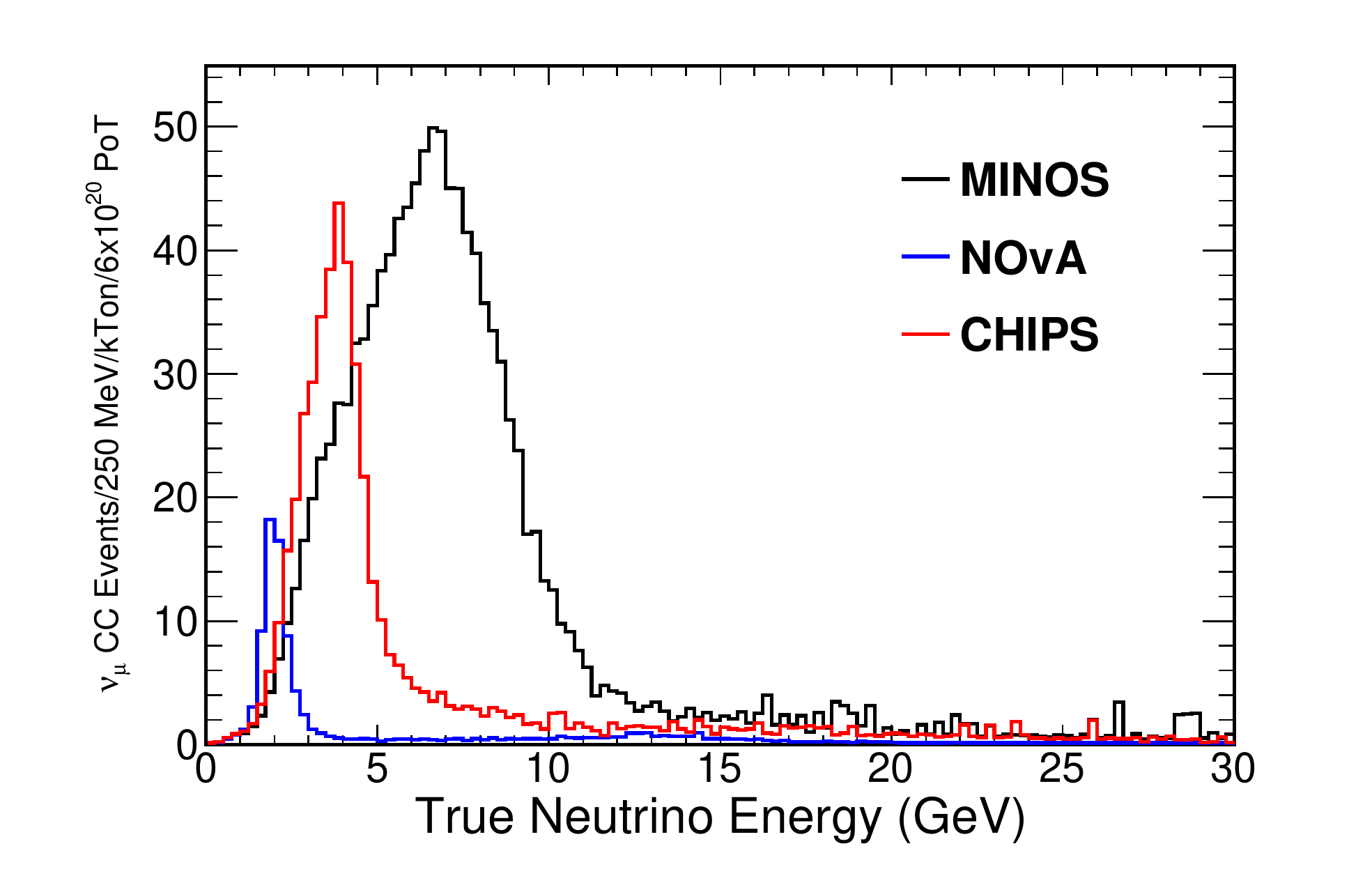} 
\end{minipage}
\begin{minipage}{0.49\textwidth}
\includegraphics[width=\textwidth]{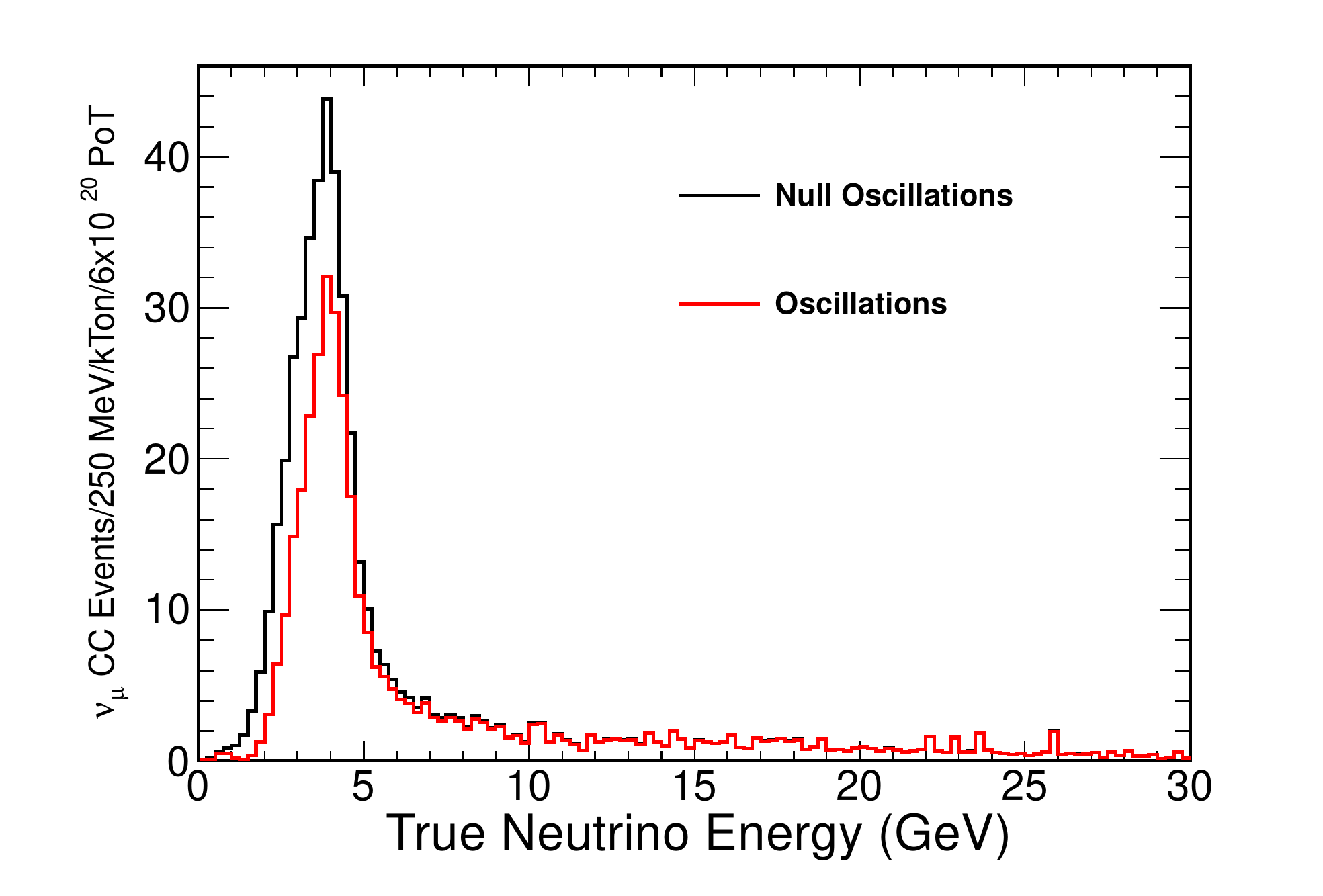}
\end{minipage}
\caption{(Left) True energy distribution of \numu{}-CC events at the MINOS, \nova{}, and \chips{} far detector locations, assuming no oscillations and \unit[1]{kton-year} of exposure. (Right) The true energy distribution of \numu{}-CC events that would be seen at \chips{} in one kiloton year with (red) and without (black) neutrino oscillations.}
\label{enufig}
\end{figure}

The MINOS, \nova{} and MINERvA experiments each have extensive simulations of the NuMI beam. By taking advantage of the fact that neutrino production from decaying hadrons is isotropic in the center of mass frame, and that the existing simulations store neutrino parent information, we can reweight the existing MC to give a neutrino flux at any location in the beam~\cite{Milburn}.  Furthermore we scan over a region of interest to construct a map of flux characteristics such as peak energy. Cross section and oscillation information can also be included to give a clear, intuitive impression of the oscillation sensitivity at a given location.  Figure~\ref{eventratemesabi} shows the computed \numu{}-CC event rate integrated over all energies for various locations in northern Minnesota.  Figure~\ref{enufig} shows the predicted \numu{}-CC event energy spectra at different detector locations in northern Minnesota.

\subsection{Detector Simulation}
\begin{figure}[h]
  \centering
  \includegraphics[width=0.6\textwidth]{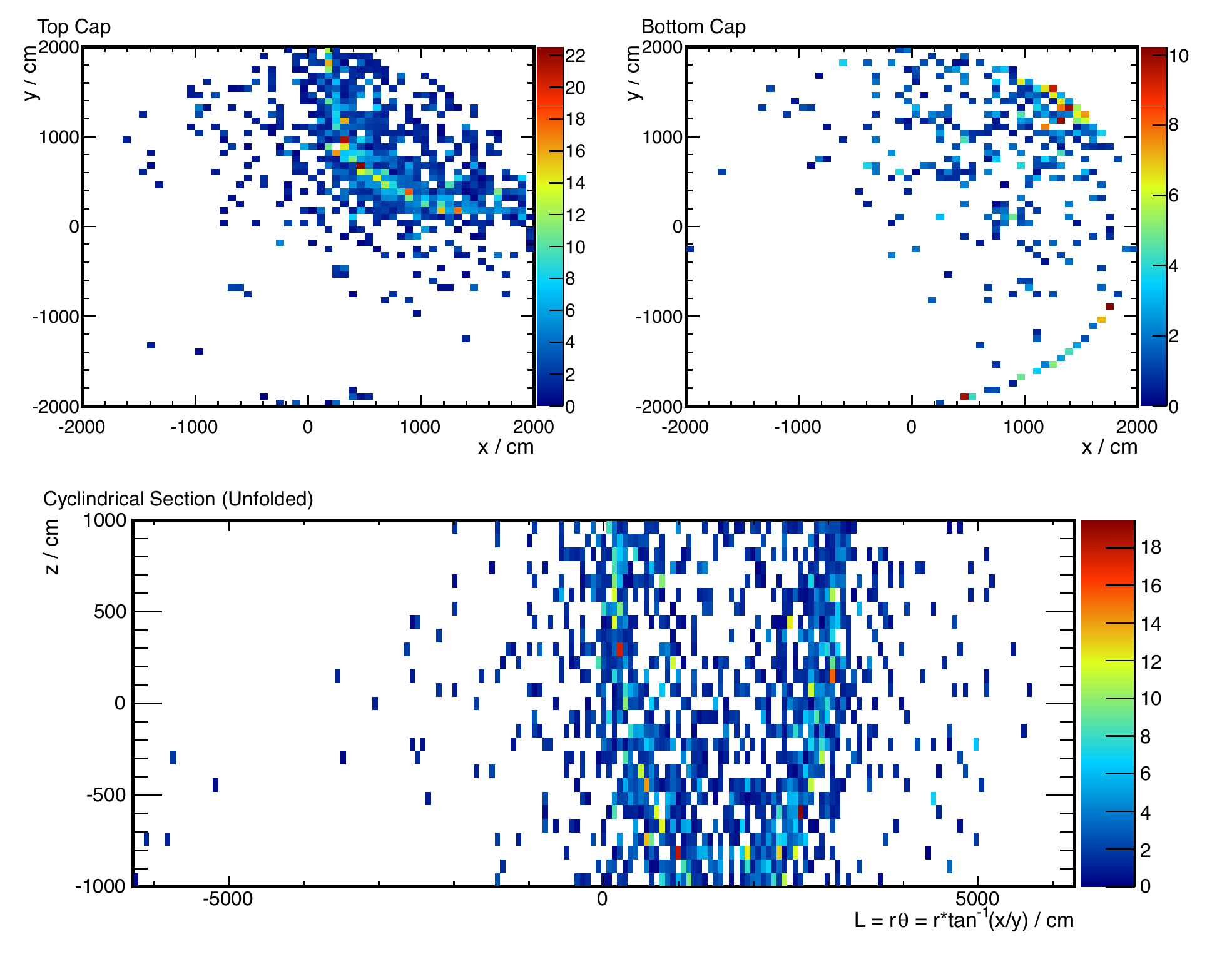}
  \caption{An event display of a $1.6\,$GeV CC $\nu_e$ interacting in the center of the detector. The top endcap (left) and bottom endcap (right) views are shown above the larger unfolded cylindrical section. Each bin shows a single PMT and the color shows the collected charge in PE.}
\label{fig:chipsNuE}
\end{figure}

The detector simulation is performed by a GEANT4~\cite{GEANT4} simulation called WCSim.  WCSim was developed to study water \CER{} detector options for the LBNE project. The simulation outputs a list of hits from PMTs.  An initial \chips{} geometry was added to WCSim describing a cylindrical detector of radius \unit[20]{m} and height \unit[20]{m}. It is instrumented with 10\% coverage using \unit[10]{\inch} HQE PMTs. Figure~\ref{fig:chipsNuE} shows a \unit[1.6]{GeV} CC $\nu_e$ interaction generated with the GENIE~\cite{GENIE} event generator occurring at the center of the detector. The fuzzy ring is the typical signature of an electron in a water \CER{} detector.

\subsection{Reconstruction}
\begin{figure}[h]
  \centering
  \includegraphics[width=\textwidth]{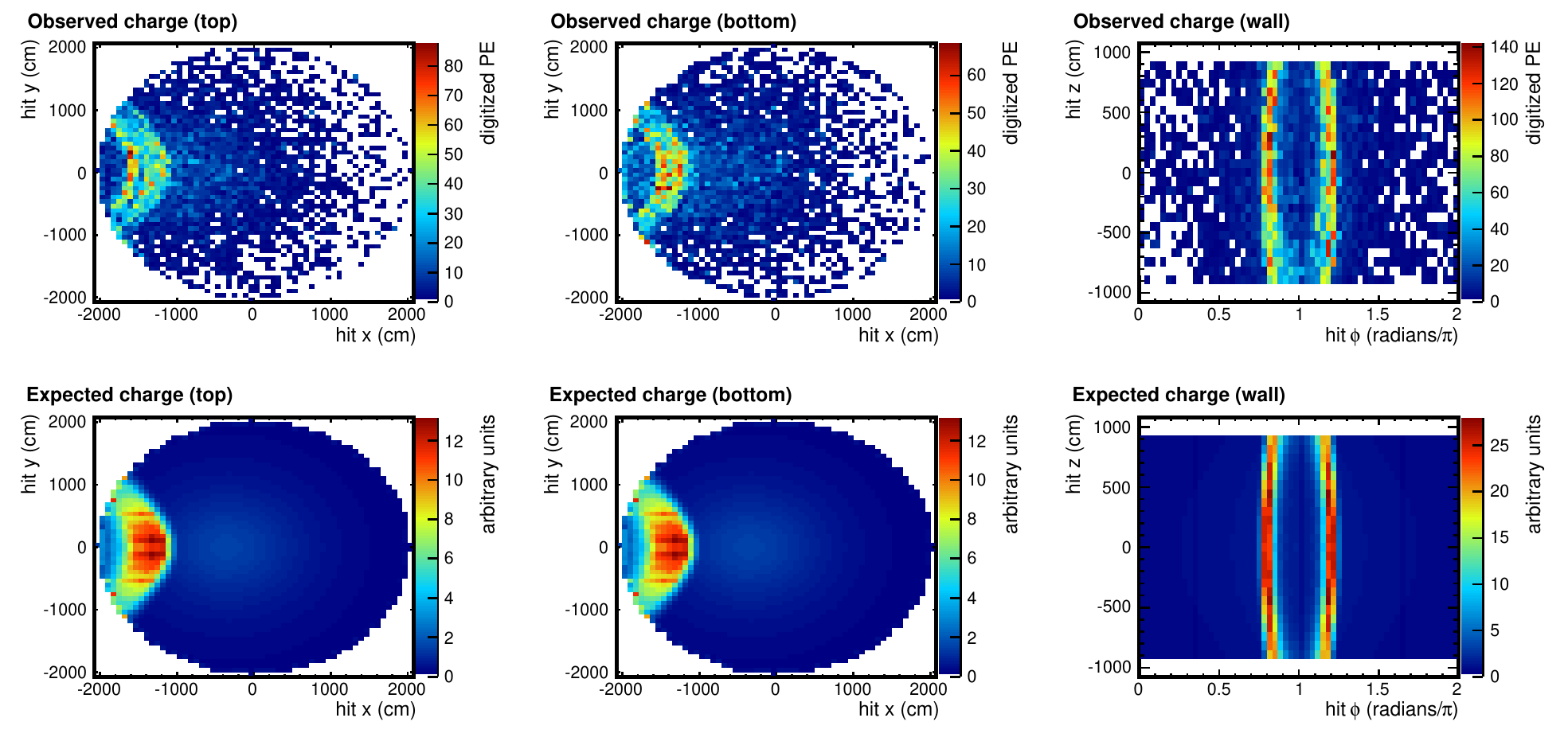}
  \caption{Comparison of the measured (top) and expected (bottom) charge distributions, for the top (left) and bottom (center) endcaps, and the unfolded cylinder wall (right). The distributions are for a muon track with 1.5 GeV of kinetic energy, created at the center of a 20 m radius by 20 m high cylindrical detector, propagating along the x axis towards the curved wall of the cylinder. The units of the measured charge are digitized photoelectrons, while the predicted charge is in arbitrary units.}
\label{chidist}
\end{figure}

A major goal of the reconstruction work is to find an optimized HQE photodetector number and layout.  The planned reconstruction method is based on an algorithm developed for the MiniBooNE experiment~\cite{MiniBooNE,rbpthesis}, modified to remove the scintillation light component. The algorithm generates a likelihood for each PMT to register a given charge at a given time, for a pre-defined set of track parameters. Minimizing this likelihood with respect to the track parameters provides the reconstructed track objects. The method is readily extendable to multiple tracks such as those from NC $\pi^{0}$ decays.  Figure~\ref{chidist} shows an example of the expected and observed charge distributions that go into forming this likelihood.  A version tested on Super-K reported a 60\% reduction~\cite{SuperK} in the NC background compared to the standard ring reconstruction method.  This improvement is not incorporated into the \globes{} physics reach calculations.

A preliminary implementation of the algorithm is already in development. The charge component of the likelihood is determined by combining the probability for a propagating particle to emit light in the direction of the PMT with the probability for this light to reach the PMT and produce a recorded signal.  This depends on the particle type and its emission profile, the geometry of the track and PMT, detector properties such as the absorption and scattering of light, and the effects of digitization at the PMT.

\begin{figure}[h]
  \centering
  \includegraphics[width=0.6\textwidth]{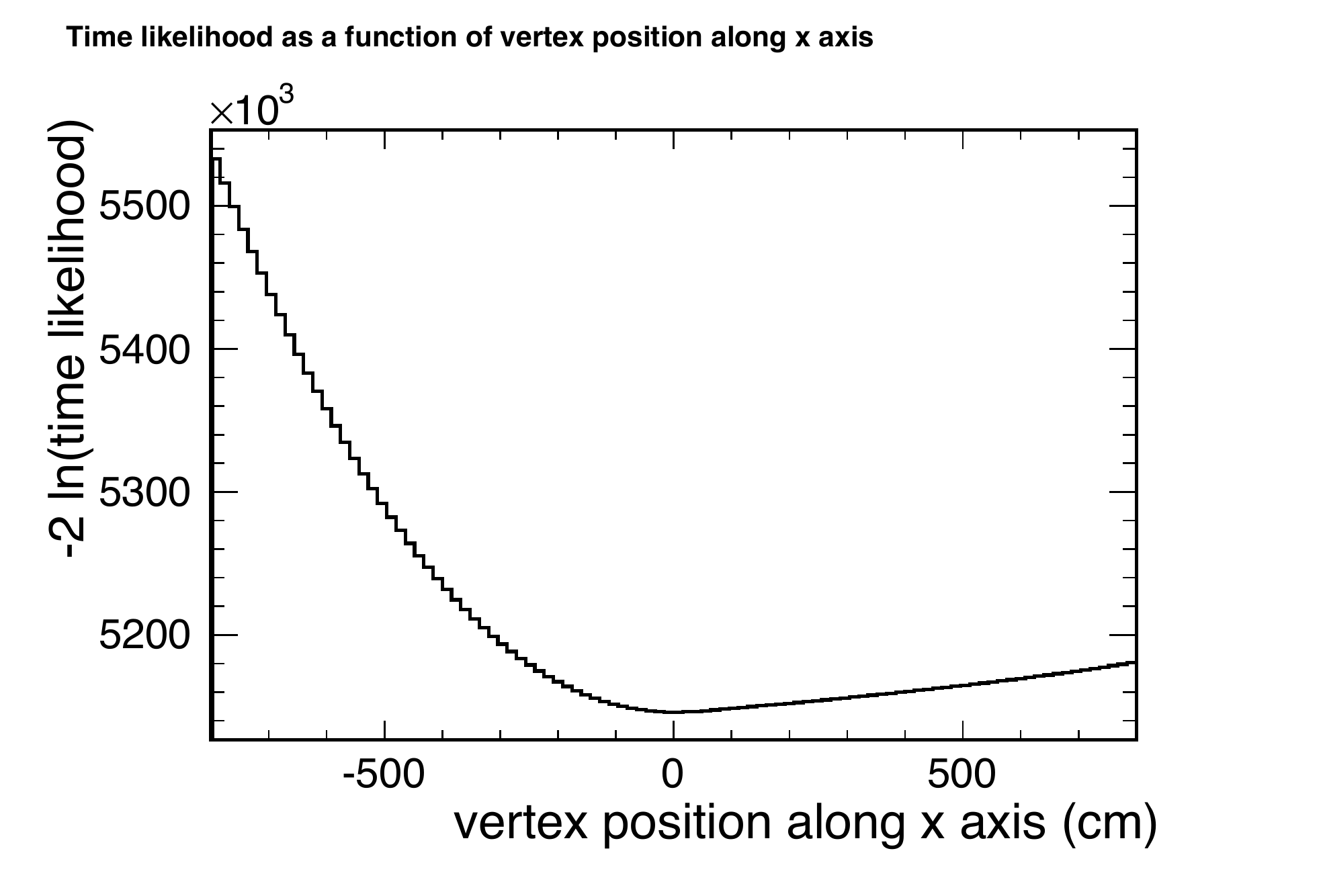}
  \caption{Example of a one dimensional time likelihood.  The simulated event is a muon with 1.5 GeV of kinetic energy, created at the center of a \unit[20]{m} radius by \unit[20]{m} high cylindrical detector, propagating along the x axis, towards the curved wall of the cylinder.  The likelihood is plotted against the vertex position along the x axis, assuming a \unit[1.5]{GeV} muon track for which all other parameters are known.}
\label{timedist}
\end{figure}

To calculate the time likelihood, the registered PMT hit times are corrected by subtracting the expected hit time of a photon emitted at the mid-point of the candidate track, and a series of fits are performed on the resulting distribution.  First, the distributions are separated into bins of charge, and a fit is carried out using a Gaussian (to model direct Cherenkov light) plus a Gaussian convolved with an exponential (to model indirect scattered light).  Polynomial fits are then used to determine these fit parameters as a function of predicted charge, and further fits are performed to express these coefficients as a function of energy, allowing the time likelihood to be calculated for arbitrary track energy and PMT charge combinations.  The overall log likelihood surface is produced by adding the charge and time surfaces.  Figure~\ref{timedist} shows an example of the time likelihood.

\section{R\&D Program} 
\label{randdsection}
This LOI also outlines a path of development towards a cost effective, \unit[100]{kton} water Cherenkov detector in a neutrino beam.  A \$10M investment over the next 4-5 years could provide a \unit[10]{kton} detector in the NuMI beam which would improve the \dcp~reach of \nova{} substantially.  The cost per kton for the initial prototype would be \$1-2M/kton, whereas the goal, over the ensuing decade, would be to reduce this cost by up to an order of magnitude, chiefly by reductions in photodetector costs.  

An R\&D proposal will be submitted to funding agencies this fall.  There are a number of issues which need to be tested on a smaller scale before the full \unit[10]{kton} prototype can be deployed.  While the PMTs, HV and readout are all reasonably well developed, and the purification plant technology is well understood, the detector structure needs to be prototyped in order to produce a full conceptual design for a \unit[10]{kton} fiducial volume detector.  The R\&D program is summarized in the Table~\ref{randdsched}, but the main issues which need to be resolved in the first years are: 

\begin{itemize}[itemsep=-1mm]
\item{Verify liner construction}
\item{Select materials for Wentworth pit water} 
\item{Design liner support structure} 
\item{Plan layout of purification pipes}
\item{Verify Ice Defense Systems}
\item{Measure magnitude of deep currents}
\item{Verify layout of cables and readout of PMT signals}
\item{Design modular support structure for PMT housings}  
\item{Design generic acrylic PMT pressure housing}
\end{itemize}

\noindent Once these critical path items have been developed, building on previous work carried out for LBNE where possible, the full \unit[10]{kton} prototype detector could be built in one season, but procurement of the PMTs is likely to be the item which dominates the build time.

Some work has already started. The University of Minnesota Duluth Large Lakes Observatory group have already deployed instruments in the Wentworth pit to monitor the deep currents before the winter. The design of the purification system has already been designed by the South Coast Water company in Santa Ana, CA, a company with significant expertise in water Cherenkov purification systems.  An engineering consultant firm based in Minnesota, Barr Engineering (http://barr.com), is currently assisting in the design and implementation of the \chips{} detector.  It is already clear that Barr Engineering is capable of handling all of the engineering aspects of the \chips{} detector. They have extensive experience designing projects for water-filled mines and hazardous waste treatment in northern Minnesota~\cite{PolyMet2012}. This experience is well-aligned with the requirements for water system design for the \chips{} detector.

Barr Engineering has also completed numerous projects involving fabricating and binding large-scale geotextiles in landfill and mining applications, including storm water structures, pile revetment walls, and other water-related structures. They have significant experience designing support structures such as dams, retaining walls/structures, and bridges~\cite{FloodManage, MouseRiver, TroutBrook}. Their expertise with such structures will be invaluable when designing and implementing the support structure for the \chips{} detector. Barr Engineering is also capable of supplying structures on the shore for housing the data acquisition, power, and water filtration systems.  These structures would either be prefabricated structures from Barr Engineering (if adequate models were available) or the work would be subcontracted out to one of Barr's partners.

Initial work has already been carried out to create a CAD model of the conceptual design to aid future engineering efforts. Additionally, basic flow calculations have been evaluated to approximate the forces induced by the flow of the lake water around the outer surface of the detector. These calculations have shown that the forces induced, though somewhat large due to the large scale of the detector, will be easily manageable with a traditional steel cable system.

\section{Summary}
The \chips{} concept outlined in this Letter of Intent could represent a step change in our ability to make precision neutrino measurements using the FNAL intense neutrino beams planned for the near and further future.  A \unit[100]{kton} fiducial mass \chips{} in NuMI would provide a $\sim$12-25$^{o}$ accuracy on $\delta_{CP}$ and an increase in the mass hierarchy reach of a factor 2, in combination with \nova{} and T2K.  
As an ultimate goal, the \chips{} detector could be redeployed off-axis in the LBNE beam line, to complement the on-axis Liquid Argon detector, enabling results on a faster timescale than presently expected.  

\section*{Acknowledgements} 
The authors are grateful for insightful discussions with Albrecht Karle, Jim Haugen, Perry Sandstrom (UW) and Terry Benson (PSL).  Further thanks to Hank Sobel for sharing his expertise on large water Cherenkov detectors and water purification systems and to Profs. Jay Austin and Liz Austin-Minor (UMD) for water data acquisition and analysis.  Finally, thanks to Cliffs Natural Resources for data on the mine pit water and for access to the site for suitability studies.
 
\clearpage

\begin{landscape}
\begin{table}[p]

{\scriptsize
\begin{tabular}{|p{0.7in}ccp{1.7in}p{1.7in}p{2.0in}|}
\hline
\centering{Dates}&Funding&Source&\centering{Design}&\centering{Procurement}&Actions\\
\hline

\begin{minipage}[t]{0.7in} 
\centering Year 0\\Present-\\Dec. 2014
\end{minipage}&\$100k&Universities&&&
\begin{minipage}[t]{2.0in}
\begin{itemize}[itemsep=-1.5mm,leftmargin=0cm,itemindent=0cm,labelwidth=\itemindent,labelsep=0cm]
\item[-]{Measure deep currents (UM Duluth)}
\item[-]{Procure and deploy small circular floating dock}
\item[-]{Construct 5mx5m cylindrical liner}
\item[-]{Construct 5mx5m cylindrical structural support}
\item[-]{Retrofit 10 Hamamatsu \unit[10]{in} PMT, Ice Cube DOMs}
\item[-]{Procure DAQ crate and off-shelf module for 10 channels}
\item[-]{Procure and deploy small purification system in on-shore shipping container}
\item[-]{Monitor cosmic rate over winter}
\end{itemize}
\end{minipage}\vspace{0.8mm}\\
\hline

\begin{minipage}[t]{0.7in} 
\centering Year 1\\Jan. 2014-\\Dec. 2014
\end{minipage}&\$100k&FNAL&
\begin{minipage}[t]{1.7in}
\begin{itemize}[itemsep=-1.5mm,leftmargin=0cm,itemindent=0cm,labelwidth=\itemindent,labelsep=0cm]
\item[-]{Design 10kt structure/liner}
\item[-]{Design generic acrylic surround for future photon detectors}
\item[-]{Design DAQ system}
\end{itemize}
\end{minipage}&&
\begin{minipage}[t]{2.0in}
\begin{itemize}[itemsep=-1.5mm,leftmargin=0cm,itemindent=0cm,labelwidth=\itemindent,labelsep=0cm]
\item[-]{Produce CHIPS-10kt Conceptual Design Report}
\end{itemize}
\end{minipage}\vspace{0.8mm}\\
\hline

\begin{minipage}[t]{0.7in} 
\centering Year 2\\Jan. 2015-\\Dec. 2015
\end{minipage}&\$1M&DOE/NSF/STFC&&
\begin{minipage}[t]{1.7in}
\begin{itemize}[itemsep=-1.5mm,leftmargin=0cm,itemindent=0cm,labelwidth=\itemindent,labelsep=0cm]
\item[-]{Procure fraction of purification system}
\item[-]{Procure liner and structure}
\item[-]{Procure floating dock and support cables}
\item[-]{Procure DAQ}
\item[-]{Procure 10-20 PMTs+housing}
\end{itemize}
\end{minipage}&
\begin{minipage}[t]{2.0in}
\begin{itemize}[itemsep=-1.5mm,leftmargin=0cm,itemindent=0cm,labelwidth=\itemindent,labelsep=0cm]
\item[-]{Raise and inspect liner, structure}
\item[-]{Deploy full size liner, structure, purification plant}
\item[-]{Deploy previous PMTs}
\item[-]{Commission DAQ, monitor beam}
\item[-]{Monitor attenuation with LED system}
\end{itemize}
\end{minipage}\vspace{0.8mm}\\
\hline

\begin{minipage}[t]{0.7in} 
\centering Year 3\\Jan. 2016-\\Dec. 2016
\end{minipage}&\$2M&DOE/NSF/STFC&
\begin{minipage}[t]{1.7in}
\begin{itemize}[itemsep=-1.5mm,leftmargin=0cm,itemindent=0cm,labelwidth=\itemindent,labelsep=0cm]
\item[-]{Consider new ideas for light detection}
\end{itemize}
\end{minipage}&
\begin{minipage}[t]{1.7in}
\begin{itemize}[itemsep=-1.5mm,leftmargin=0cm,itemindent=0cm,labelwidth=\itemindent,labelsep=0cm]
\item[-]{Procure 500 PMTs (baseline design)}
\item[-]{Procure second tranche of purification system}
\item[-]{Procure housings for PMTs}
\end{itemize}
\end{minipage}&
\begin{minipage}[t]{2.0in}
\begin{itemize}[itemsep=-1.5mm,leftmargin=0cm,itemindent=0cm,labelwidth=\itemindent,labelsep=0cm]
\item[-]{Raise and inspect liner and structure}
\item[-]{Deploy 500 PMTs, purification plant}
\item[-]{Commission readout, monitor beam}
\item[-]{Reconstruction development and signal identification (handful of events expected)}
\item[-]{Develop Calibration System}
\end{itemize}
\end{minipage}\vspace{0.8mm}\\
\hline

\begin{minipage}[t]{0.7in} 
\centering Year 4\\Jan. 2017-\\Dec. 2017
\end{minipage}&\$2-3M&DOE/NSF/STFC&
\begin{minipage}[t]{1.7in}
\begin{itemize}[itemsep=-1.5mm,leftmargin=0cm,itemindent=0cm,labelwidth=\itemindent,labelsep=0cm]
\item[-]{Consider new ideas for light detection}
\item[-]{Produce TDR for 25kT FV CHIPS full size modules}
\end{itemize}
\end{minipage}&
\begin{minipage}[t]{1.7in}
\begin{itemize}[itemsep=-1.5mm,leftmargin=0cm,itemindent=0cm,labelwidth=\itemindent,labelsep=0cm]
\item[-]{Procure \$2.5M PMT equivalent units (baseline 1000 tubes)}
\item[-]{Procure third tranche of purification system}
\item[-]{Procure housings for PMTs}
\end{itemize}
\end{minipage}&
\begin{minipage}[t]{2.0in}
\begin{itemize}[itemsep=-1.5mm,leftmargin=0cm,itemindent=0cm,labelwidth=\itemindent,labelsep=0cm]
\item[-]{Raise and inspect liner and structure}
\item[-]{Deploy 1000 baseline PMT equivalent units and second section of purification plant}
\item[-]{Commission readout, monitor beam}
\item[-]{Reconstruction development and signal identification (2 handfuls of events expected with full bottom endcap)}
\item[-]{Calibration}
\end{itemize}
\end{minipage}\vspace{0.8mm}\\
\hline

\begin{minipage}[t]{0.7in} 
\centering Year 5\\Jan. 2018-\\Dec. 2018
\end{minipage}&\$2-3M&DOE/NSF/STFC&
\begin{minipage}[t]{1.7in}
\begin{itemize}[itemsep=-1.5mm,leftmargin=0cm,itemindent=0cm,labelwidth=\itemindent,labelsep=0cm]
\item[-]{Consider new ideas for light detection}
\end{itemize}
\end{minipage}&
\begin{minipage}[t]{1.7in}
\begin{itemize}[itemsep=-1.5mm,leftmargin=0cm,itemindent=0cm,labelwidth=\itemindent,labelsep=0cm]
\item[-]{Procure \$2.5M worth of photon detectors (1000 baseline units)}
\end{itemize}
\end{minipage}&
\begin{minipage}[t]{2.0in}
\begin{itemize}[itemsep=-1.5mm,leftmargin=0cm,itemindent=0cm,labelwidth=\itemindent,labelsep=0cm]
\item[-]{Raise and inspect liner and structure}
\item[-]{Deploy photon detectors on walls of cylinder and complete 10kt detector}
\item[-]{Commission readout, monitor beam}
\item[-]{Reconstruction development and signal identification (100 events expected)}
\item[-]{Calibration}
\item[-]{Physics Results}
\end{itemize}
\end{minipage}\vspace{0.8mm}\\
\hline

\end{tabular}
}
\caption{5 year research and development plan towards a \unit[10]{kton} detector in the Wentworth Pit.}
\label{randdsched}
\end{table}
\end{landscape}

\section{Appendix 1: Pit Water Content}
\label{waterapx}
Results of water tests are summarized in Table~\ref{tab:WaterQuality1} with 1$\sigma$ uncertainties.  Table~\ref{tab:WaterQuality2} summarizes additional results from the profile measurements.

\begin{table}[htbp]
\centering
\begin{minipage}{0.49\textwidth}
\begin{tabular}{|lr|}
\hline
pH    & 8.3 $\pm$ 0.29  \\
Total Hardness & \unit[400 $\pm$ 78]{mg/L} \\
Total Alkalinity & \unit[300 $\pm$ 57]{mg/L}  \\
Turbidity & \unit[$0.7 \pm 0.5$]{NTU}\\
Total Dissolved Solids &\unit[500 $\pm$ 100]{mg/L}\\
Total Suspended Solids &\unit[2.5 $\pm$ 1.4]{mg/L} \\
\hline
\end{tabular}
\end{minipage}
\begin{minipage}{0.49\textwidth}
\begin{tabular}{|lr|}
\hline
Sulfate & \unit[120 $\pm$ 25]{mg/L}  \\
Magnesium & \unit[70 $\pm$ 14]{mg/L}  \\
Calcium & \unit[43 $\pm$ 8.8]{mg/L}  \\
Chloride & \unit[37 $\pm$ 8.5]{mg/L}  \\
Potassium & \unit[13.5 $\pm$ 0.42]{mg/L}  \\
Strontium & \unit[150 $\pm$ 19]{$\mu$g/L} \\
Arsenic & \unit[1.7 $\pm$ 0.6]{$\mu$g/L} \\
Mercury & \unit[0.8 $\pm$ 0.21]{ng/L} \\
\hline
\end{tabular}%
\end{minipage}
\caption{Data from surface water quality tests conducted in the Wentworth Mine Pit with 1$\sigma$ uncertainties. Data courtesy of Cliffs Natural Resources.}
\label{tab:WaterQuality1}
\end{table}

\begin{table}[hbtp]
\centering
\begin{tabular}{|lrlr|}
  \hline
  Sodium&\unit[38]{mg/L}&Copper&$<$\unit[5]{$\mu$g/L}\\
  Chemical Oxygen Demand (COD)&\unit[10.3]{mg/L}&Molybdenum&\unit[4.3]{$\mu$g/L}\\
  Sulfide&$<$\unit[5.0]{mg/L}&Barium&\unit[2.63]{$\mu$g/L}\\
  Biochemical Oxygen Demand (BOD)&$<$\unit[2.4]{mg/L}&Nickel&$<$\unit[2]{$\mu$g/L}\\
  Total Organic Carbon (TOC)&\unit[1.7]{mg/L}&Chromium&$<$\unit[1]{$\mu$g/L}\\
  Total organic Nitrogen&$<$\unit[1]{mg/L}&Selenium&$<$\unit[1]{$\mu$g/L}\\
  Nitrogen as Total Kjeldahl&\unit[0.72]{mg/L}&Antimony&$<$\unit[0.5]{$\mu$g/L}\\
  Fluoride&\unit[0.57]{mg/L}&Lead&\unit[0.5]{$\mu$g/L}\\
  Gasoline Range Organic (GRO)&$<$\unit[0.4]{mg/L}&Tin&$<$\unit[0.5]{$\mu$g/L}\\
  Nitrogen as Nitrate+Nitrite&\unit[180]{$\mu$g/L}&Beryllium&$<$\unit[0.2]{$\mu$g/L}\\
  Strontium&\unit[132]{$\mu$g/L}&Cadmium&$<$\unit[0.2]{$\mu$g/L}\\
  Bromide&\unit[120]{$\mu$g/L}&Cobalt&$<$\unit[0.2]{$\mu$g/L}\\
  Boron&\unit[101]{$\mu$g/L}&Silver&$<$\unit[0.2]{$\mu$g/L}\\
  Phosphorus&\unit[100]{$\mu$g/L}&Thallium&$<$\unit[0.2]{$\mu$g/L}\\
  Nitrogen as Ammonia&$<$\unit[100]{$\mu$g/L}&Zinc &$<$\unit[6]{$\mu$g/L}\\
  Iron&$<$\unit[50]{$\mu$g/L}&Gross Alpha&\unit[3.5]{pCi/L}\\
  Surfactants&$<$\unit[40]{$\mu$g/L}&Radon&\unit[3.4]{pCi/L}\\
  Aluminum&$<$\unit[25]{$\mu$g/L}&Radium 226&\unit[0.26]{pCi/L}\\
  Lithium&$<$\unit[10]{$\mu$g/L}&Radium 228&\unit[0.08]{pCi/L}\\
  Manganese&$<$\unit[10]{$\mu$g/L}&Uranium&\unit[0.52]{pCi/L}\\
  Titanium&$<$\unit[10]{$\mu$g/L}&Total Fibers&$<$\unit[0.20]{million fibers/L}\\
  Vanadium&$<$\unit[10]{$\mu$g/L}&&\\ 
  \hline 
\end{tabular}%
\caption{Data from two broad-spectrum water quality tests conducted in 2011.}
\label{tab:WaterQuality2}
\end{table}

\clearpage

\bibliographystyle{unsrt}
\bibliography{chips}

\end{document}